\documentclass[12pt,titlepage,twoside]{diss}
\usepackage{amsmath}
\usepackage{amssymb}
\usepackage{amscd}

\usepackage{babel}

\usepackage{latexsym}
\usepackage{epsfig}
\usepackage{bbm}

\usepackage{winter}    % citation style

\newtheorem{defi}{Definition}[chapter]
\newtheorem{lemma}[defi]{Lemma}
\newtheorem{satz}[defi]{Theorem}
\newtheorem{cor}[defi]{Corollary}
\newtheorem{bem}[defi]{Remark}

\newtheorem{exempel}[defi]{Example}

\newtheorem{conj}[defi]{Conjecture}

\newcommand{\qed}{\hfill $\Box$}
\newcommand{\tr}{{\operatorname{Tr}\,}}
\newcommand{\id}{{\operatorname{id}}}
\newcommand{\supp}{{\operatorname{supp}\,}}
\newcommand{\bra}[1]{{\langle{#1}|}}
\newcommand{\ket}[1]{{|{#1}\rangle}}
\newcommand{\C}{{\mathbb{C}}}
\newcommand{\R}{{\mathbb{R}}}
\newcommand{\N}{{\mathbb{N}}}
\newcommand{\alg}[1]{{\mathfrak{#1}}}
\newcommand{\fset}[1]{{\mathcal{#1}}}
\newcommand{\Rho}{{\sf P}}
\newcommand{\Sym}{{\operatorname{Sym}}}
\newcommand{\1}{{\mathbbm{1}}}
\newcommand{\im}{{\operatorname{im}}}
\newcommand{\lcsupp}{{\operatorname{l.c.supp}}}

\newenvironment{expl}[1][{}]{\begin{exempel}[#1]\normalfont}{\end{exempel}}
\newenvironment{expl_noname}{\begin{exempel}\normalfont}{\end{exempel}}
\newlength{\blank}
\newenvironment{beweis}[1][{\hspace{-\blank}}]{{\noindent\emph{Proof\hspace{\blank}{#1}.\ }}}{\hfill $\Box$\vskip 0.5\baselineskip}

\renewcommand{\thechapter}{\Roman{chapter}}

\begin{document}

\pagestyle{myheadings}
\markboth{}{}

\addtocounter{tocdepth}{-1}
\addtocounter{secnumdepth}{-2}

% titelblatt

\title{\vspace{4cm}{\scshape\Huge
        Coding Theorems of\\
        Quantum Information Theory}}
\author{Andreas Winter
        \vspace{2cm}\\
        Dissertation zur Erlangung des Doktorgrades,\\
        vorgelegt der Fakult\"at f\"ur Mathematik,\\
        Universit\"at Bielefeld
        \vspace{3cm}}
\date{13. April 1999}
% begonnen am soundsovielten maerz 1999

\maketitle

\pagenumbering{roman}
\setcounter{page}{2}

% hier kommt abstract und dank auf einer seite:

\phantom{OMMMMMMM}
\thispagestyle{empty}
\vspace{1cm}
\centerline{\bf Abstract}
\vspace{1cm}
Coding theorems and (strong) converses for memoryless quantum communication
channels and quantum sources are proved: for the quantum source the
coding theorem is reviewed, and the strong converse proven. For classical
information transmission via quantum channels we give a new proof of the
coding theorem, and prove the strong converse, even under the extended
model of nonstationary channels. As a by--product we obtain a
new proof of the famous {\sc Holevo} bound.
Then multi--user systems are investigated, and the
capacity region for the quantum multiple access channel is determined.
The last chapter contains a preliminary discussion of some models of compression
of correlated quantum sources, and a proposal for a program to obtain
operational meaning for quantum conditional entropy. An appendix features
the introduction of a notation and calculus of entropy in quantum systems.

\vspace{1cm}
\centerline{\bf Acknowledgements}
\vspace{1cm}
This work grew out of an attempt to learn about quantum mechanics and
information theory at the same time. My innocence in the one field and
almost--ignorance in the other enabled me to proceed in both in the
kind of na{\"\i}ve way that is necessary for success.
\par
Thanks are due to Prof.~Rudolf Ahlswede, who became my Doktorvater out
of his desire to explore quantum information theory, and thus I was his
student: I am indebted to him for his teaching and constant support.
To Prof.~Martin Aigner (FU Berlin) who, after being my
Diplom advisor, obtained for me the opportunity to study with Prof.~Ahlswede.
Also to Prof.~Volker Strassen (Uni Konstanz) whom I regard as a teacher,
though he never was in a formal sense, for widening my mathematical
interests, and for educating me in combinatorics and information
theory, without which preparation I would have been lost at Bielefeld.
Not least I want to thank my collegue Peter L{\"o}ber for many discussions
during the last one and a half years, and sparing me the ill fate of
isolation by working in close neighbourhood: mathematically (as one
also pursuing problems of quantum communication), and physically
(by sharing the bureau room).
Finally, the justification of the present work's pages was done by
hand based on explanations by Prof.~Eberhard K\"onig (FU Berlin), which
hereby I acknowledge.

\clearpage

% inhalt

\thispagestyle{myheadings}
\tableofcontents
\clearpage

\pagenumbering{arabic}

% hier gehts richtig los:
% einleitung
% QDissertation: Introduction

\chapter*{Introduction}
\addcontentsline{toc}{chapter}{Introduction}
\label{chap:intro}

\thispagestyle{myheadings}

In the present thesis problems of information in quantum systems are
discussed, mainly in the context of coding problems of various kind.
Thus we follow a line of research initiated by \cite{shannon:theory},
where informational--operational meaning was lent to terms like
entropy, information, capacity, building on models of a stochastic
nature. This is where quantum theory enters, which is generally
understood to be a stochastic theory (starting with \cite{born:statistical},
now in any modern textbook, e.g. \cite{peres:quantum:theory}). A
stochastic theory however of a novel type: it was soon understood that
the statistical predictions of quantum theory cannot be described in
ordinary (``classical'') stochastic theories (\cite{epr:paradox},
\cite{bell:inequality}), and this is formally mirrored in the
necessity to introduce a ``noncommutative probability''.
\par
These observations led physicists during the 1960s to speculate
about the role of quantum probabilism in information theory:
cf.~the works of \cite{gordon:bound}, \cite{levitin:bound}, and
\cite{forney:bound}. \cite{holevo:bound} however is to be credited
with founding an appropriate mathematical theory (after a first
step by \cite{stratonovich:qinf}) and proving
the justly named {\sc Holevo} bound on quantum channel capacities.
This work was extended subsequently by \cite{holevo:superadditivity}.
Apart from this and formulating the definite model
(\cite{holevo:channels}, relying on earlier clarifying
work on quantum stochastics by \cite{ludwig:grundlagen}, {\sc Holevo},
and {\sc Davies \& Lewis}) efforts
concentrated on the analysis of specific restrictive or highly symmetric
situations.
\par
Then progress in foundations ceased, until the stormy
revival and extension of the subject in 1994, which year saw two
important contributions: the quantum algorithm of \cite{shor:factoring}
for factoring integers, proving the power of quantum information
processing, and by \cite{schumacher:qucoding} the successful interpretation
of {\sc von Neumann} entropy as asymptotic source coding rate for
quantum information (at the same time establishing quantum information
at all as a quantity, distinguished from what is now called
``classical information''. The reader should be aware however that
it was known from the early days of quantum theory on that operationally
quantum states are ``more'' than the knowlegde we can acquire about
them. A true expression of this qualitative distinction is the
\emph{no--cloning theorem} of \cite{wootters:zurek}, stating that quantum
states cannot be duplicated, i.e. ``copied'', whereas classical data
obviously can).
\par
Both works continue to exert a tremendous
influence on the new thinking about quantum information theory.
After that soon the coding theorem complementing the {\sc Holevo}
bound was proved (\cite{hausladen:qucap:purecase},
\cite{holevo:qucapacity}, \cite{schumacher:capacity}), and today
we face a variety of classical, quantum, or mixed information models,
some of which at least we understand.
\par
The present work opens and closes with quantum information: beginning
with a review of {\sc Schumacher}'s quantum source coding, to which
we contribute the strong converse, ending with some speculations about
multiple quantum source coding. In between we deal with transmission of
classical information via quantum channels. Here our achievements include
new proofs of the channel coding theorem (which is new for nonstationary
channels), and the completely new strong
converse (independently \cite{ogawa:nagaoka} have proved the strong converse
for stationary and finite alphabet channels by a different method),
estimates on the reliability function, and --- as a by--product a new proof
of the {\sc Holevo} bound. In the third chapter we determine
the capacity region of quantum multiple access channels, using
our results on multiple quantum source coding with side information from
the fourth chapter, where also a number of simple estimates on the rate region
and some examples are discussed. Among the positive results of this
part are the weak subadditivity for the so--called \emph{coherent
information} (while the ordinary subadditivity one would expect fails),
and the determination of the rate region for multiple classical
source coding with quantum side information at the decoder. Thus we completely
skirt all questions of channel coding of quantum information and noise
protection of quantum information by quantum error correcting codes,
these issues only entering implicitely in the discussion of multiple
quantum source coding. Also we choose to stay with discrete (i.e. finite, or,
in the quantum case, finite--dimensional) and memoryless systems: this
is not an essential restriction for our results, but allows to work consistently
with techniques of a combinatorial flavor and to skip technicalities (such as
finite variance conditions etc.) which, at the present state of techniques,
could not have been avoided. The restriction is further justified by the
ignorance on many questions even in this somewhat narrow setting.
An appendix contains the necessary elements of quantum probabilistic theory
and a calculus of entropy and information in quantum systems.
It will be referred to for any concept of that field needed in the
main text.
\bigskip\par
Parts of this work have been pre--published in the author's work:
appendix~\ref{chap:quprob} is distilled from its (very inadequate) predecessor
\cite{winter:quil}, chapter~\ref{chap:source} is from \cite{winter:qcode},
and the results of chapters~\ref{chap:channel} and \ref{chap:mac} were
reported by \cite{winter:qstrong}, \cite{winter:qnonst},
\cite{winter:qestim}, and \cite{winter:qmac}.
\bigskip\par
I have tried to give due credit (or else a reasonable reference) to any
result of some importance, especially in the main text. If there is no credit
it is implicit that I am the inventor. However this does not apply to a number
of propositions of less weight, especially in the appendix,
which I found on my own but which I regard as ``folklore'',
and thus never tried to trace them back to an original inventor.

% haupttext
% QDissertation: Quantum source coding

\chapter{Quantum Source Coding}
\label{chap:source}

\thispagestyle{myheadings}

In this chapter quantum information and the notion of its compression
are introduced. To prove the corresponding coding theorem and
strong converse basic techniques are developed:
a relation between fidelity and trace norm
distance, different notions of \emph{typical subspace},
and an estimate on general $\eta$--\emph{shadows}. Finally we comment
on the relation to classical source coding.

\section{Models of quantum data compression}
\label{sec:source:models}
Fix the complex Hilbert space ${\cal H}$, $d=\dim{\cal H}<\infty$.
$\alg{L}({\cal H})$ denotes the algebra of (bounded) linear operators of
${\cal H}$, $\alg{L}({\cal H})_*$ its predual under the
trace pairing.\footnote{For these notions (algebras, states, operations,
trace pairing, trace norm, etc.)
see appendix~\ref{chap:quprob}, section {\em Quantum systems}.}
\par
A \emph{(discrete memoryless) quantum source} (q--DMS) is a pair $(\Rho,P)$ 
with a finite set $\Rho\subset\alg{L}({\cal H})_*$ of pure states on
$\alg{L}({\cal H})$ and a p.d. $P$ on $\Rho$.
The \emph{average state} of the source is $P\Rho=\sum_{\pi\in\Rho} P(\pi)\pi$.
\par
An $n$--\emph{block code} for the q--DMS $(\Rho,P)$ is a pair
$(\varepsilon_*,\delta_*)$ where
$\varepsilon_*:\Rho^n\rightarrow\alg{L}({\cal K})_*$ maps $\Rho^n$
into the states on $\alg{L}({\cal K})$ (with some Hilbert space ${\cal K}$),
and $\delta_*:\alg{L}({\cal K})_*\rightarrow\alg{L}({\cal H})_*^{\otimes n}$ is
trace preserving and completely positive (i.e. it is a physical state transformation,
see appendix~\ref{chap:quprob}).\par
We say that $(\varepsilon_*,\delta_*)$ is \emph{quantum encoding} if $\varepsilon_*$
is the restriction to $\Rho^n$ of a trace preserving and completely positive map
$\varepsilon_*:\alg{L}({\cal H})_*^{\otimes n}\rightarrow\alg{L}({\cal K})_*$.
If there is no condition on $\varepsilon_*$ we say that $(\varepsilon_*,\delta_*)$
is \emph{arbitrary encoding}.\par
For an $n$--block code $(\varepsilon_*,\delta_*)$ define
\begin{enumerate}
  \item the \emph{(average) fidelity}
      $$\bar{F}=\bar{F}(\varepsilon_*,\delta_*)=\sum_{\pi^n\in\Rho^n}
               P^n(\pi^n)\!\cdot\!\tr((\delta_*\varepsilon_*\pi^n)\pi^n),$$
  \item the \emph{(average) distortion}
      $$\bar{D}=\bar{D}(\varepsilon_*,\delta_*)=\sum_{\pi^n\in\Rho^n}
                   P^n(\pi^n)\!\cdot\!\frac{1}{2}\| \delta_*\varepsilon_*\pi^n-\pi^n \|_1\ ,$$
  \item the \emph{entanglement fidelity} (see \cite{schumacher:F:e})
      $$F_e=F_e(\varepsilon_*,\delta_*)
           =\tr\left(((\delta_*\varepsilon_*\otimes\id)\Psi_{P\Rho}^{\otimes n})
                                                     \Psi_{P\Rho}^{\otimes n}\right),$$
      where $\Psi_{P\Rho}$ is a purification of $P\Rho$,
      i.e. it is a pure state on an extended system (by tensor product with some space
      ${\cal H}_0$), and $P\Rho=\Psi_{P\Rho}|_{\alg{L}({\cal H})}$
      (cf. \cite{schumacher:F:e} who proves that $F_e$
      does not depend on the purification chosen). Note that
      this makes sense only if $(\varepsilon_*,\delta_*)$ is quantum encoding.
\end{enumerate}
Observe that generally $\rho^n=\rho_1\otimes\cdots\otimes\rho_n$
denotes a product state of $n$ factors, while
$\rho^{\otimes n}=\rho\otimes\cdots\otimes\rho$ is the $n$--fold
tensor power of $\rho$.
\begin{satz}
  \label{satz:criteria}
  $$\bar{D}^2\leq 1-\bar{F}\leq \bar{D}\text{  and  }1-\bar{F}\leq 1-F_e\ .$$
\end{satz}
\begin{beweis}
  For the last inequality see \cite{schumacher:F:e}. The first double inequality
  follows from lemma~\ref{lemma:mixed:state} below by linearity, and by convexity
  of the square function.
\end{beweis}
\paragraph{A digression on fidelity}
First note that both $D(\rho,\sigma)=\dfrac{1}{2}\|\rho-\sigma\|_1$ and
$1-F(\rho,\sigma)=1-\tr(\rho\sigma)$ obey a triangle inequality:
$$\|\rho_1\otimes\rho_2-\sigma_1\otimes\sigma_2\|_1\leq
                                  \|\rho_1-\sigma_1\|_1+\|\rho_2-\sigma_2\|_1$$
and
$$1-F(\rho_1\otimes\rho_2,\sigma_1\otimes\sigma_2)\leq
                                    1-F(\rho_1,\sigma_1)+1-F(\rho_2,\sigma_2).$$
\begin{lemma}[Pure state]
  \label{lemma:pure:state}
  Let $\rho=\ket{\psi}\bra{\psi}$ and $\sigma=\ket{\phi}\bra{\phi}$ pure states. Then
  $$1-F(\rho,\sigma)=D(\rho,\sigma)^2\ .$$
\end{lemma}
\begin{beweis}
  W.l.o.g. we may assume $\ket{\psi}=\alpha\ket{0}+\beta\ket{1}$ and
  $\ket{\phi}=\alpha\ket{0}-\beta\ket{1}$ ($|\alpha|^2+|\beta|^2=1$).
  A straightforward calculation shows
  $F=(|\alpha|^2-|\beta|^2)^2$, and $D=2|\alpha\beta|$. Now
  \begin{equation*}\begin{split}
    1-F &=1-(|\alpha|^2-|\beta|^2)^2\\
        &=(1+|\alpha|^2-|\beta|^2)(1-|\alpha|^2+|\beta|^2)\\
        &=4|\alpha\beta|^2=D^2\ .
  \end{split}\end{equation*}
\end{beweis}
\begin{lemma}[Mixed state]
  \label{lemma:mixed:state}
  Let $\sigma$ an arbitrary mixed state (and $\rho$ pure as above). Then
  $$D\geq 1-F\geq D^2\ .$$
\end{lemma}
\begin{beweis}
  Write $\sigma=\sum_j q_j\pi_j$ with pure states $\pi_j$. Then
  \begin{equation*}\begin{split}
    1-F(\rho,\sigma) &=\sum_j q_j\left(1-F(\rho,\pi_j)\right)=\sum_j q_jD(\rho,\pi_j)^2\\
                     &\geq \left(\sum_j q_jD(\rho,\pi_j)\right)^{\!\! 2}
                        \geq D(\rho,\sigma)^2\ .
  \end{split}\end{equation*}
  Conversely: extend $\rho$ to the observable $(\rho,\1-\rho)$
  and consider the quantum operation
  $$\kappa_*:\sigma\longmapsto \rho\sigma\rho+(\1-\rho)\sigma(\1-\rho).$$
  Then (with monotonicity of $\|\cdot\|_1$ under quantum operations, see
  appendix~\ref{chap:quprob}, section {\em Quantum systems})
  $$2D=\|\rho-\sigma\|_1\geq\|\kappa_*\rho-\kappa_*\sigma\|_1=\|\rho-\kappa_*\sigma\|_1$$
  (since $\rho=\kappa_*\rho$). Hence with $F=\tr(\sigma\rho)$
  \begin{equation*}\begin{split}
    2D &\geq\big{\|}(1-F)\rho-\tr(\sigma(\1-\rho))\pi\big{\|}_1\\
       &=(1-F)+(1-F)=2(1-F)
  \end{split}\end{equation*}
  for a state $\pi$ supported in $\1-\rho$.
\end{beweis}
Observe that the inequalities of this lemma still hold if only $\sum_j q_j\leq 1$.
To close our digression we want to note two useful lemmata concerning
``good'' measurements:
\begin{lemma}[Tender operator]
  \label{lemma:tender:operator}
  Let $\rho$ be a state, and $X$ a positive operator with $X\leq\1$
  and $1-\tr(\rho X)\leq\lambda\leq 1$. Then
  $$\left\|\rho-\sqrt{X}\rho\sqrt{X}\right\|_1\leq\sqrt{8\lambda}\ .$$
\end{lemma}
\begin{beweis}
  Let $Y=\sqrt{X}$ and write $\rho=\sum_k p_k\pi_k$ with one--dimensional
  projectors $\pi_k$ and weights $p_k\geq 0$. Now
  \begin{equation*}\begin{split}
    \|\rho-Y\rho Y\|_1^2 &\leq \left(\sum_k p_k\|\pi_k-Y\pi_k Y\|_1\right)^2\\
                         &\leq \sum_k p_k\|\pi_k-Y\pi_k Y\|_1^2\\
                         &\leq 4\sum_k p_k(1-\tr(\pi_k Y\pi_k Y))\\
                         &\leq 8\sum_k p_k(1-\tr(\pi_k Y))\\
                         &=    8(1-\tr(\rho Y))\\
                         &\leq 8(1-\tr(\rho X))\leq 8\lambda
  \end{split}\end{equation*}
  by triangle inequality, convexity of $x\mapsto x^2$,
  lemma~\ref{lemma:mixed:state}, $1-x^2\leq 2(1-x)$, and $X\leq Y$.
\end{beweis}
\begin{lemma}[Tender measurement]
  \label{lemma:tender:measurement}
  Let $\rho_a$ ($a\in\fset{A}$) a family of states on ${\cal H}$, and $D$ an observable
  indexed by $\fset{B}$. Let $\varphi:\fset{A}\longrightarrow\fset{B}$ a
  map and $\lambda>0$ such that
  $$\forall a\in\fset{A}\qquad 1-\tr(\rho_a{D}_{\varphi(a)})\leq \lambda$$
  (i.e. the observable identifies $\varphi(a)$ from $\rho_a$ with maximal
  error probability $\lambda$). Then the canonically corresponding
  quantum operation
  $$D_{\text{\rm int}*}:\rho\longmapsto \sum_{b\in\fset{B}} \sqrt{{D}_b}\rho\sqrt{{D}_b}$$
  disturbes the states $\rho_a$ only a little:
  $$\forall a\in\fset{A}\qquad
                      \|\rho_a-D_{\text{\rm int}*}\rho_a\|_1\leq \sqrt{8\lambda}+\lambda.$$
  Furthermore the total observable operation\footnote{See also
    appendix~\ref{chap:quprob}, section {\em Common tongue}, for
    $D_{\text{int}}$ and $D_{\text{tot}}$.}
  $$D_{\text{\rm tot}*}:\rho\longmapsto
                              \sum_{b\in\fset{B}} [b]\otimes\sqrt{{D}_b}\rho\sqrt{{D}_b}$$
  satisfies
  $$\forall a\in\fset{A}\qquad \|[\varphi(a)]\otimes\rho_a
                              -D_{\text{\rm tot}*}\rho_a\|_1\leq \sqrt{8\lambda}+\lambda.$$
\end{lemma}
\begin{beweis}
  An easy calculation:
  \begin{equation*}\begin{split}
    \|\rho_a-D_{{\rm int}*}\rho_a\|_1
                     &\leq \|\rho_a-\sqrt{{D}_{\varphi(a)}}\rho_a\sqrt{{D}_{\varphi(a)}}\|_1
                          +\sum_{b\neq \varphi(a)} \|\sqrt{{D}_b}\rho_a\sqrt{{D}_b}\|_1 \\
                     &=    \|\rho_a-\sqrt{{D}_{\varphi(a)}}\rho_a\sqrt{{D}_{\varphi(a)}}\|_1
                          +\sum_{b\neq \varphi(a)} \tr(\rho_a{D}_b) \\
                     &\leq \sqrt{8(1-\tr(\rho_a{D}_{\varphi(a)}))}
                          +1-\tr(\rho_a{D}_{\varphi(a)})\\
                     &\leq \sqrt{8\lambda}+\lambda,
  \end{split}\end{equation*}
  using triangle inequality and lemma~\ref{lemma:tender:operator}.
  The second part (which actually implies the first) is similar.
\end{beweis}
\begin{bem}
  \label{bem:average:tenderness}
  If we modify the statement of the lemma to that the average error
  in identifying $\varphi(a)$ from $\rho_a$ should be at most $\bar{\lambda}$
  (relative a distribution on $\fset{A}$), then also the distortion bound of
  the lemma holds --- on average.
\end{bem}

\bigskip\par
Let us return to the source coding schemes:
The n--block code $(\varepsilon_*,\delta_*)$ is called an
$(n,\lambda)_{\bar{F}}$--\emph{code}
if $1-\bar{F}(\varepsilon_*,\delta_*)\leq\lambda$. Similarly an
$(n,\lambda)_{F_e}$--code is defined.
The n--block code $(\varepsilon_*,\delta_*)$ is called an
$(n,\lambda)_{\bar{D}}$--\emph{code}
if $\bar{D}(\varepsilon_*,\delta_*)\leq\lambda$.
\par
The \emph{rate} of an n--block code $(\varepsilon_*,\delta_*)$ is defined
as $R(\varepsilon_*,\delta_*)=\frac{1}{n}\log\dim{\cal K}$.\footnote{Here
  and in the sequel $\log$ is always understood to base $2$, as well as $\exp$.
  The unit of this rate is usually called \emph{qubit} (short
  for quantum bit: the states of a two--level quantum system
  $\alg{L}(\C^2)$).}
\par
From the previous theorem it is clear that the most restrictive model
is where we have to find an $(n,\lambda)_{F_e}$--code with quantum encoding,
whereas the most powerful model is where we have to find an
$(n,\lambda)_{\bar{F}}$--code with arbitrary encoding
(equivalently we may use $\bar{D}$).
Now we define for a q--DMS $(\Rho,P)$
\begin{enumerate}
  \item the $\lambda$--(quantum,$F_e$)--\emph{rate} as
        $$R_{q,F_e}(\lambda)=\limsup_{n\rightarrow\infty}\min\{ 
                 R(\varepsilon_*,\delta_*):(\varepsilon_*,\delta_*)\text{ an }\\
                                (n,\lambda)_{F_e}\text{--code with qu. encoding}\},$$
  \item the $\lambda$--(quantum,$\bar{F}$)--\emph{rate} as
        $$R_{q,\bar{F}}(\lambda)=\limsup_{n\rightarrow\infty}\min\{ 
                 R(\varepsilon_*,\delta_*):(\varepsilon_*,\delta_*)\text{ an }\\
                            (n,\lambda)_{\bar{F}}\text{--code with qu. encoding}\},$$
  \item the $\lambda$--(arbitrary,$\bar{F}$)--\emph{rate} as
        $$R_{a,\bar{F}}(\lambda)=\limsup_{n\rightarrow\infty}\min\{ 
                 R(\varepsilon_*,\delta_*):(\varepsilon_*,\delta_*)\text{ an }\\
                           (n,\lambda)_{\bar{F}}\text{--code with arb. encoding}\}.$$
\end{enumerate}
Despite our lot of definitions the situation turns out to be quite simple:
\begin{satz}
  \label{satz:schumacher:qucoding}
  For all $\lambda\in(0,1)$ the three $\lambda$--rates of the q--DMS $(\Rho,P)$
  are equal to the {\sc von Neumann} entropy of the ensemble $(\Rho,P)$:
  $$R_{q,F_e}(\lambda)=R_{q,\bar{F}}(\lambda)=R_{a,\bar{F}}(\lambda)=H(P\Rho),$$
  where $H(\rho)=-\tr(\rho\log\rho)$ (see appendix~\ref{chap:quprob},
  section {\em Entropy and divergence}).
\end{satz}
\begin{beweis}
  Between the first two members of the chain we have ``$\geq$'' by
  theorem~\ref{satz:criteria},
  between the second and the third ``$\geq$'' is obvious. 
  $R_{q,F_e}(\lambda)\leq H(P\Rho)$ follows from the
  coding theorem~\ref{satz:noiseless:coding}.
  Finally $R_{a,\bar{F}}(\lambda)\geq H(P\Rho)$ follows from
  the strong converse theorem~\ref{satz:strong:converse}.
\end{beweis}

\section{Typical subspaces and shadows}
\label{sec:source:shadows}
Let $P$ a p.d. on the set $\fset{X}$, with $|\fset{X}|=a<\infty$.
Define for $\alpha>0$ the set
$$\fset{T}^n_{V,P,\alpha}=\{x^n\in\fset{X}^n:\forall x\in\fset{X}\ 
                         |N(x|x^n)-nP(x)|\leq\alpha\sqrt{P(x)(1-P(x))}\sqrt{n}\}$$
of \emph{variance--typical sequences} with constant $\alpha$
(in the sense of \cite{wolfowitz:coding}),
where $N(x|x^n)=|\{i:x_i=x\}|$. For a sequence $x^n$ the empirical distribution
$P_{x^n}$ on $\fset{X}$ (i.e. $P_{x^n}(x)=\frac{1}{n}N(x|x^n)$)
is called \emph{type} of $x^n$.
\par
It is easily seen that there are at most $(n+1)^a$ types; this kind
of reasoning is generally called \emph{type counting}.
\begin{lemma}[cf. \cite{wolfowitz:coding}]
  \label{lemma:variance:typical:sequences}
  For every p.d. $P$ on $\fset{X}$ and $\alpha>0$
  $$P^{\otimes n}(\fset{T}^n_{V,P,\alpha})\geq 1-\frac{a}{\alpha^2}$$
  $$|\fset{T}^n_{V,P,\alpha}| \leq \exp\left(nH(P)+Kd\alpha\sqrt{n}\right).$$
\end{lemma}
\begin{beweis}
  $\fset{T}^n_{V,P,\alpha}$ is the intersection of $a$ events, namely
  for each $x\in\fset{X}$ that the mean of the independent Bernoulli
  variables $X_i$ with value $1$ iff $x_i=x$ has a deviation from its
  expectation $P(x)$ at most $\alpha\sqrt{P(x)(1-P(x))}/\sqrt{n}$.
  By {\sc Chebyshev}'s inequality each of these has probability at
  least $1-1/\alpha^{2}$.
  \par
  The cardinality estimate is like in the proof of the
  following lemma~\ref{lemma:variance:typical:proj}.
\end{beweis}
Now construct \emph{variance--typical projectors} $\Pi^n_{V,\rho,\alpha}$
using typical sequences: for a diagonalization
$\rho=\sum_j q_j\pi_j$ let $s_j=\sqrt{q_j(1-q_j)}$ and
$$\fset{T}^n_{V,\rho,\alpha}=\{(j_1,\ldots,j_n):\forall j\ 
                                      |N(j|j^n)-nq_j|\leq s_j\alpha\sqrt{n}\},$$
and define
$$\Pi^n_{V,\rho,\alpha}=\sum_{(j_1,\ldots,j_n)\in\fset{T}^n_{V,\rho,\alpha}}
                                       \pi_{j_1}\otimes\cdots\otimes\pi_{j_n}\ .$$
For a state $\rho$ define $\mu(\rho)$ as the minimal nonzero eigenvalue
of $\sqrt{\rho(\1-\rho)}$ and $N(\rho)=\dim\supp\sqrt{\rho(\1-\rho)}$, finally 
$K=2\frac{\log e}{e}$.
Then one has
\begin{lemma}
  \label{lemma:variance:typical:proj}
  For every state $\rho$ and $n>0$
  \begin{align*}
    \tr(\rho^{\otimes n}\Pi^n_{V,\rho,\alpha}) &\geq 1-\frac{d}{\alpha^2} \\
    \tr(\rho^{\otimes n}\Pi^n_{V,\rho,\alpha}) &\geq 1-2N(\rho)e^{-2\mu(\rho)^2\alpha^2}\ ,
  \end{align*}
  and with $\Pi^n=\Pi^n_{V,\rho,\alpha}$
  $$\Pi^n\exp \left(-nH(\rho)-Kd\alpha\sqrt{n}\right)\leq \Pi^n\rho^{\otimes n}\Pi^n\leq
                                        \Pi^n\exp\left(-nH(\rho)+Kd\alpha\sqrt{n}\right)$$
  $$\tr\Pi^n_{V,\rho,\alpha}\leq\exp\left(nH(\rho)+Kd\alpha\sqrt{n}\right).$$
  Every $\eta$--\emph{shadow} $B$ of $\rho^{\otimes n}$ (this means $0\leq B\leq\1$ and
  $\tr(\rho^{\otimes n}B)\geq\eta$) satifies
  $$\tr B\geq\left(\eta-2N(\rho)e^{-2\mu(\rho)^2\alpha^2}\right)
                        \exp\left(nH(\rho)-Kd\alpha\sqrt{n}\right).$$
\end{lemma}
\begin{beweis}
  The first estimate is the {\sc Chebyshev} inequality, as before:
  the trace is the probability of a set of variance--typical sequences of
  eigenvectors of the $\rho_i'$ in the product of the measures
  given by the eigenvalue lists. Similarly the second estimate
  is the well known inequality of \cite{hoeffding:inequality}.
  The third estimate is the key: to prove it let
  $\pi^n=\pi_{j_1}\otimes\cdots\otimes\pi_{j_n}$
  one of the eigenprojections of $\rho^{\otimes n}$ which contributes to
  $\Pi^n_{V,\rho,\alpha}$. Then
  $$\tr(\rho^{\otimes n}\pi^n)=q_{j_1}\cdots q_{j_n}=\prod_j q_j^{N(j|j^n)}\ .$$
  Taking logs and using the defining relation for the $N(j|j^n)$
  we find
  \begin{equation*}\begin{split}
    \left|\sum_j \!\!-N(j|j^n)\log{q_j} \!-\! nH(\rho)\right|
                                    &\leq \sum_j \!-\log{q_j}|N(j|j^n)-nq_j|\\
                                    &\leq \sum_j -\alpha\sqrt{n}\sqrt{q_j}\log{q_j}\\
                                    &=2\alpha\sqrt{n}\sum_j -\sqrt{q_j}\log{\sqrt{q_j}}\\
                                    &\leq 2d\frac{\log e}{e}\alpha\sqrt{n}\ .
  \end{split}\end{equation*}
  The rest follows from the following lemma~\ref{lemma:abstract:shadow:bound}.
\end{beweis}
\begin{lemma}[Shadow bound]
  \label{lemma:abstract:shadow:bound}
  Let $0\leq\Lambda\leq\1$ and $\rho$ a state such that for some
  $\lambda,\mu_1,\mu_2>0$
  $$\tr(\rho\Lambda)\geq 1-\lambda\text{ and }
                      \mu_1\Lambda\leq\sqrt{\Lambda}\rho\sqrt{\Lambda}\leq\mu_2\Lambda.$$
  Then $(1-\lambda)\mu_2^{-1}\leq\tr\Lambda\leq\mu_1^{-1}$
  and for $0\leq B\leq\1$ with $\tr(\rho B)\geq\eta$ one has
  $\tr B\geq\left(\eta-\sqrt{8\lambda}\right)\mu_2^{-1}$.  
  If $\rho$ and $\Lambda$ commute this can be improved to
  $\tr B\geq\left(\eta-\lambda\right)\mu_2^{-1}$.
\end{lemma}
\begin{beweis}
  The bounds on $\tr\Lambda$ follow by taking traces in the inequalities
  in $\sqrt{\Lambda}\rho\sqrt{\Lambda}$ and using
  $1-\lambda\leq\tr(\rho\Lambda)\leq 1$.
  For the $\eta$--shadow $B$ observe
  \begin{equation*}\begin{split}
    \mu_2\tr B
       &\geq \tr\left(\mu_2\Lambda B\right)
           \geq \tr\left(\sqrt{\Lambda}\rho\sqrt{\Lambda} B\right) \\
       &=\tr(\rho B)-\tr\left(\left(\rho-\sqrt{\Lambda}\rho\sqrt{\Lambda}\right)B\right)
           \geq \eta-\left\|\rho-\sqrt{\Lambda}\rho\sqrt{\Lambda}\right\|_1\ .
  \end{split}\end{equation*}
  If $\rho$ and $\Lambda$ commute the trace norm can obviously be estimated
  by $\lambda$, else we have to invoke the tender operator
  lemma~\ref{lemma:tender:operator} to bound it by $\sqrt{8\lambda}$.
\end{beweis}
For the benefit of discussions in later chapters let us mention here
two other notions of typical projector:
\paragraph{Entropy typical projectors}
Let $\rho_1,\ldots,\rho_n$ states on $\alg{L}({\cal H})$, with diagonalizations
$\rho_i=\sum_j q_{j|i}\pi_{ij}$ with one--dimensional projectors $\pi_{ij}$.
Let $\delta>0$, and define
$$\fset{T}^n_{H,\rho^n,\delta}=\{(j_1,\ldots,j_n):
   \left|\sum_{i=1}^n \!-\log q_{ij_i}\!-\sum_{i=1}^n \!H(\rho_i)\right|\leq\delta\sqrt{n}\}.$$
Define the \emph{entropy--typical projector}\footnote{This is essentially what
  \cite{schumacher:qucoding} calls \emph{typical subspace}.}
\emph{of} $\rho^n$ \emph{with constant} $\delta$ as
$$\Pi^n_{H,\rho^n,\delta}=\sum_{(j_1,\ldots,j_n)\in\fset{T}^n_{H,\rho^n,\delta}}
                                 \pi_{1j_1}\otimes\cdots\otimes\pi_{nj_n}\ .$$
Then we have the following
\begin{lemma}
  \label{lemma:entropy:typical:proj}
  There is a constant $K$ depending only on $d$ (in fact one may choose
  $K\leq\max\{(\log 3)^2,(\log d)^2\}$) such that for arbitrary states $\rho_1,\ldots\rho_n$
  $$\tr(\rho^n\Pi^n_{H,\rho^n,\delta})\geq 1-\frac{K}{\delta^2}\ .$$
\end{lemma}
\begin{beweis}
  This is just Chebyshev's inequality applied to the random variables $X_i=-\log q_{\cdot|i}$
  for the diagonalizations $\rho_i=\sum_j q_{j|i}\pi_{ij}$. Observe that
  $K$ may be any bound for the variance of the $X_i$.
\end{beweis}
Concerning its size we have
\begin{lemma}
  \label{lemma:entropy:typical:proj:size}
  For the entropy--typical projector
  $$\left(1-\frac{K}{\delta^2}\right)\exp\left(\sum_{i=1}^n H(\rho_i)-\delta\sqrt{n}\right)
         \leq\tr\Pi^n_{H,\rho^n,\delta}
         \leq\exp\left(\sum_{i=1}^n H(\rho_i)+\delta\sqrt{n}\right).$$
  Conversely, if $B$ is an $\eta$--shadow of $\rho^n$ then
  $$\tr B\geq \left(\eta-\frac{K}{\delta^2}\right)
                       \exp\left(\sum_{i=1}^n H(\rho_i)-\delta\sqrt{n}\right).$$
\end{lemma}
\begin{beweis}
  Observe that by definition of $\Pi^n=\Pi^n_{H,\rho^n,\delta}$
  $$\Pi^n\exp\left(-\sum_{i=1}^n H(\rho_i)-\delta\sqrt{n}\right)
           \leq\Pi^n\rho^n\Pi^n
           \leq\Pi^n\exp\left(-\sum_{i=1}^n H(\rho_i)+\delta\sqrt{n}\right).$$
  Now the lemma follows by the shadow bound lemma~\ref{lemma:abstract:shadow:bound}.
\end{beweis}

\paragraph{Constant typical projectors}
Let $\rho$ a state with diagonalization $\rho=\sum_j q_j\pi_j$,
and $\delta>0$, then define
$$\fset{T}^n_{C,\rho,\delta}=\{(j_1,\ldots,j_n):
                            \forall j\ \big| N(j|j^n)-nq_j \big|\leq\delta\sqrt{n}\},$$
and the \emph{constant--typical projector}
\begin{equation*}\begin{split}
  \Pi^n_{C,\rho,\delta} &=\sum_{j^n\in\fset{T}^n_{C,\rho^n,\delta}}
                                 \pi_{j_1}\otimes\cdots\otimes\pi_{j_n}\\
                        &=\sum_{j^n\text{ with }\|\sum_{i=1}^n
                                     \pi_{j_i}-n\rho\|_\infty\leq\delta\sqrt{n}}
                                 \pi_{j_1}\otimes\cdots\otimes\pi_{j_n}\ .
\end{split}\end{equation*}
Then one has
\begin{lemma}[Weak law]
  \label{lemma:weak:law}
  Let $\tilde{\rho},\rho_1,\ldots,\rho_n$ states of a system
  and $\delta,\epsilon>0$ such that
  $$\left\|\frac{1}{n}\sum_{i=1}^n \rho_i-\tilde{\rho}\right\|_\infty\leq\epsilon.$$
  Then
  $$\tr(\rho^n\Pi^n_{C,\tilde{\rho},\delta+\epsilon\sqrt{n}})\geq 1-\frac{1}{\delta^2}\ .$$
\end{lemma}
\begin{beweis}
  Consider the diagonalization $\tilde{\rho}=\sum_j q_j\pi_j$, and the
  conditional expectation map
  $$\kappa_*:\sigma\longmapsto\sum_j \pi_j\sigma\pi_j\ .$$
  Defining $\rho_i'=\kappa_*(\rho_i)$ we claim that
  $$\Pi^n_{C,\frac{1}{n}\sum_{i=1}^n \rho_i',\delta}\leq
                                       \Pi^n_{C,\tilde{\rho},\delta+\epsilon\sqrt{n}}\ .$$
  Indeed observe that we have
  $$\left\|\frac{1}{n}\sum_{i=1}^n \rho_i'-\tilde{\rho}\right\|_\infty\leq
      \left\|\frac{1}{n}\sum_{i=1}^n \rho_i -\tilde{\rho}\right\|_\infty\leq\epsilon.$$
  Thus for $j^n=(j_1,\ldots,j_n)$ with
  $$\left\|\sum_{i=1}^n \pi_{j_i}-\sum_{i=1}^n \rho_i'\right\|_\infty\leq\delta\sqrt{n}$$
  we have by triangle inequality
  $$\left\|\sum_{i=1}^n \pi_{j_i}-n\tilde{\rho}\right\|_\infty
                                           \leq(\delta+\epsilon\sqrt{n})\sqrt{n}\ .$$
  So we can estimate
  \begin{equation*}\begin{split}
    \tr(\rho^n\Pi^n_{C,\tilde{\rho},\delta+\epsilon\sqrt{n}})
            &\geq\tr(\rho^n\Pi^n_{C,\frac{1}{n}\sum_{i=1}^n \rho_i',\delta}) \\
            &=   \tr(\rho^{\prime n}\Pi^n_{C,\frac{1}{n}\sum_{i=1}^n \rho_i',\delta}) \\
            &\geq 1-\frac{1}{\delta^2}\ ,
  \end{split}\end{equation*}
  the last line by $d$ uses of {\sc Chebyshev}'s inequality, as in the proof
  of lemma~\ref{lemma:variance:typical:proj}.
\end{beweis}
\par
Concerning the size of this projector we have
\begin{lemma}
  \label{lemma:constant:typical:size}
  For every state $\rho$ and $0<\delta\leq\frac{1}{2d}\sqrt{n}$
  $$\tr\Pi^n_{C,\rho,\delta}\leq
        (n+1)^{d}\exp\left(nH(\rho)+nd\eta\left(\frac{\delta}{\sqrt{n}}\right)\right).$$
\end{lemma}
\begin{beweis}
  The whole question reduces obviously to counting sequences of eigenvectors
  of $\rho$ with type close to the p.d.
  given by the eigenvalue list of $\rho$.
  Each sequence of type $P$ has $P^{\otimes n}$--probability
  $\exp(-nH(P))$. Thus there are at most $\exp(nH(P))$ of these. Since
  there at most $(n+1)^{d}$ many types, and by the
  continuity of entropy (lemma~\ref{lemma:H:continuous})
  the statement follows.
\end{beweis}
The constant typical projectors will be used as shadows of
whole \emph{sets} (namely of
states which satisfy the ``average'' condition of the weak
law lemma~\ref{lemma:weak:law}).

\section{{\sc Schumacher}'s quantum coding}
\label{sec:source:qcoding}
Let $\alpha>0$. The {\sc Schumacher} \emph{scheme with constant} $\alpha$
for the  q--DMS $(\Rho,P)$ is the following family of $n$--block
codes $(\varepsilon_*,\delta_*)$ with quantum encoding: define
$\Pi^n=\Pi^n_{V,P\Rho,\alpha}$ and the Hilbert space ${\cal K}=\im\Pi^n$,
and
\begin{align*}
  \varepsilon_*:\alg{L}({\cal H})_*^{\otimes n} &\longrightarrow \alg{L}({\cal K})_*\\
                                         \sigma &\longmapsto \Pi^n\sigma\Pi^n
                                           +\frac{1-\tr(\sigma\Pi^n)}{\dim{\cal K}}\1\\
  \delta_*:\alg{L}({\cal K})_* &\longrightarrow \alg{L}({\cal H})_*^{\otimes n}\\
                        \sigma &\longmapsto \sigma\ .
\end{align*}
\begin{bem}
  Essentially the above scheme was first defined
  by \cite{schumacher:qucoding}, with a slightly
  different definition of $\Pi^n$. The great contribution
  of \cite{schumacher:qucoding} was to notice the possibility and
  importance of having a \emph{typical subspace}, and the
  following theorem is just a variation of the original argument.
  Subsequently there appeared minor modifications and
  refinements (\cite{jozsa:schumacher} and \cite{jozsa:3horodecki}),
  but all rely on one or another notion of typical subspace of ${\cal H}^{\otimes n}$.
\end{bem}
\begin{satz}
  \label{satz:noiseless:coding}
  The {\sc Schumacher} scheme has rate
  $$R(\varepsilon_*,\delta_*)\leq H(P\Rho)+\frac{Kd\alpha}{\sqrt{n}}$$
  and entanglement fidelity
  $$F_e(\varepsilon_*,\delta_*)\geq 1-4N(P\Rho)e^{-2\mu(P\Rho)^2\alpha^2}.$$
\end{satz}
\begin{beweis}
  The rate estimate is immediate from
  lemma~\ref{lemma:variance:typical:proj}. For the fidelity
  consider a purification of $P\Rho=\sum_j q_j\ket{\varphi_j}\bra{\varphi_j}$,
  e.g. the projector of
  $\ket{\psi}=\sum_j \sqrt{q_j}\ket{\varphi_j}\otimes\ket{\varphi_j}$
  on $\alg{L}({\cal H}^{\otimes 2})$. With that the fidelity estimate follows
  easily from the shadow
  lemma~\ref{lemma:variance:typical:proj}.
\end{beweis}
By slightly changing the definition of the subspace used we arrive at the
{\sc JHHH}--\emph{scheme} of \cite{jozsa:3horodecki}: just take for
$\Pi^n$ the projector
$$\Pi^n_{H(\cdot)\leq R}=\lcsupp_{H(\nu)\leq R} \Pi^n_{V,\nu,0}$$
(the least common support) with some rate $R\geq 0$.
Then in \cite{jozsa:3horodecki} it is proved that this gives universally
good compression of all sources $(\Rho,P)$ with
$H(P\Rho)<R$. For one thing (see \cite{jozsa:3horodecki})
$$\tr\Pi^n_{H(\cdot)\leq R}\leq (n+1)^{d^2+d}\exp(nR),$$
and for the fidelity one has
\begin{satz}
  \label{satz:universal:compression}
  Let $(\varepsilon_*,\delta_*)$ the {\sc JHHH}--scheme with rate $R$ and block length
  $n$. Then for every q--DMS $(\Rho,P)$
  $$F_e(\varepsilon_*,\delta_*)\geq
                    1-2(n+1)^d\exp\left(-n\cdot\min_{H(\nu)\geq R} D(\nu\|P\Rho)\right).$$
\end{satz}
\begin{beweis}
  First note that by direct calculation for $\nu$ codiagonal with a state
  $\rho$ we have
  $$\Pi^n_{V,\nu,0}\rho^{\otimes n}\Pi^n_{V,\nu,0}
                 =\Pi^n_{V,\nu,0}\exp\left(-nD(\nu\|\rho)-nH(\nu)\right)$$
  (see lemma~\ref{lemma:exact:type:weight}). Fix a diagonalization
  $P\Rho=\sum_j q_j\pi_j$ and observe
  $$\Pi^n_{H(\cdot)\leq R}\geq\sum_{\nu\in\C[\pi_1,\ldots,\pi_d],H(\nu)\leq R}
                                                                  \Pi^n_{V,\nu,0}\ .$$
  Using the simple facts that $\Pi_{V,\nu,0}\neq 0$ only if
  $\nu\in\frac{1}{n}\N[\pi_j|j]$, and $\tr\Pi^n_{V,\nu,0}\leq\exp(nH(\nu))$,
  we find as in the previous theorem
  \begin{equation*}\begin{split}
    1-\bar{F}(\varepsilon_*,\delta_*) &\leq 2\sum_{\nu\in\frac{1}{n}\N[\pi_j|j],H(\nu)>R}
                                                   \exp(-nD(\nu\|P\Rho))\\
                              &\leq 2(n+1)^d\exp\left(-n\cdot\min_{H(\nu)\geq R}
                                                   D(\nu\|P\Rho)\right),
  \end{split}\end{equation*}
  where the last estimate is by type counting: there are at most
  $(n+1)^d$ different $\nu$ diagonal in the basis $\{\pi_j|j\}$
  and $\Pi^n_{V,\nu,0}\neq 0$.
\end{beweis}

\section{Strong converse}
\label{sec:source:converse}
The first proofs by \cite{schumacher:qucoding} and \cite{jozsa:schumacher}
for the optimality of the {\sc Schumacher}
scheme where valid only under the additional assumption that $\delta_*$ is of the
form $\delta_*(\sigma)=U\sigma U^*$ for a unitary embedding $U$ of
${\cal K}$ into ${\cal H}^{\otimes n}$. Also they achieved the bound $H(P\Rho)$ only
in the limit of $\lambda\rightarrow 0$ (so they proved a \emph{weak converse}).
The proof of \cite{barnum:et:al} removed the restriction on $\delta_*$,
but still yields only a weak converse. Also it works with some surprising and difficult
fidelity estimates, involving even mixed state fidelity, see \cite{jozsa:fidelity}
(We may note that they seem to be related to our inequalities of
theorem~\ref{satz:criteria}). We should also mention the
work of \cite{allahverdyan:saakian:converse} where a weak converse was proved for
quantum encodings and using entanglement fidelity (compare
our theorem~\ref{satz:lower:blind:F_e} with $s=1$). The criticism of the authors
on \cite{barnum:et:al} however is unjustified: by the above discussion
(proof of theorem~\ref{satz:schumacher:qucoding}) their result is
weaker than that of \cite{barnum:et:al}.
Then \cite{horodecki:qucoding} noticed that considering $\bar{D}$
instead of $\bar{F}$ drastically simplifies the proof. His argument is as follows:
\par
Assume that we are given a code $(\varepsilon_*,\delta_*)$ with arbitrary encoding
in the states on a $k$--dimensional Hilbert space and
$$\bar{D}=\frac{1}{2}\sum_{\pi^n\in\Rho^n} P^n(\pi^n)\|\pi^n-\delta_*\varepsilon_*\pi^n\|_1
                        \leq\lambda\leq\frac{1}{4}\ .$$
So by {\sc Markov}'s inequality there is a subset $\fset{C}\subset\Rho^n$
with $P^n(\fset{C})\geq 1-2\sqrt{\lambda}$ and
$$\forall\pi^n\in\fset{C}\quad
                \|\pi^n-\delta_*\varepsilon_*\pi^n\|_1\leq\sqrt{\lambda}\ .$$
Now form the state $\sigma=\sum_{\pi^n\in\Rho^n} P^n(\pi^n)\varepsilon_*\pi^n$,
then by {\sc Uhlmann}'s monotonicity of the
quantum I--divergence (theorem~\ref{satz:monoton})
$$\forall\pi^n\in\Rho^n\quad D(\varepsilon_*\pi^n\|\sigma)\geq
                              D(\delta_*\varepsilon_*\pi^n\|\delta_*\sigma).$$
Averaging we obtain
$$\sum_{\pi^n\in\Rho^n} P^n(\pi^n)D(\varepsilon_*\pi^n\|\sigma)\geq
         \sum_{\pi^n\in\Rho^n} P^n(\pi^n)D(\delta_*\varepsilon_*\pi^n\|\delta_*\sigma).$$
Now it is straightforward to calculate the l.h.s. of this to
$H(\sigma)-\sum_{\pi^n\in\Rho^n} P^n(\pi^n)H(\varepsilon_*\pi^n)$, whereas
the r.h.s. evaluates similarly to
$H(\delta_*\sigma)-\sum_{\pi^n\in\Rho^n} P^n(\pi^n)H(\delta_*\varepsilon_*\pi^n)$.
Since $2\lambda\leq 1/2$ and $\sqrt{\lambda}\leq 1/2$ we can use a
continuity property of $H$ (see lemma~\ref{lemma:H:continuous}):
\par
$\|\delta_*\sigma-(P\Rho)^{\otimes n}\|_1\leq 2\lambda$ implies
$$\left|H(\delta_*\sigma)-nH(P\Rho)\right|\leq -2\lambda\log\frac{2\lambda}{d^n}\ ,$$
and (for $\pi^n\in\fset{C}$) $\|\delta_*\varepsilon_*\pi^n-\pi^n\|_1\leq\sqrt{\lambda}$ implies
$$\left|H(\delta_*\varepsilon_*\pi^n)-H(\pi^n)\right|
              \leq -\sqrt{\lambda}\log\frac{\sqrt{\lambda}}{d^n}\ .$$
Combining we get the chain of inequalities
\begin{equation*}\begin{split}
  \log k &\geq H(\sigma)\\
         &\geq H(\sigma)-\sum_{\pi^n\in\Rho^n} P^n(\pi^n)H(\varepsilon_*\pi^n)\\
         &\geq H(\delta_*\sigma)
           -\sum_{\pi^n\in\Rho^n} P^n(\pi^n)H(\delta_*\varepsilon_*\pi^n)\\
         &\geq nH(P\Rho)-\sum_{\pi^n\in\Rho^n} P^n(\pi^n)H(\pi^n)
           -2\sqrt{\lambda}\log d^n+2\lambda\log\frac{2\lambda}{d^n}
            +\sqrt{\lambda}\log\frac{\sqrt{\lambda}}{d^n}\\
         &=    nH(P\Rho)-n(2\lambda+3\sqrt{\lambda})\log d
           +2\lambda\log 2\lambda+\sqrt{\lambda}\log\sqrt{\lambda}\ .
\end{split}\end{equation*}
Thus we proved
\begin{satz}[Weak converse]
  For every q--DMS $(\Rho,P)$
  $$\liminf_{\lambda\rightarrow 0} R_{a,\bar{F}}(\lambda)=
    \liminf_{\lambda\rightarrow 0} R_{a,\bar{D}}(\lambda)\geq H(P\Rho).$$
  \phantom{.}\qed
\end{satz}
But in fact much more is true:
\begin{satz}[Strong converse]
  \label{satz:strong:converse}
  Let $(\Rho,P)$ a q--DMS and
  $(\varepsilon_*,\delta_*)$ an $(n,\lambda)_{\bar{F}}$--code with arbitrary
  encoding, and $\alpha>0$. Then
  $$\dim{\cal K}\geq \left(1-\lambda-4\sqrt{N(P\Rho)}
                       e^{-\mu(P\Rho)^2\alpha^2}\right)\cdot
                          \exp\left(nH(P\Rho)-Kd\alpha\sqrt{n}\right).$$
\end{satz}
\begin{beweis}
  Let $B=\delta_*(\1_{\cal K})$ and $\Pi^n=\Pi^n_{V,P\Rho,\alpha}$.
  Since $\varepsilon_*\pi^n\leq\1_{{\cal K}}$ for every $\pi^n\in\Rho^n$
  it is clear that $\delta_*\varepsilon_*\pi^n\leq B$. Thus
  \begin{equation*}\begin{split}
    \tr\left(B\cdot\Pi^n\pi^n\Pi^n\right)
                      &\geq\tr\left((\delta_*\varepsilon_*\pi^n)\Pi^n\pi^n\Pi^n\right)\\
                      &   =\tr\left((\delta_*\varepsilon_*\pi^n)\pi^n\right)
                    -\tr\left(\delta_*\varepsilon_*\pi^n(\pi^n-\Pi^n\pi^n\Pi^n)\right)\\
                      &\geq\tr\left((\delta_*\varepsilon_*\pi^n)\pi^n\right)
                           -\|\pi^n-\Pi^n\pi^n\Pi^n\|_1\\
                      &\geq\tr\left((\delta_*\varepsilon_*\pi^n)\pi^n\right)
                           -\sqrt{8(1-\tr\pi^n\Pi^n)}
  \end{split}\end{equation*}
  (the last estimate by lemma~\ref{lemma:tender:operator}).
  Averaging over $P^{\otimes n}$ we find, with the shadow
  lemma~\ref{lemma:variance:typical:proj}
  and concavity of the square root:
  \begin{equation*}\begin{split}
    \tr\left(\Pi^n(P\Rho)^{\otimes n}\Pi^n B\right)
                   &\geq \bar{F}-\sqrt{8\left(1-\tr(P\Rho)^{\otimes n}\Pi^n\right)}\\
                   &\geq 1-\lambda-4\sqrt{N(P\Rho)}e^{-\mu(P\Rho)^2\alpha^2}\ .
  \end{split}\end{equation*}
  Since by lemma~\ref{lemma:variance:typical:proj}
  $$\Pi^n(P\Rho)^{\otimes n}\Pi^n\leq\Pi^n\exp\left(-nH(P\Rho)+Kd\alpha\sqrt{n}\right)$$
  we conclude
  $$\tr B\geq\tr\left(B\Pi^n\right)
          \geq\left(1-\lambda-4\sqrt{N(P\Rho)}e^{-\mu(P\Rho)^2\alpha^2}\right)\cdot
                                           \exp\left(nH(P\Rho)-Kd\alpha\sqrt{n}\right),$$
  and with $\dim{\cal K}=\tr\1_{{\cal K}}=\tr B$ the proof is complete.
\end{beweis}
\begin{cor}
  \label{cor:error:convergence}
  Let $E_n=o(n)$ and $\lambda_n\leq 1-e^{-E_n}$. Then for every sequence
  $(\varepsilon_{n*},\delta_{n*})$ of $(n,\lambda_n)_{\bar{F}}$--codes with
  arbitrary encoding for the q--DMS $(\Rho,P)$
  $$\liminf_{n\rightarrow\infty} R(\varepsilon_{n*},\delta_{n*})\geq H(P\Rho).$$
  \phantom{.}\qed
\end{cor}
\begin{bem}
  The proof of the above theorem is remarkable in that
  it employs a positive operator which is not necessarily bounded by $\1$
  (this is why we could not directly apply the shadow).
  Even though it has consequently no interpretation as a physical
  measurement (maybe it has one as a \emph{quantity}),
  it can be analyzed to give information about the coding scheme.
\end{bem}

\section{Relation to classical source coding}
\label{sec:source:general}
Consider a slight variation of our initial model: $\Rho$ is now a set
of pure states on a finite dimensional C${}^*$--algebra $\alg{A}$
(which is a direct sum of full matrix algebras $\alg{L}({\cal H})$),
and consider only $\bar{F}$ as a fidelity measure.
A major (and extremal) example is a classical source, i.e.
$\alg{A}=\C\fset{X}$ is commutative, with a finite set $\fset{X}$,
and w.l.o.g. $\Rho=\fset{X}$ (all possible pure states).
The general case may be seen as an interpolation between this and
the quantum case $\alg{A}=\alg{L}({\cal H})$.
\par
Observe that since $P\Rho\in\alg{A}_*$ we find the typical projectors
$\Pi^n$ in $\alg{A}^{\otimes n}$ (note that for $\alg{A}=\C\fset{X}$
such a projector is given just by a set of typical sequences from
$\fset{X}^n$). This means that the {\sc Schumacher} and
{\sc JHHH}--schemes make sense by just replacing $\alg{L}({\cal H})$
in the definitions by $\alg{A}$, without changing the fidelity values
(note again that for $\alg{A}=\C\fset{X}$ the average fidelity is just
the classical success probability). The strong converse need not be modified
at all as $\alg{L}({\cal H})$ is already the most ``spacious'' algebra
imaginable. Thus we arrive (with obvious definitions) at
\begin{satz}
  \label{satz:general:source:coding}
  For all $\lambda\in(0,1)$ the arbitary and quantum encoding rates
  of the discrete memoryless source $(\Rho,P)$ on the C${}^*$--algebra
  $\alg{A}$ are equal to the {\sc von Neumann} entropy of the ensemble
  $(\Rho,P)$:
  $$R_{q,\bar{F}}(\lambda)=R_{a,\bar{F}}(\lambda)=H(P\Rho).$$
  \phantom{.}\qed
\end{satz}

\section{Open questions}
\label{sec:source:open}

\paragraph{Dimension}
Why stay with finite dimensional spaces? In fact there is no obstruction to defining
sensibly a {\sc Schumacher} scheme, indeed the original paper of
\cite{schumacher:qucoding} had no dimension restriction, instead (implicitly)
requiring bounded variance of the information density, i.e. in the present
setting the condition $\tr\left(\rho(\log\rho)^2\right)<\infty$.
Then the typical projector of choice is the entropy typical one,
and in fact the reader may as an exercise translate the coding theorem
and our strong converse to this situation.

\paragraph{Memory}
It appears that no one has formalized the concept of
coding a ``quantum {\sc Markov} chain''.

\paragraph{Lossless coding}
It might be worthwhile to try and to convert the
techniques of {\sc Huffman} coding, and especially of arithmetic coding
of the source to quantum sources. See \cite{braunstein:qhuffman}
for a discussion.

\paragraph{Rate distortion theory} 
Develop further a rate distortion theory: the start to this was made
by \cite{bendjaballah:rate:dist}, and a short note
of \cite{barnum:rate:distortion}.

\paragraph{Refined resource analysis}
A not yet investigated (and perhaps most interesting)
problem is, how much ``quantum'' one actually
needs to compress the source $(\Rho,P)$: whereas $\dim{\cal K}$ is shown
by theorem~\ref{satz:general:source:coding} to be a good resource measure, it
is oblivious to the difference between an orthogonal ensemble (for whose coding
a commutative algebra, i.e. a classical system, suffices), and a highly
non--orthogonal one (which presumably needs all the quantum resources, i.e.
possibilities of superpositions, of $k$ degrees of freedom). As a measure of this
``quantum'' resource I propose the following:
\par
A coding scheme is a pair
$(\varepsilon_*,\delta_*)$ with
\begin{align*}
  \varepsilon_*: &\Rho^n\longrightarrow\alg{K}_*
                                        \ \text{  a mapping,}\\
  \delta_*:      &\alg{K}_*\longrightarrow\alg{A}_*^{\otimes n}
                                        \ \text{  a quantum operation,}
\end{align*}
where $\alg{K}$ is a finite dimensional C${}^*$--algebra. Quantum and arbitrary
encoding schemes are as before. Observe that $\tr\1_{\alg{K}}$ takes now
the place of the previous $\dim{\cal K}$. Define the, say,
\emph{rate of superposition} as
$$r(\varepsilon_*,\delta_*)
           =\frac{1}{n}\left(\log\dim_{\C}\alg{K}-\log\tr\1_{\alg{K}}\right).$$
Observe that $0\leq r(\varepsilon_*,\delta_*)\leq \frac{1}{n}\log\tr\1_{\alg{K}}$,
with $r(\varepsilon_*,\delta_*)=0$ iff $\alg{K}$ is commutative.
\par
Now define for $\lambda\in(0,1)$, $R\geq 0$ the
$\lambda$--rates of superposition with arbitrary and quantum encoding:
\begin{equation*}\begin{split}
  r_{a,\bar{F}}(\lambda,R) &=\limsup_{n\rightarrow\infty}
         \min\{r(\varepsilon_*,\delta_*):\ (\varepsilon_*,\delta_*)\text{ an }
               (n,\lambda)_{\bar{F}}\text{--code (arb. enc.), }
                                              R(\varepsilon_*,\delta_*)\leq R\},\\
  r_{q,\bar{F}}(\lambda,R) &=\limsup_{n\rightarrow\infty}
         \min\{r(\varepsilon_*,\delta_*):\ (\varepsilon_*,\delta_*)\text{ an }
               (n,\lambda)_{\bar{F}}\text{--code (qu. enc.), }
                                              R(\varepsilon_*,\delta_*)\leq R\}.
\end{split}\end{equation*}
It is obvious that $r_{a,\bar{F}}$ and $r_{q,\bar{F}}$ are nonincreasing functions
of $R$, and that both are upper bounded by $H(P\Rho)$.
The problem is now to analyze $r_{a,\bar{F}}$ and $r_{q,\bar{F}}$
depending on $\lambda$ and $R$.
\par
\begin{itemize}
  \item It is clear that $r_{a,\bar{F}}(\lambda,R)=0$ if $R$ is large enough
    ($R=H(P)$ suffices). It would be
    interesting to determine the exact threshold, the value at
    $R=H(P\Rho)$ and the behavior between these points.
    In any case, I conjecture that
    $r_{a,\bar{F}}(\lambda,R)$ does not depend on $\lambda\in(0,1)$.
  \item I conjecture further that $r_{q,\bar{F}}$ depends neither on $\lambda\in(0,1)$
    nor on $R>H(P\Rho)$. If this is true $r_{q,\bar{F}}$ is an interesting
    ensemble property of $(\Rho,P)$.
\end{itemize}

% QDissertation: Quantum channel coding

\chapter{Quantum Channel Coding}
\label{chap:channel}

\thispagestyle{myheadings}

In this chapter we introduce the notion of a quantum channel. From the
beginning we focus on  the
\emph{product state capacity for transmission of classical information},
and prove coding theorem and strong converse, even for nonstationary
channels. In the finite stationary case we can sharpen our rate estimates
and derive some bounds for the reliability function.
As a corollary to our strong converse we obtain another proof of the
{\sc Holevo} bound.

\section{Quantum channels and codes}
\label{sec:channel:qchannel}
The following definition is after \cite{holevo:channels}:
a (discrete memoryless) \emph{quantum channel} (q--DMC)
is a completely positive, trace preserving mapping $\varphi_*$ 
from the states on a C${}^*$--algebra $\alg{A}$ into
the states on $\alg{L}({\cal H})$, where $d=\dim{\cal H}$ is assumed to
be finite.
\par
A \emph{nonstationary q--DMC} is a sequence $(\varphi_{n*})_{n\in\N}$
of q--DMCs, with a global Hilbert space ${\cal H}$. This extends the concept
of q--DMCs which are contained as \emph{constant} sequences.
\par
An $n$--\emph{block code} for a nonstationary quantum channel
$(\varphi_{n*})_n$ is a pair $(f,D)$, where $f$ is a mapping from a finite
set $\fset{M}$ into $\alg{S}(\alg{A}_1)\times\cdots\times\alg{S}(\alg{A}_n)$,
and $D$ is an observable on $\alg{L}({\cal H})^{\otimes n}$
indexed by $\fset{M}'\supset\fset{M}$,
i.e. a partition of $\1$ into positive operators $D_m$, $m\in\fset{M}'$.
The (maximum) error probability of the code is defined as
$$e(f,D)=\max\{1-\tr(\varphi_*^{\otimes n}(f(m))D_m):m\in\fset{M}\}.$$
We call $(f,D)$ an $(n,\lambda)$--\emph{code}, if $e(f,D)\leq\lambda$.
The \emph{rate} of an $n$--block code is defined as $\frac{1}{n}\log|\fset{M}|$.
Finally define $N(n,\lambda)$ as the maximal size (i.e. $|\fset{M}|$) of
an $(n,\lambda)$--code.
\begin{bem}
  Observe that we did not allow all joint states of the system
  $\alg{A}_1\otimes\cdots\otimes\alg{A}_n$ as code words,
  but only \emph{product states}. This is the restriction under which the
  current theory was done. It is unknown if the following
  theorem~\ref{satz:coding:converse} is still true in the more general model:
  maybe higher capacities can be achieved
  there, see the discussion of \cite{schumacher:capacity}.
\end{bem}
With our restriction we may without harm identify a channel mapping $\varphi_*$
with its image $\alg{W}_\varphi=\varphi_*(\alg{S}(\alg{A}))$
in the set of states on $\alg{L}({\cal H})$ (for
then the image of an input state under $\varphi^n_*$ is a product state on
$\alg{L}({\cal H})^{\otimes n}$).
\par
Generalizing, a nonstationary quantum channel
is now a sequence $(\alg{W}_n)_n$ of arbitrary (measureable) subsets of states on
a fixed $\alg{L}({\cal H})$.
In this spirit we reformulate the definition of
an $n$--block code as a pair $(f,D)$ with a mapping
$f:\fset{M}\rightarrow\alg{W}_1\times\cdots\times\alg{W}_n$\footnote{Where we identify
  $(\rho_1,\ldots,\rho_n)$ with $\rho^n=\rho_1\otimes\cdots\otimes\rho_n$.}
and $D$ as before.
The main result of the present chapter
(to be proved in the following sections) is
\begin{satz}
  \label{satz:coding:converse}
  Let $(\alg{W}_1,\alg{W}_2,\ldots)$ a nonstationary q--DMC, and
  $$C(\alg{W_i})=\sup_{P\text{ p.d. on }\alg{W}_i} I(P;\alg{W}_i)$$
  (with $I(P;\alg{W})=H(P\alg{W})-H(\alg{W}|P)$, see
  remark~\ref{bem:holevo:bound}). Then for every $\lambda\in(0,1)$
  $$\left|\frac{1}{n}\log N(n,\lambda)-\frac{1}{n}\sum_{i=1}^{n}C(\alg{W}_i)\right|
                                          \rightarrow 0\text{  as }n\rightarrow\infty.$$
\end{satz}
\begin{beweis}
  Combine the coding theorem~\ref{satz:maximal:codes}
  and the strong converse theorem~\ref{satz:channel:strong:converse}.
\end{beweis}
This theorem justifies the name \emph{capacity} (of the channel $\alg{W}$)
for the quantity $C(\alg{W})$, even in the strong sense of
\cite{wolfowitz:coding}. Observe that this theorem is a
quantum generalization of a theorem by \cite{ahlswede:nichtstationaer}.
\begin{bem}
  \label{bem:classical:is:contained}
  It should be clear that the same (including proofs) applies if the output
  system $\alg{L}({\cal H})$ is replaced by a $*$--subalgebra $\alg{A}$.
\end{bem}

\section{Maximal code construction}
\label{sec:channel:maximal}
\begin{satz}[Maximal codes]
  \label{satz:maximal:codes}
  For $0<\tau,\lambda<1$ there is a constant $K'$ and $\delta>0$ such that
  for every nonstationary q--DMC $(\alg{W}_i)_i$, distributions $P_i$ on $\alg{W}_i$ and
  $\fset{A}\subset\alg{W}^n$ with $P^n(\fset{A})\geq\tau$
  there exists an $(n,\lambda)$--code
  $(f,D)$ with the properties
  $$\forall m\in\fset{M}\quad
               f(m)\in\fset{A}\text{ and }\tr D_m\leq\tr\Pi^n_{H,f(m),\delta}\ ,$$
  \begin{equation*}\begin{split}
    \log|\fset{M}| &\geq H(P^n\alg{W}^n)-H(\alg{W}^n|P^n)-K'\sqrt{n}\\
                   &= \sum_{i=1}^n \left(H(P_i\alg{W}_i)-H(\alg{W}_i|P_i)\right)-K'\sqrt{n}\ .
  \end{split}\end{equation*}
\end{satz}
\begin{beweis}
  On every $\alg{W}_i$ the entropy $H$ is a random variable with
  expectation $H(\alg{W}_i|P_i)$ and variance bounded by $(\log d)^2$.
  Define $\delta=\max\{\sqrt{2/\lambda},\sqrt{2/\tau}\log d\}$, then by
  {\sc Chebyshev}'s inequality the set
  $$\fset{A}'\!=\{\rho^n\!\in\!\fset{A}:\!
       \left|H(\rho^n)-\!\sum_{i=1}^n \!H(\alg{W}_i|P_i)\right|\!\leq\!\delta\sqrt{n}\}$$
  has probability $P^n(\fset{A}')\geq \tau/2$.
  Now let $(f,D)$ a maximal $(n,\lambda)$--code with
  $$\forall m\in\fset{M}\quad f(m)\in\fset{A}'\text{ and }
                                                     \tr D_m\leq\tr\Pi^n_{H,f(m),\delta}\ .$$
  Define $B=\sum_{m\in\fset{M}} {D}_m$.
  We claim that with $\eta=\min\{1-\lambda,\lambda^2/{32}\}$
  $$\forall\rho^n\in\fset{A}'\quad \tr(\rho^nB)\geq\eta.$$
  This is clear for codewords, and true for the other states because
  otherwise we could extend our code by the codeword $\rho^n$ with corresponding
  observable operator
  $$D=\sqrt{\1-B}\Pi^n_{H,\rho^n,\delta}\sqrt{\1-B}$$
  which clearly satisfies the trace bound
  (note that $B+D\leq\1$):
  to see this apply lemma~\ref{lemma:tender:operator} to obtain
  $$\|\rho^n-\sqrt{\1-B}\rho^n\sqrt{\1-B}\|_1\leq\sqrt{8\eta}\leq \frac{\lambda}{2}\ .$$
  Thus
  \begin{equation*}\begin{split}
    \tr(\rho^n\sqrt{\1-B}\Pi^n_{H,\rho^n,\delta}\sqrt{\1-B})
       &=\tr\Bigl(\rho^n\Pi^n_{H,\rho^n,\delta}\Bigr)\!
         -\tr\!\left((\rho^n-\sqrt{\1-B}\rho^n\sqrt{\1-B})\Pi^n_{H,\rho^n,\delta}\right)\\
       &\geq\left(1-\frac{\lambda}{2}\right)-\frac{\lambda}{2}=1-\lambda.
  \end{split}\end{equation*}
  So $B$ is an $\eta$--shadow of $\fset{A}'$, and consequently
  $$\tr(P^n\alg{W}^n B)\geq \eta\tau/2\ .$$
  By lemma~\ref{lemma:entropy:typical:proj:size} there is $K$ with
  $$\tr B\geq \exp\left(\sum_{i=1}^n H(P_i\alg{W}_i)-K\sqrt{n}\right).$$
  On the other hand
  $$\tr B=\sum_{m\in\fset{M}} \tr{D}_m
            \leq \sum_{m\in\fset{M}} \tr\Pi^n_{H,f(m),\delta}
            \leq |\fset{M}|\exp\left(\sum_{i=1}^n H(\alg{W}_i|P_i)+2\delta\sqrt{n}\right),$$
  the last inequality again by
  lemma~\ref{lemma:entropy:typical:proj:size}, and we are done.
\end{beweis}
\begin{bem}
  We can strengthen the theorem to that all the $D_m$ are projectors. The proof
  goes through unchanged but for the construction of the code extension: there
  we take the support of the above $D$. The trace estimate holds because the trace
  of a projector is the dimension of its range.
\end{bem}
\begin{bem}
  The above coding theorem --- for stationary channels and with slightly weaker
  bounds --- was first proved by \cite{holevo:qucapacity} (and independently
  by \cite{schumacher:capacity}), building on ideas of
  \cite{hausladen:qucap:purecase} for the \emph{pure state channel}.
\end{bem}

\section{Strong converse}
\label{sec:channel:converse}
\begin{satz}[Strong Converse]
  \label{satz:channel:strong:converse}
  For every $\lambda\in(0,1)$ and $\epsilon>0$ there is $n_0=n_0(\lambda,\epsilon)$
  such that for every $n\geq n_0$ and every nonstationary q--DMC $(\alg{W}_i)_i$
  $$\log N(n,\lambda)\leq \sum_{i=1}^n C(\alg{W}_i) +n\epsilon.$$
\end{satz}
Before proving this we need to follow a short technical digression:

\paragraph{Approximation of channels}
We have continuum many states on $\alg{L}({\cal H})$ to deal with,
and even more channels, so we
introduce a simple approximation scheme: a partition $\alg{Z}$
of $\alg{S}(\alg{L}({\cal H}))$
into $t$ sections $\alg{Z}_1,\ldots,\alg{Z}_t$ each having $\|\cdot\|_1$--diameter
at most $\theta>0$
is called $\theta$--\emph{fine}. The relation of the parameters $t$ and $\theta$ is:
\begin{lemma}
  \label{lemma:partition}
  For any $\theta>0$ there is a $\theta$--fine partition of $\alg{S}(\alg{L}({\cal H}))$
  into $t\leq C\theta^{-d^2}$ sections, with a constant $C$ depending only on
  $d$.
\end{lemma}
\begin{beweis}
  The set of states is $\|\cdot\|_1$--isometric to the set of positive
  $d\times d$--matrices with trace one. This is obviously a compact set of real
  dimension $d^2-1$. It is contained in the set of all selfadjoint matrices
  with the real and imaginary parts of all its entries in the interval $[-1,1]$
  which is geometrically a $d^2$--dimensional cube. Now obviously we may decompose
  this cube into ${(2\sqrt{2}d^3)^{d^2}}{\theta^{-d^2}}$ cubes
  of edge length $\theta/(d^3\sqrt{2})$. We claim that
  for two states $\rho,\rho'$ in the same small cube $\|\rho-\rho'\|_1\leq\theta$.
  But this follows from the fact that a matrix with all entries absolutely
  bounded by $\epsilon$ has all its eigenvalues bounded by $d^2\epsilon$, which is
  straightforward (and rather crude).
\end{beweis}
\par
We close the digression with two definitions:
the $\alg{Z}$--\emph{type} of a state $\rho^n$ is the empirical distribution on
sections in which $\alg{Z}_j$ has weight proportional to the number of $\rho_i\in\alg{Z}_j$.
The $\alg{Z}$--\emph{class} of a channel $\alg{W}_i$ is the set of sections $\alg{Z}_j$
which have nonempty intersection with $\alg{W}_i$.\par
Obviously the number of $\alg{Z}$--types is bounded by $(n+1)^t$, the number
of $\alg{Z}$--classes is bounded by $2^t$.
\bigskip\par
\begin{beweis}[of theorem~\ref{satz:channel:strong:converse}]
  Let $(f,D)$ an $(n,\lambda)$--code. Consider a $\theta$--fine partition $\alg{Z}$ of 
  $\alg{S}(\alg{L}({\cal H}))$ into $t$ sections and choose
  representatives $\sigma_j\in\alg{Z}_j$. For every ($\alg{Z}$--)class $\gamma$ let $I_\gamma$
  the set of indices $i\in[n]$ with $\alg{W}_i$ of class $\gamma$.
  Consider the ($\alg{Z}$--)types of
  the restrictions $f(m)^{I_\gamma}$ of the codewords to the positions $I_\gamma$.
  For each $\gamma$ with $I_\gamma\neq\emptyset$
  there is a type $P_\gamma$ occuring in a fraction of
  at least $(|I_\gamma|+1)^{-t}$ of the codewords. Successively choosing subcodes
  we arrive at a code $\fset{M}'$ with at least $|\fset{M}|\cdot(n+1)^{-t2^t}$ codewords
  and $f(m)^{I_\gamma}$ of type $P_\gamma$ for all $m\in\fset{M}'$,
  whenever $I_\gamma\neq\emptyset$.\par
  For each $i$, $i\in I_\gamma$ choose states $\tilde{\rho}_{ij}\in\alg{W}_i\cap\alg{Z}_j$
  and define a distribution $P_i$ on $\alg{W}_i$ by
  $P_i(\tilde{\rho}_{ij})=P_\gamma(j)$. Finally let
  $\tilde{\rho}_{i\gamma}=P_i\alg{W}_i=\sum_j P_\gamma(j)\tilde{\rho}_{ij}$
  and $\tilde{\sigma}_\gamma=\sum_j P_\gamma(j)\tilde{\sigma}_j$.
  \par
  For classes $\gamma$ with $|I_\gamma|\geq n2^{-2t}$ (which we call
  \emph{good}) define (with some $\delta>0$)
  $$\Pi_\gamma=\Pi^{I_\gamma}_{C,\tilde{\sigma}_\gamma,\delta+\theta\sqrt{|I_\gamma|}}
                    \quad\text{ in }\alg{L}({\cal H})^{\otimes I_\gamma}.$$
  For bad $\gamma$ define $\Pi_\gamma=\1$ in $\alg{L}({\cal H})^{\otimes I_\gamma}$.
  Then by the weak law lemma~\ref{lemma:weak:law} for \emph{every} $\gamma$
  $$\forall m\in\fset{M}'\quad \tr(f(m)^{I_\gamma}\Pi_\gamma)\geq 1-\frac{1}{\delta^2}$$
  and thus defining $\Pi_0=\bigotimes_\gamma \Pi_\gamma$ we obtain
  $$\forall m\in\fset{M}'\quad \tr(f(m)\Pi_0)\geq 1-\frac{2^t}{\delta^2}\ .$$
  Now assume that $n2^{-2t}$ is large enough and $\theta$ is small enough
  so that according to
  lemmata~\ref{lemma:constant:typical:size} and~\ref{lemma:H:continuous} we have for
  good $\gamma$
  $$\tr \Pi_\gamma \leq\exp\left(|I_\gamma|(H(\tilde{\sigma}_\gamma)+\epsilon)\right)
                   \leq\exp\left(\sum_{i\in I_\gamma}
                      H(\tilde{\rho}_{i\gamma})+2|I_\gamma|\epsilon\right).$$
  Hence we get (collecting the contributions of good and bad classes)
  $$\tr\Pi_0\leq
     \exp\left(\sum_{i=1}^n H(\tilde{\rho}_{i\gamma})+2n\epsilon+n2^{-t}\log d\right).$$
  Now consider the code $(f',D')$ with $f'=f|_{\fset{M}'}$ and $D_m'=\Pi_0 D_m\Pi_0$ for
  $m\in\fset{M}'$. By the above considerations and
  lemma~\ref{lemma:tender:operator}
  it is an $(n,\lambda+\sqrt{8}2^{t/2}\delta^{-1})$--code.
  Assuming $\sqrt{8}2^{t/2}\delta^{-1}\leq\frac{1-\lambda}{2}$,
  by lemma~\ref{lemma:entropy:typical:proj:size} we get
  $$\tr D_m'\geq \exp\left(\sum_{i=1}^n H(\alg{W}_i|P_i)-n\epsilon\right)$$
  if $n$ is large enough. So we arrive at
  $$\tr\Pi_0 \geq \sum_{m\in\fset{M}'} D_m'
             \geq |\fset{M}'|\exp\left(\sum_{i=1}^n H(\alg{W}_i|P_i)-n\epsilon\right),$$
  and thus
  \begin{equation*}\begin{split}
    |\fset{M}| &\leq (n+1)^{t2^t}\exp\left(\sum_{i=1}^n 
                                      H(P_i\alg{W}_i)-H(\alg{W}_i|P_i))
                                                 +3n\epsilon+n2^{-t}\log d\right)\\
               &\leq \exp\left(\sum_{i=1}^n
                             (H(P_i\alg{W}_i)-H(\alg{W}_i|P_i))+5n\epsilon\right)\\
               &\leq \exp\left(\sum_{i=1}^n C(\alg{W}_i)+5n\epsilon\right)
  \end{split}\end{equation*}
  if we can adjust our parameters accordingly:
  choose for example $t=\lceil\frac{1}{3}\log n\rceil$ with
  $\theta\leq\left(\frac{3C}{\log n}\right)^{d^{-2}}$ (which is possible by
  lemma~\ref{lemma:partition}), $\delta=n^{1/3}$, and let $n$ large enough.
\end{beweis}
\begin{bem}
  The \emph{weak converse} is already a consequence of the information bound
  of \cite{holevo:bound}, see theorem~\ref{satz:holevo:bound}, together
  with subadditivity of quantum mutual information (corollary~\ref{cor:info:subadd})
  and the classical {\sc Fano} inequality (see theorem~\ref{satz:fano:inequality}).
\end{bem}

\section{Refined analysis for stationary channels}
\label{sec:channel:stationary}
From this point on we restrict ourselves to the finite and stationary case.
\par
Let $W:\fset{X}\rightarrow\alg{S}(\alg{L}({\cal H}))$
a finite q--DMC, mapping $x\in\fset{X}$ to the state $W_x$,
with a set $\fset{X}$, say of cardinality $|\fset{X}|=a<\infty$,
for a fixed complex Hilbert space
${\cal H}$ of finite dimension $d$ (i.e., in slight variation to the previous
sections, we label the set of channel states by $\fset{X}$).
We will have occasion to consider
other channels, say $V$, implicitely all with the same $\fset{X}$.
Note that we drop here the subscript $*$ for state maps, to be closer
to the notation in use in the literature.
\par
For an $n$--block code $(f,D)$ for $W$ we will here interpret
$f$ as a mapping from the finite set $\fset{M}$ into $\fset{X}^n$.
The (maximum) error probability of the code then reads as
$$e(f,D)=\max\{1-\tr(W_{f(m)}D_m):m\in\fset{M}\}.$$
(For $f(m)=x^n\in\fset{X}^n$ we adopt the convention
$W_{f(m)}=W_{x^n}=W_{x_1}\otimes\cdots\otimes W_{x_n}$).
The \emph{rate} of an $n$--block code is defined as $\frac{1}{n}\log|\fset{M}|$.
Recall that $N(n,\lambda)$ is the maximal size (i.e. $|\fset{M}|$) of
an $(n,\lambda)$--code, and define
$$e_{\min}(n,R)=\min\{e(f,D):(f,D)\text{ is }
                                n\text{--block code, }|\fset{M}|\geq\exp(nR)\}.$$
Finally for states $\rho$ and $\nu$, and another channel
$V$ and p.d. $P$ on $\fset{X}$ let
\begin{align*}
  D(\nu\|\rho) &=\tr\bigl(\nu(\log\nu-\log\rho)\bigr)\\
  D(V\| W|P)   &=\sum_{x\in\fset{X}} P(x)D(V_x\| W_x),
\end{align*}
the (conditional) quantum I--divergence, see appendix~\ref{chap:quprob},
section {\em Entropy and divergence}.
\par
The rewards of our restriction are stronger estimates on $N(n,\lambda)$,
and --- more interestingly ---
upper and lower bounds on $e_{\min}(n,R)$, which lead to nontrivial
lower and upper bounds on the reliability function of the channel.
This extends results of \cite{burnashev:holevo:reliability} from pure state to general
channels, and thus gives (partial) answers to two problems
posed by \cite{holevo:overview}.
\par
\paragraph{Some more typicalities}
We begin with an extension of lemma~\ref{lemma:variance:typical:proj}: define
the \emph{conditional variance--typical projectors} $\Pi^n_{V,{W},\delta}(x^n)$
with $x^n\in\fset{X}^n$ to be
$$\Pi^n_{V,{W},\delta}(x^n)=\bigotimes_{x\in\fset{X}} \Pi^{I_x}_{V,W_x,\delta}\ ,$$
where $I_x=\{i\in[n]:x_i=x\}$.
\begin{lemma}
  \label{lemma:cond:variance:typical:proj}
  For every $x^n\in\fset{X}^n$ of type $P$, and with $\Pi^n=\Pi^n_{V,{W},\delta}(x^n)$
  $$\tr W_{x^n}\Pi^n\geq 1-\frac{ad}{\delta^2}$$
  $$\Pi^n\exp\left(-nH({W}|P)-Kd\sqrt{a}\delta\sqrt{n}\right)
         \leq \Pi^n W_{x^n}\Pi^n
         \leq \Pi^n\exp\left(-nH({W}|P)+Kd\sqrt{a}\delta\sqrt{n}\right)$$
  \begin{align*}
    \tr\Pi^n_{V,{W},\delta}(x^n) &\leq\exp\left(nH({W}|P)+Kd\sqrt{a}\delta\sqrt{n}\right)\\
    \tr\Pi^n_{V,{W},\delta}(x^n) &\geq\left(1-\frac{ad}{\delta^2}\right)
                                       \exp\left(nH({W}|P)-Kd\sqrt{a}\delta\sqrt{n}\right).
 \end{align*}
 Every $\eta$--shadow $B$ of $W_{x^n}$ satifies
 $$\tr B\geq\left(\eta-\frac{ad}{\delta^2}\right)
                                    \exp\left(nH({W}|P)-Kd\sqrt{a}\delta\sqrt{n}\right).$$
\end{lemma}
\begin{beweis}
  The first inequality is just $a$ times the estimate from
  lemma~\ref{lemma:variance:typical:proj}.
  The estimate for $\Pi^n_{V,{W},\delta}(x^n)W_{x^n}\Pi^n_{V,{W},\delta}(x^n)$
  follows from piecing together the estimates for the
  $\Pi^{I_x}_{V,W_x,\delta}$ in the same lemma (using
  $\sum_{x\in\fset{X}} \sqrt{P(x)}\leq \sqrt{a}$). The rest follows
  from the shadow bound lemma~\ref{lemma:abstract:shadow:bound}.
\end{beweis}
From this we get the following
\begin{lemma}
  \label{lemma:weak:law:exact}
  Let $\delta>0$ and $x^n\in\fset{X}^n$ of type $P$. Then
  $$\tr(W_{x^n}\Pi^n_{V,P{W},\delta\sqrt{a}})\geq 1-\frac{ad}{\delta^2}\ .$$
\end{lemma}
\begin{beweis}
  Diagonalize $P{W}=\sum_j q_j\pi_j$, and let
  $\kappa_*:\alg{L}({\cal H})_*\rightarrow\alg{L}({\cal H})_*$ the conditional
  expectation be defined by $\kappa_*(\sigma)=\sum_j \pi_j\sigma\pi_j$.
  We claim that
  $$\Pi^n_{V,P{W},\delta\sqrt{a}}\geq \Pi^n_{V,\kappa_*{W},\delta}(x^n).$$
  Indeed let $\pi_{j_1}\otimes\cdots\otimes\pi_{j_n}$ one of the product
  states constituting
  $\bigotimes_{x\in\fset{X}} \Pi^{I_x}_{V,\kappa_*(W_x),\delta}$, i.e.
  with $\kappa_*(W_x)=\sum_j q_{j|x}\pi_j$
  $$\forall x\!\in\!\fset{X}\,\forall j\ \left| N(j|j^{I_x})-q_{j|x}|I_x| \right|\leq
                             \delta\sqrt{|I_x|}\sqrt{q_{j|x}(1-q_{j|x})}.$$
  Hence (with $|I_x|=P(x)n$)
  \begin{equation*}\begin{split}
    | N(j|j^n)-q_jn |
         &\leq\sum_{x\in\fset{X}} \left| N(j|j^{I_x})-q_{j|x}|I_x| \right| \\
         &\leq\sum_{x\in\fset{X}} \delta\sqrt{n}\sqrt{P(x)}\sqrt{q_{j|x}(1-q_{j|x})}\\
         &\leq\delta\sqrt{a}\sqrt{n}\sqrt{\sum_{x\in\fset{X}} P(x)q_{j|x}(1-q_{j|x})}\\
         &\leq\delta\sqrt{a}\sqrt{n}\sqrt{q_j(1-q_j)},
  \end{split}\end{equation*}
  the last inequality by concavity of the map $x\mapsto x(1-x)$,
  and $q_j=\sum_{x\in\fset{X}} P(x)q_{j|x}$.
  \par
  Thus we can estimate
  \begin{equation*}\begin{split}
    \tr(W_{x^n}\Pi^n_{V,P{W},\delta\sqrt{a}})
        &=\tr\left((\kappa_*^{\otimes n}W_{x^n})\Pi^n_{V,P{W},\delta\sqrt{a}}\right)\\
        &\geq \tr\left((\kappa_*^{\otimes n}W_{x^n})\Pi^n_{V,\kappa_*{W},\delta}
                  (x^n)\right)\\
        &\geq 1-\frac{ad}{\delta^2}\ ,
  \end{split}\end{equation*}
  the last line by lemma~\ref{lemma:cond:variance:typical:proj}.
\end{beweis}
\par
Of particular interest are the variance--typical projectors with $\delta=0$, i.e.
the $\Pi^n_\rho=\Pi^n_{V,\rho,0}$ and $\Pi^n_W(x^n)=\Pi^n_{V,W,0}(x^n)$,
which we call \emph{exact types}.\par
For the following fix diagonalizations $\rho=\sum_j q_j\pi_j$
and $W_x=\sum_j q_{j|x}\pi_{xj}$. The commutative algebras
$\C[\pi_j|j]$ and $\C[\pi_{xj}|j]$ (which are maximal commutative subalgebras
of the \emph{commutants} $\C[\rho]'$ and $\C[W_x]'$) will be important below.
\begin{lemma}
  \label{lemma:exact:type:weight}
  For $\nu\in\C[\rho]'$ we have
  $$\Pi^n_{\nu}\rho^{\otimes n}\Pi^n_{\nu}=\Pi^n_{\nu}
                          \exp\left(-nD(\nu\|\rho)-nH(\nu)\right)$$
  $$(n+1)^{-d}\exp(nH(\nu))\leq\tr\Pi^n_{\nu}\leq\exp(nH(\nu)).$$
  For $V_x\in\C[W_x]'$ and $x^n\in\fset{X}^n$ of type $P$
  $$\Pi^n_{V}(x^n)W_{x^n}\Pi^n_{V}(x^n)=
            \Pi^n_{V}(x^n)\exp\left(-nD({V}\|{W}|P)-nH({V}|P)\right)$$
  $$(n+1)^{-ad}\exp(nH(V|P))\leq\tr\Pi^n_{V}(x^n)\leq\exp(nH(V|P)).$$
\end{lemma}
\begin{beweis}
  The first equation is straightforward. To estimate $\tr\Pi^n_\nu$ let
  $\rho=\nu$ and note that
  $$(n+1)^{-d}\leq\tr(\nu^{\otimes n}\Pi^n_\nu)\leq 1.$$
  There the upper bound is trivial, whereas the lower bound is by
  type counting, i.e. observing that in the decomposition
  $\1=\sum_{\hat{\nu}\in\C[\pi_j|j]} \Pi^n_{\hat{\nu}}$
  there appear at most $(n+1)^d$ nonzero terms, and the fact
  that for such $\hat{\nu}$ the quantity $\tr(\nu^{\otimes n}\Pi^n_{\hat{\nu}})$
  is maximized with $\hat{\nu}=\nu$ (Compare \cite{csiszar:koerner}, lemma 1.2.3).
  The second part of the lemma follows from the first by collecting
  positions of equal letters in $x^n$.
\end{beweis}
\begin{cor}
  \label{cor:exact:type:weight}
  If $\nu\in\C[\rho]'$ and $\Pi^n_\nu\neq 0$ then
  $$(n+1)^{-d}\exp(-nD(\nu\|\rho))
                      \leq\tr(\rho^{\otimes n}\Pi^n_\nu)\leq\exp(-nD(\nu\|\rho)).$$
  \phantom{.}\hfill\qed
\end{cor}
\par
Define for a state $\rho$, channel ${W}$, $x^n\in\fset{X}^n$ of type $P$,
and a real number $L$:
\begin{align*}
  \Pi^n_{\rho,H(\cdot)\leq L}       &=\sum_{\nu\in\C[\pi_j|j],H(\nu)\leq L} \Pi^n_{\nu}\\
  \Pi^n_{\rho,H(\cdot)\geq L}       &=\sum_{\nu\in\C[\pi_j|j],H(\nu)\geq L} \Pi^n_{\nu}\\
  \Pi^n_{{W},H(\cdot|P)\leq L}(x^n) &=\sum_{V_x\in\C[\pi_{xj}|j],H(V|P)\leq L} \Pi^n_{V}(x^n)\\
  \Pi^n_{{W},H(\cdot|P)\geq L}(x^n) &=\sum_{V_x\in\C[\pi_{xj}|j],H(V|P)\geq L} \Pi^n_{V}(x^n).
\end{align*}
\begin{lemma}
  \label{lemma:universal:proj:weight}
  For $\rho$, $W$, $x^n\in\fset{X}^n$ of type $P$, and $L$ as above
  $$\tr\left(\Pi^n_{W,H(\cdot|P)\leq L}(x^n)\right)\leq (n+1)^{ad}\exp(nL)$$
  $$\tr\left(W_{x^n}\Pi^n_{W,H(\cdot|P)\leq L}(x^n)\right)
            \geq 1-(n+1)^{ad}\exp\left(-n\cdot\!\inf_{H(V|P)>L}D(V\| W|P)\right)$$
  $$\tr\left(\rho^{\otimes n}\Pi^n_{\rho,H(\cdot)\geq L}\right)
         \geq 1-(n+1)^d\exp\left(-n\cdot\!\min_{H(\nu)\leq L}D(\nu\|\rho)\right).$$
\end{lemma}
\begin{beweis}
  The inequalities all follow from lemma~\ref{lemma:exact:type:weight} and
  corollary~\ref{cor:exact:type:weight} together with type counting.
\end{beweis}
\begin{lemma}
  \label{lemma:universal:shadow}
  For $\rho$ and $L$ as above
  \begin{equation*}\begin{split}
    \Pi^n_{\rho,H(\cdot)\geq L}\rho^{\otimes n}\Pi^n_{\rho,H(\cdot)\geq L}
                   &\leq\Pi^n_{\rho,H(\cdot)\geq L}\exp\left(-n\cdot\!
                             \min_{H(\nu)\geq L}(H(\nu)+D(\nu\|\rho))\!\right)\\
                   &=   \Pi^n_{\rho,H(\cdot)\geq L}\exp\left(-nL-n\cdot\!
                             \min_{H(\nu)=L} D(\nu\|\rho)\right)\\
                   &\leq\Pi^n_{\rho,H(\cdot)\geq L}\exp\left(-nL-n\cdot\!
                             \min_{H(\nu)\leq L} D(\nu\|\rho)\right).
  \end{split}\end{equation*}
  For an $\eta$--shadow $B$ of $\rho^{\otimes n}$
  $$\tr B\geq\left(\eta-(n+1)^d\exp(-n\cdot\!\min_{H(\nu)\leq L} D(\nu\|\rho))\right)
                      \cdot\exp\left(nL+n\cdot\!\min_{H(\nu)\leq L} D(\nu\|\rho)\right).$$
\end{lemma}
\begin{beweis}
  The first estimate is directly from lemma~\ref{lemma:exact:type:weight}. To see that
  the required $\min$ is assumed at the boundary of the (convex) region where
  $H(\nu)\geq L$ observe that the minimized quantity is linear in $\nu$.\par
  For the $\eta$--shadow $B$: note that by
  lemma~\ref{lemma:universal:proj:weight} with
  $\Pi^n=\Pi^n_{\rho,H(\cdot)\geq L}$
  $$\tr\!\left(\rho^{\otimes n}\Pi^n B\Pi^n\right)
              \geq \eta-(n+1)^d\exp\left(-n\cdot\!\min_{H(\nu)\leq L} D(\nu\|\rho)\right)$$
  and the rest follows by the estimate on
  $\Pi^n\rho^{\otimes n}\Pi^n$.
\end{beweis}

\paragraph{Code bounds up to $\mathbf{O(\sqrt{n})}$ terms}
Our first result is a variation of theorem~\ref{satz:maximal:codes}:
\begin{satz}
  \label{satz:code:lowerbound}
  For every $\lambda\in(0,1)$ there is a constant
  $K(a,d,\lambda)$ such that for every q--DMC ${W}$
  $$N(n,\lambda)\geq \exp\left(nC({W})-K(a,d,\lambda)\sqrt{n}\right).$$
\end{satz}
\begin{beweis}
  Let $P$ a p.d.~on $\fset{X}$ with $C({W})=H(P{W})-H({W}|P)$.
  Let $(f,D)$ a maximal $(n,\lambda)$--code with the property
  $$\forall m\in\fset{M}\qquad
          f(m)\in\fset{T}^n_{V,P,\sqrt{2a}},\ \tr D_m\leq \tr\Pi^n_{V,{W},\delta}(f(m)),$$
  with $\delta=\sqrt{\frac{2ad}{\lambda}}$. In particular (by
  lemma~\ref{lemma:cond:variance:typical:proj})
  $$\tr D_m\leq\exp\left(nH(W|P)+(Kd\sqrt{a}\delta+Ka\sqrt{2a}\log d)\sqrt{n}\right).$$
  Let $B=\sum_{m\in\fset{M}} D_m$, we claim that for all $x^n\in\fset{T}^n_{V,P,\sqrt{2a}}$
  $$\tr(W_{x^n}B)\geq\eta=\min\{1-\lambda,\lambda^2/32\}.$$
  This is clear if $x^n$ is a code word, and true else, for otherwise we could
  extend our code with the word $x^n$ and decoding operator
  $$D'=\sqrt{\1-B}\Pi^n_{V,W,\delta}(x^n)\sqrt{\1-B}\ .$$
  This is exactly as in the proof of theorem~\ref{satz:maximal:codes}.
  Thus we arrive at
  $$\tr\left((PW)^{\otimes n}B\right)\geq \eta/2$$
  which by lemma~\ref{lemma:variance:typical:proj} implies the estimate
  $$\tr B\geq\left(\frac{\eta}{2}-\frac{d}{\delta_0^2}\right)
                                  \exp\left(nH(PW)-Kd\delta_0\sqrt{n}\right).$$
  Choosing $\delta_0=\sqrt{\frac{4d}{\eta}}$ the proof is complete.
\end{beweis}
The next theorem improves upon our previous converse,
theorem~\ref{satz:channel:strong:converse}:
\begin{satz}
  \label{satz:code:upperbound}
  For every $\lambda\in(0,1)$ there is a constant
  $K(a,d,\lambda)$ such that for every q--DMC ${W}$
  and every $(n,\lambda)$--code $(f,D)$
  $$|\fset{M}|\leq (n+1)^a\exp\left(nC({W})+K(a,d,\lambda)\sqrt{n}\right).$$
\end{satz}
\begin{beweis}
  We will prove even more: under the additional assumption that all code words are
  of the same type $P$ (such codes are called \emph{constant composition}) one has
  $$|\fset{M}|\leq \exp\left(nI(P;W)+K(a,d,\lambda)\sqrt{n}\right)$$
  (from which the theorem clearly follows).
  To see this modify the decoder as follows: let
  $$D_m'=\Pi^n_{V,PW,\delta}D_m\Pi^n_{V,PW,\delta}$$
  with $\delta=\frac{\sqrt{32ad}}{1-\lambda}$. Then $(f,D')$ is
  an $(n,\frac{1+\lambda}{2})$--code:
  \begin{equation*}\begin{split}
    \tr(W_{f(m)}D_m')&=\tr(W_{f(m)}\Pi^n_{V,PW,\delta}D_m\Pi^n_{V,PW,\delta})\\
                     &=\tr(W_{f(m)}D_m)-\tr\left((W_{f(m)}-
                        \Pi^n_{V,PW,\delta}W_{f(m)}\Pi^n_{V,PW,\delta})D_m\right)\\
                     &\geq 1-\lambda-\frac{1-\lambda}{2}
  \end{split}\end{equation*}
  (the last line by lemma~\ref{lemma:weak:law:exact}
  and the tender operator lemma~\ref{lemma:tender:operator}).
  Now from lemma~\ref{lemma:cond:variance:typical:proj}
  \begin{equation*}\begin{split}
    \tr D_m' &\geq \left(\frac{1-\lambda}{2}-\frac{ad}{\delta^2}\right)
                               \exp\left(nH({W}|P)-Kd\sqrt{a}\delta\sqrt{n}\right)\\
             &\geq \frac{1-\lambda}{4}
                               \exp\left(nH({W}|P)-Kd\sqrt{a}\delta\sqrt{n}\right).
  \end{split}\end{equation*}
  On the other hand $\sum_{m\in\fset{M}} D_m'\leq\Pi^n_{V,PW,\delta}$, hence
  by lemma~\ref{lemma:variance:typical:proj}
  $$\sum_{m\in\fset{M}} \tr D_m'\leq\exp\left(nH(PW)+Kd\delta\sqrt{n}\right)$$
  and we are done.
\end{beweis}

\paragraph{Reliability function}
For the finite q--DMC ${W}$ with capacity $C({W})$
the \emph{reliability function} $E(R)$ is defined by
$$E(R)=\liminf_{n\rightarrow\infty,\ \delta\rightarrow 0} 
                                     -\frac{1}{n}\log e_{\min}(n,R-\delta).$$
From the previous section we see that $E(R)=0$ for $R>C({W})$.
On the other hand define the \emph{greedy} bound
$$E_{\text{g}}(R,P)=\max\{\min\{\mu_{\text{i}}(L,P),
                            \frac{1}{2}\mu_{\text{c}}(L',P)\}:R\leq L'-L\},$$
with the \emph{individual exponent} (which may be $+\infty$)
$$\mu_{\text{i}}(L,P)=\inf\{D(V\| W|P):H(V|P)>L\},$$
and the \emph{collective exponent} (which is finite)
$$\mu_{\text{c}}(L',P)=\min\{D(\rho\| PW):H(\rho)\leq L'\}.$$
Then we have
\begin{satz}
  \label{satz:error:upper}
  For $n>0$, a type $P$, and $R<I(P,W)$ there exist constant composition
  $n$--block codes $(f,D)$ of type $P$ with
  $$|\fset{M}|\geq (n+1)^{d-ad}\exp(nR)$$
  and error probability
  $$e(f,D)\leq 8(n+1)^{ad}\exp(-nE_{\text{g}}(R,P))$$
  if $n\geq n_0(a,d,P)$.
\end{satz}
\begin{beweis}
  Let $L,L'$ a pair of numbers with $R\leq L'-L$ and
  $$E_{\text{g}}(R,P)=\min\{\mu_{\text{i}}(L,P),\frac{1}{2}\mu_{\text{c}}(L',P)\}.$$
  It is easily seen that we may assume
  $\mu_{\text{i}}(L,P)\geq\frac{1}{2}\mu_{\text{c}}(L',P)$.
  Also that in this case $L'<H(PW)$ and $L\geq H(W|P)$, in particular
  $E_{\text{g}}(R,P)>0$.\par
  Define $\lambda=8(n+1)^{ad}\exp(-nE_{\text{g}}(R,P))$ and assume $n$
  to be large enough such that $\eta=\frac{\lambda^2}{32}\leq 1-\lambda$.
  Let $(f,D)$ a maximal $(n,\lambda)$--code with the additional requirement
  $$\forall m\in\fset{M}\qquad \tr D_m\leq (n+1)^{ad}\exp(nL).$$
  We claim that with $B=\sum_{m\in\fset{M}} D_m$
  $$\forall x^n\text{ of type }P\qquad \tr(W_{x^n}B)\geq\eta.$$
  For else we could extend our code by an exceptional $x^n$ and corresponding
  decoding operator
  $$D'=\sqrt{\1-B}\Pi^n_{W,H(\cdot|P)\leq L}(x^n)\sqrt{\1-B}\ .$$
  The argument is as in the proof of theorem~\ref{satz:maximal:codes}:
  observe that $\Pi^n_{W,H(\cdot|P)\leq L}(x^n)$,
  and hence $D'$, satisfies the trace requirement, and
  $$\tr\!\left(W_{x^n}\Pi^n_{W,H(\cdot|P)\leq L}(x^n)\right)
                 \geq 1-(n+1)^{ad}\exp(-n\mu_{\text{i}}(L,P)).$$
  Consequently
  $$\tr\left((PW)^{\otimes n}B\right)\geq \eta(n+1)^{-a}$$
  and by lemma~\ref{lemma:universal:shadow}
  \begin{equation*}\begin{split}
    \tr B &\geq \left(\eta(n+1)^{-a}-(n+1)^d\exp(-n\mu_{\text{c}}(L',P))\right)
                                           \cdot\exp(nL'+n\mu_{\text{c}}(L',P))\\
          &\geq (n+1)^d\exp(nL'),
  \end{split}\end{equation*}
  from which the estimate on $|\fset{M}|$ follows immediately.
\end{beweis}
\begin{cor}
  \label{cor:reliability:lower}
  For $0\leq R\leq C({W})$
  $$E(R)\geq E_{\text{g}}(R)=\max_{P\text{ p.d.: }R\leq I(P;W)} E_{\text{g}}(R,P).$$
  \phantom{.}\qed
\end{cor}
%
% folgender beweis war noetig mit frueherer definition von E(R):
%
%\begin{beweis}
%  It suffices to check lower semicontinuity of $E_{\text{g}}$ in $P$, i.e.
%  $$\liminf_{Q\rightarrow P} E_{\text{g}}(R,Q)\geq E_{\text{g}}(R,P)$$
%  (the ingredients are lower semicontinuity of $\mu_{\text{i}}$ and
%  $\mu_{\text{c}}$) and continuity in $R$ from the right, i.e.
%  $$\lim_{\delta\searrow 0} E_{\text{g}}(R+\delta,P)=E_{\text{g}}(R,P)$$
%  (which is quite obvious).
%\end{beweis}
%
\par
Conversely, defining the \emph{sphere packing} bound
$$E_{\text{sp}}(R,P)=\min_{V\text{ channel: }I(P;V)\leq R} D(V\| W|P)$$
we have
\begin{satz}
  \label{satz:error:lower}
  For $R\geq 0$ and $n>0$ let $(f,D)$ a constant composition $n$--block code
  (of type $P$) with
  $$|\fset{M}|\geq \exp(n(R+\delta)).$$
  Then for the error probability
  $$e(f,D)\geq \frac{1}{2}\exp(-nE_{\text{sp}}(R,P)(1+\delta))$$
  if $n\geq n_0(a,d,\delta)$.
\end{satz}
\begin{beweis}
  We can directly apply the original idea of \cite{haroutunian:sphere}:
  consider a channel $V:\fset{X}\rightarrow\alg{S}(\alg{L}({\cal H}))$ with
  $I(P;V)\leq R$. From the proof of the strong converse
  theorem~\ref{satz:code:upperbound} we see
  that $e(f,D)\geq 1-\frac{\delta}{2}$ if $n$ is large enough (we assume
  $\delta<1$). I.e. for some message $m\in\fset{M}$ and $S_m=\1-D_m$
  $$\tr(V_{f(m)}S_m)\geq 1-\frac{\delta}{2}$$
  Now generally for two states $\rho,\sigma$ and complementary positive 
  operators $S,D$ (i.e. $S+D=\1$) one has
  $$\tr(\rho S)\log\frac{\tr(\rho S)}{\tr(\sigma S)}+
             \tr(\rho D)\log\frac{\tr(\rho D)}{\tr(\sigma D)}\leq D(\rho\|\sigma).$$
  This follows immediately from the monotonicity of quantum
  I--divergence, theorem~\ref{satz:monoton}, applied to the completely positive,
  trace preserving map
  \begin{align*}
    \alg{L}({\cal H})_* &\longrightarrow \C^2 \\
                 \alpha &\longmapsto     \tr(\alpha S)e_1+\tr(\alpha D)e_2\ .
  \end{align*}
  From this we get by elementary operations
  $$\tr(\sigma S)\geq \exp\left(-\frac{D(\rho\|\sigma)+h(\tr(\rho S))}{\tr(\rho S)}\right).$$
  Applying this to $\rho=V_{f(m)}$, $\sigma=W_{f(m)}$ and $S=S_m$, $D=D_m$
  we find
  \begin{equation*}\begin{split}
    \tr(W_{f(m)}S_m) &\geq\exp\left(-\frac{nD(V\| W|P)+h\left(1-\frac{\delta}{2}\right)}
                                                              {1-\frac{\delta}{2}}\right)\\
                     &\geq \frac{1}{2}\exp(-nD(V\| W|P)(1+\delta))
  \end{split}\end{equation*}
  if only $\delta$ is small enough (which is no real restriction). Now we choose
  $V$ such that $D(V\| W|P)$ is minimal.
\end{beweis}
\begin{cor}
  \label{cor:reliability:upper}
  For $0\leq R\leq C({W})$ (with the possible exception
  of the leftmost finite value of $E_{\text{sp}}$)
  $$E(R)\leq E_{\text{sp}}(R)=\max_{P\text{ p.d.}}E_{\text{sp}}(R,P).$$
\end{cor}
\begin{beweis}
  To apply the theorem we have just to note the continuity of $E_{\text{sp}}$ in $R$,
  which follows from its convexity.
\end{beweis}
\begin{bem}
  The proof obviously also works for infinite input alphabet, if only we have
  a strong converse which indeed we have, by the previous section.
\end{bem}
\begin{bem}
  The reader may wish to apply the techniques of the previous proofs to show that
  $e(f,D)$ tends to $1$ exponentially for rates above the capacity. The results however yield
  nothing of interest beyond the analysis of \cite{ogawa:nagaoka}.
\end{bem}

\section{{\sc Holevo} bound}
\label{sec:channel:holevo}
An interesting application of our converse theorem~\ref{satz:code:upperbound}
is in a new, and completely elementary, proof of the
{\sc Holevo} bound (theorem~\ref{satz:holevo:bound}):
\par
For a q--DMC $W:\fset{X}\rightarrow\alg{S}(\alg{L}({\cal H}))$,
a p.d. $P$ on $\fset{X}$ and $D$ an observable on
$\alg{L}({\cal H})$, say indexed
by $\fset{Y}$, the composition $D_*\circ W:\fset{X}\rightarrow\fset{Y}$
is a classical channel.
\par
\cite{holevo:bound} considered
$C_1=\max_{P,D} I(P;D_*\circ W)$
(the capacity if one is restricted to tensor product observables!)
and proved $C_1\leq C(W)$. For us this is now clear, since all codes
for the classical channel $D_*\circ W$ (whose maximal rates are asymptotically
just $C_1$) can be interpreted as special channel codes for $W$.
\par
But we can show even a little more, namely {\sc Holevo}'s original
\emph{information bound} $I(P;D_*\circ W)\leq I(P;W)$ (from which the
capacity estimate clearly follows).
\bigskip\par\noindent
\begin{beweis}
  Assume the opposite, $I(P;D_*\circ W)>I(P;W)$. Then by the well known
  classical coding theorem (\cite{shannon:theory} --- alternatively
  theorem~\ref{satz:maximal:codes} which by
  remark~\ref{bem:classical:is:contained} generalizes the
  classical case) there is to every
  $\delta>0$ an infinite sequence of $(n,1/2)$--codes with codewords
  chosen from
  $\fset{T}^n_{V,P,\sqrt{2a}}$ for the channel $D_*\circ W$ with
  rates exceeding $I(P;D_*\circ W)-\delta$. Restricting to a single
  type of codewords we find constant composition codes (of type $P_n$)
  with rate exceeding $I(P;D_*\circ W)-2\delta$ (if $n$ is large enough).
  \par
  As already explained these are special channel codes for $W$, so
  by theorem~\ref{satz:code:upperbound} (proof) their rates are bounded by
  $I(P_n;W)+\delta$ (again, $n$ large enough), hence
  $$I(P;D_*\circ W)-2\delta \leq I(P_n;W)+\delta.$$
  Collecting inequalities we find
  $$I(P;W) < I(P;D_*\circ W) \leq I(P_n;W)+3\delta.$$
  But since $P_n\rightarrow P$ by assumption and by the continuity
  of $I$ in $P$ (see lemma~\ref{lemma:H:continuous}),
  since furthermore $\delta$ is arbitrarily small, we end up with
  $$I(P;W) < I(P;D_*\circ W) \leq I(P;W),$$
  a contradiction.
\end{beweis}

\section{Open questions}
\label{sec:channel:open}
We left open a number of problems:

\paragraph{Entangled input}
Is it possible to exceed the rate $C^{(1)}=C(\varphi_*)=\max_P I(P;\varphi_*)$ by using
block codes where not only product states but arbitrary (entangled)
states are allowed as ``codewords''? We conjecture that the ``ultimate''
classical information capacity of $\varphi_*$,
$$\tilde{C}=\limsup_{n\rightarrow\infty} \frac{1}{n}\max_P I(P;\varphi_*^{\otimes n})$$
equals $C^{(1)}$ (compare \cite{schumacher:capacity}).

\paragraph{Computations}
Closely related is the issue of constructing a feasible algorithm to
numerically compute the quantity $C^{(1)}$, maybe by an adaption of
{\sc Arimoto}'s algorithm for computing the capacity of a classical
channel (cf. ideas of \cite{nagaoka:algorithm}).
This could be used for experimental tests of whether
$C^{(n)}=\frac{1}{n}\max_P I(P;\varphi_*^{\otimes n})$ exceeds $C^{(1)}$.

\paragraph{Abstract approach}
In the proofs so far we relied heavily on the product structure of the $n$--fold
channel. For reasons of better understanding of the foundations, as well as for
having a unified framework for proof, it is desireable to have ``abstract'' coding
theorems and converses at ones disposal. What this means is that time structure (blocks,
in our case even products) is not used: after all the $n$--fold use of a channel
is just a channel with larger alphabet. This is e.g. how {\sc Fano}'s inequality
is used in weak converses. For something closer to our present setting
compare \cite{wolfowitz:coding}, chapter 7.
\begin{itemize}
  \item Prove an abstract coding theorem in this spirit!
  \item Prove the abstract converse, by exhibiting a usable ``packing lemma'',
    as is known in the classical theory.
\end{itemize}

\paragraph{Blowing up}
Prove a blowing up lemma as in the classical theory
(commutative $\alg{A}$), due to \cite{ahlswede:gacs:koerner}!
I suggest the following definition:
\par
Let $\alg{A}=\alg{L}({\cal H})$ a C${}^*$--algebra with $q=\dim_{\C}\alg{A}$, and
$\Pi\in\alg{A}^{\otimes n}$ a projector. Define the \emph{blow--up}
of $\Pi$ as
$$\Gamma\Pi=\lcsupp\{A_{(i)}\Pi A_{(i)}^*:\ 
                                       1\leq i\leq n,\ A\in\alg{A},\ A^*A\leq\1\}$$
where $A_{(i)}=\1^{\otimes(i-1)}\otimes A\otimes\1^{\otimes(n-i)}$.
The $l^{\text th}$ \emph{blow--up} of $\Pi$ is $\Gamma^l\Pi$, defined as
$$\Gamma^l\Pi=\lcsupp\{A_{(I)}\Pi A_{(I)}^*:\ 
                     I\subset[n],\ |I|=l,\ A\in\alg{A}^{\otimes l},\ A^*A\leq\1\}$$
where $A_{(I)}=\1^{\otimes([n]\setminus I)}\otimes A$ (in the right order).
\par
In loose words: $\Gamma^l\Pi$ is the least common support of all images of $\Pi$
under all quantum operations confined to $l$ positions (factors in the tensor product).
\begin{lemma}
  The blowing up operation has the following properties:
  \begin{enumerate}
    \item $\Gamma^l\Pi$ is a projector.
    \item $\Gamma^l$ is the $l$--fold iteration of $\Gamma$.
    \item For $0\leq l\leq l'$ one has $\Pi\leq\Gamma^l\Pi\leq\Gamma^{l'}\Pi$.
    \item $\tr\Gamma^l\Pi\leq (qn)^l\cdot\tr\Pi$.
  \end{enumerate}
\end{lemma}
\begin{beweis}
  Points (1) and (3) are obvious. For (2) and (4) write $\Pi=\sum_{\pi\in\Rho} \pi$
  for a set $\Rho$ of (necessarily orthogonal) minimal idempotents. Clearly
  $\tr\Pi=|\Rho|$. Then
  $$\Gamma^l\Pi=\lcsupp\{A_{(I)}\pi A_{(I)}^*:\ \pi\in\Rho,\ 
                     I\subset[n],\ |I|=l,\ A\in\alg{A}^{\otimes l},\ A^*A\leq\1\}$$
  and the supporting subspace\footnote{In
    ${\cal H}_1\oplus\cdots\oplus{\cal H}_m$, which we think of
    $\alg{A}=\bigoplus_{i=1}^m \alg{L}({\cal H}_i)$ to live on!}
  of this is
  $$\sum_{\pi=\ket{\psi}\bra{\psi}\in\Rho}\operatorname{span}\{A_{(I)}\ket{\psi}:\ 
                     I\subset[n],\ |I|=l,\ A\in\alg{A}^{\otimes l},\ A^*A\leq\1\}.$$
  But $\alg{A}$ has a linear basis $(A_1,\ldots,A_q)$ which produces by
  tensor products a basis of length $q^l$ of $\alg{A}^{\otimes l}$. This shows
  (2), and since there are at most $n^l$ many $I\subset[n]$ of cardinality
  $l$ we get (4).
\end{beweis}
\begin{conj}
  \label{conj:blowing:up}
  Let $W$ a fixed q--DMC, $m_W$ the smallest non--zero eigenvalue of the
  $W_x$, $x^n\in\fset{X}^n$, and $B$ a projector. Then
  $$\tr(W_{x^n}\Gamma^l B)\geq
            \Phi\left(\Phi^{-1}(\tr(W_{x^n}B))+a\frac{l-1}{\sqrt{n}}\right),$$
  with $a=c\dfrac{m_W}{\sqrt{-\ln m_W}}$, where $c>0$ is a universal constant
  and $\Phi:\R\rightarrow[0,1]$ is the Gaussian distribution function:
  $\Phi(x)=\frac{1}{\sqrt{2\pi}}\int_{-\infty}^{x} e^{-t^2/2}\text{d}t$.
\end{conj}
Among the possible applications would be the transition from weak to strong
converses (after {\sc Ahlswede \& Dueck}, cf. \cite{csiszar:koerner},
chapter~2.1).

\paragraph{Reliability function}
We proved the sphere packing bound and a lower bound
on the reliability function which at least shows its positivity
for rates below the capacity.
For the pure state channel this is matched by random coding and expurgated
lower bounds of \cite{burnashev:holevo:reliability}.
Unfortunately in this case our sphere packing bound is trivial!
\par
We leave as open problems:
the proof of a random coding lower bound in the general case
(which should enable us to \emph{determine} the reliability
function above a critical rate), and (at least in the pure
state case) to find a suitable modification of the sphere packing bound
(as the present formulation does
not take into account possible noncommutativity).

% QDissertation: Quantum multiple access channel

\chapter{Quantum Multiple Access Channels}
\label{chap:mac}

\thispagestyle{myheadings}

The multiway channel with $s$ senders and $r$ receivers in
classical information theory was already studied by \cite{shannon:MAC}.
\cite{ahlswede:MAC} and \cite{ahlswede:MWC} first determined its capacity
region. For a good overview on multiuser communication
theory one should consult \cite{elgamal:cover}.
In the present chapter we will define the corresponding quantum channel (after
recent work by \cite{allahverdyan:saakian:qmac}), extending
the results of the previous chapter: we bound the capacity region (in the
limit of vanishing error probability), and
--- for the multiple access channel, i.e. one receiver ---
we are able to prove the corresponding coding theorem.

\section{Quantum multiway channels and capacity region}
\label{sec:mac:capacity}
This is the simplest situation of multi--user communication in general:
consider $s$ independent senders, sender $i$ using an alphabet $\fset{X}_i$, say with an
a priori probability distribution $P_i$. We describe this by the quantum
state $\sigma_i=\sum_{x_i\in\fset{X}_i} P_i(x_i)x_i$ on the commuative
C${}^*$--algebra $\alg{X}_i=\C\fset{X}_i$ 
generated by the $x_i$ which are mutually orthogonal idempotents (to distinguish these
as generators of this algebra we will sometimes write $[x_i]$).
The channel is then a map
$$W:\fset{X}_1\times\cdots\times\fset{X}_s\rightarrow\alg{S}(\alg{Y})$$
with a (finite dimensional) C${}^*$--algebra $\alg{Y}$, which connects the
input $(x_1,\dots,x_s)$ with the output $W_{x_1\dots x_s}$. By linear extension we
may view $W$ as a completely positive, trace preserving map from
$\alg{X}_{1*}\otimes\cdots\otimes\alg{X}_{s*}$ to $\alg{Y}_*$.
The receivers are modelled by compatible $*$--subalgebras $\alg{Y}_j$
(see appendix~\ref{chap:quprob}, section {\em Quantum systems}).
\par
If all the $W_{x_1\ldots x_s}$ commute with each other
(hence have a common diagonalization) the channel is called \emph{classical}.
\par
For fixed a priori distributions we have the \emph{channel state}
$$\gamma=\sum_{\forall i\ x_i\in\fset{X}_i} P_1(x_1)\cdots P_s(x_s)
                        [x_1]\otimes\cdots\otimes[x_s]\otimes W_{x_1\ldots x_s}$$
on $\alg{X}_1\otimes\cdots\otimes\alg{X}_s\otimes\alg{Y}$.
\par
For a subset $J\subset[s]$ denote $P_J=\bigotimes_{i\in J} P_i$, i.e.
$P_J(x_i|i\in J)=\prod_{i\in J} P_i(x_i)$, and
${\cal X}(J)=\prod_{i\in J}{\cal X}_i$ (similarly
$\alg{X}(J)=\bigotimes_{i\in J}\alg{X}_i$).
\par
Further define a \emph{reduced channel} $P_{J^c}W:{\cal X}(J)\rightarrow\alg{S}(\alg{Y})$ by
$$P_{J^c}W:(x_i|i\in J)\longmapsto
   \sum_{\forall i\in J^c:\ x_i\in\fset{X}_i} P_{J^c}(x_i|i\in J^c)W_{x_1\ldots x_s}\ .$$
Note that
$$\tr_{\alg{X}(J^c)}\gamma=\sum_{\forall i\in J:\ x_i\in\fset{X}_i}
                        P_J(x_i|i\in J)[x_i|i\in J]\otimes (P_{J^c}W)_{(x_i|i\in J)}\ .$$
An $n$--\emph{block code} is a collection
$(f_1,\ldots,f_s,D_1,\ldots,D_r)$ of maps $f_i:\fset{M}_i\rightarrow\fset{X}_i^n$
and decoding observables $D_j\subset\alg{Y}_j^{\otimes n}$, indexed
by $\fset{M}_1'\times\cdots\times\fset{M}_s'\supset\fset{M}_1\times\cdots\times\fset{M}_s$.
There are $r$ (average) \emph{error probabilities} 
of the code, the probability that the receiver $j$ guesses wrongly
any one of the sent words, taken over the uniform distribution on the codebooks:
$$\bar{e}_j(f_1,\ldots,f_s,D_j)=1-\frac{1}{|\fset{M}_1|\cdots|\fset{M}_s|}
      \sum_{\forall i:m_i\in\fset{M}_i}\tr\left({W^{\otimes n}(f(m_1),\ldots,f(m_s))
             D_{j,m_1\ldots m_s}}\right).$$
We call $(f_1,\ldots,f_s,D_1,\ldots,D_r)$ an $(n,\bar{\lambda})$--code
if all $\bar{e}_j(f_1,\ldots,f_s,D_j)$ are at most $\bar{\lambda}$.
\par
The \emph{rates} of the code are the $R_i=\frac{1}{n}\log|\fset{M}_i|$.
A tuple $(R_1,\ldots,R_s)$ is said to be \emph{achievable}, if for any
$\bar{\lambda},\delta>0$ there
exists for any large enough $n$ an $(n,\bar{\lambda})$--code
with $i$--th rate at least $R_i-\delta$.
The set of all achievable tuples (which is clearly closed, and convex
by the \emph{time sharing principle}, cf. \cite{csiszar:koerner},
lemma 2.2.2) is called the \emph{capacity region} of the channel.

\section{Outer bounds}
\label{sec:mac:outer}
In the case $r=1$, $s=2$ the following theorem was already
stated by \cite{allahverdyan:saakian:qmac}, who also gave hints
on the proof.
\begin{satz}[Outer bounds]
  \label{satz:weak:converse}
  The capacity region of the quantum multiway channel is contained in the
  closure of all nonnegative
  $(R_1,\ldots,R_s)$ satisfying
  $$\forall J\subset[s],j\in[r]\qquad R(J)=\sum_{i\in J} R_i
            \leq\sum_u q_u I_{\gamma_u}\left(\alg{X}(J)\wedge\alg{Y}_j|\alg{X}(J^c)\right)$$
  for some channel states $\gamma_u$ (belonging to appropriate input distributions) and 
  $q_u\geq 0$, $\sum_u q_u=1$.
\end{satz}
\begin{beweis}
  Consider any $(n,\bar{\lambda})$--code
  $(f_1,\ldots,f_s,D_1,\ldots,D_r)$ with rate tuple $(R_1,\ldots,R_s)$.
  Then the uniform distribution on the codewords induces a channel state
  $\gamma$ on the block 
  $(\alg{X}_1\cdots\alg{X}_s\alg{Y})^{\otimes n}$. Its restriction
  to the $u$--th copy in this tensor power will be denoted $\gamma_u$.
  Let $j\in[r]$, $J\subset[s]$.
  By {\sc Fano} inequality in the form of corollary~\ref{cor:fano:inequality}
  we have
  $$H(\alg{X}^{\otimes n}(J)|\alg{Y}^{\otimes n}_j\alg{X}^{\otimes n}(J^c))
                                                     \leq 1+\bar{\lambda}\cdot nR(J).$$
  With
  \begin{equation*}\begin{split}
    H(\alg{X}^{\otimes n}(J)|\alg{Y}^{\otimes n}_j\alg{X}^{\otimes n}(J^c))
             &=H(\alg{X}^{\otimes n}(J))-I(\alg{X}^{\otimes n}(J)\wedge
                                     \alg{Y}^{\otimes n}_j\alg{X}^{\otimes n}(J^c))\\
             &=nR(J)-I(\alg{X}^{\otimes n}(J)\wedge
                                     \alg{Y}^{\otimes n}_j\alg{X}^{\otimes n}(J^c))
  \end{split}\end{equation*}
  we conclude (with subadditivity of mutual information,
  corollary~\ref{cor:info:subadd}) that
  \begin{equation*}\begin{split}
    (1-\bar{\lambda})R(J) &\leq \frac{1}{n}+\frac{1}{n}
         I_\gamma(\alg{X}^{\otimes n}(J)\wedge\alg{Y}^{\otimes n}_j\alg{X}^{\otimes n}(J^c))\\
                           &\leq \frac{1}{n}+\frac{1}{n}\sum_{u=1}^{n}
                                 I_{\gamma_u}(\alg{X}(J)\wedge\alg{Y}_j\alg{X}(J^c)).
  \end{split}\end{equation*}
\end{beweis}
\begin{bem}
  \label{bem:ahlswede:cap}
  In the case of classical channels the region described in the theorem is
  the exact capacity region (i.e. all the rates there are achievable),
  as was first proved by \cite{ahlswede:MAC} and \cite{ahlswede:MWC}.
\end{bem}
\begin{bem}
  The numeric computation of the above regions is not yet possible from
  the given description: we need a bound on the number of different single--letter
  channel states one has to consider in the convex combinations.
  For the multiple access channel ($r=1$) this is easy: by {\sc Caratheodory}'s theorem
  $s$ will suffice. For general $r$ there are also classical bounds, which
  carry over unchanged to the quantum case (since the quantum mutual information has
  properties similar to those of classical mutual information): $r(2^s-1)$ always
  suffice, as was observed by \cite{ahlswede:rate:slicing}.
\end{bem}

\section{Coding theorem for multiple access channels}
\label{sec:mac:coding}
With the notation as before for a quantum multiway channel $W$ with one receiver
we have
\begin{satz}
  \label{satz:qmac:capacity}
  An $s$--tuple $(R_1,\ldots,R_s)$ is achievable (i.e. there is an infinite sequence of
  $(n,\bar{\lambda}_n)$--codes with $\bar{\lambda}_n\rightarrow 0$ and rate tuple
  tending to $(R_1,\ldots,R_s)$), if and only if it is in the convex hull
  of the pairs satifying (for some input distributions which induce a channel
  state $\gamma$)
  $$\forall J\subset[s]\qquad R(J)=\sum_{i\in J} R_i
                   \leq I_{\gamma}\left(\alg{X}(J)\wedge\alg{Y}|\alg{X}(J^c)\right).$$
\end{satz}
We shall prove this only in the case $s=2$, the reader should have no
difficulty to see the extension to larger numbers. In this case the
conditions reduce to
$$R_1+R_2\leq I(\alg{Y}\wedge\alg{X}_1\alg{X}_2),$$
$$R_1\leq I(\alg{Y}\wedge\alg{X}_1|\alg{X}_2),
               \qquad R_2\leq I(\alg{Y}\wedge\alg{X}_2|\alg{X}_1).$$
That these are necessary is of course theorem~\ref{satz:weak:converse}.
For proof of the achievability it is (by the time sharing principle)
sufficient to consider an extreme point of the region
described by the above inequalities for a
particular channel state. It is easily seen that w.l.o.g.
$R_1=I(\alg{X}_1\wedge\alg{Y})$, $R_2=I(\alg{X}_2\wedge\alg{Y}\alg{X}_1)$.
That this point is achievable follows immediately from 
theorem~\ref{satz:multsrc:hybrid:c:coding} and the following theorem,
applied with $\bar{R}_1=I(\alg{X}_1\wedge\alg{Y})+\delta$ and
$\bar{R}_2=I(\alg{X}_2\wedge\alg{X}_1\alg{Y})+\delta$.
\begin{satz}(Cf. \cite{csiszar:koerner}, proof of theorem 3.2.3)
  \label{satz:mac:code:constr}
  Let $\bar{\lambda},\delta>0$, $W$ a quantum multiple access channel with two
  senders, and $P_i$ probability distributions on the sender alphabets
  $\fset{X}_i$. Define the $c^2 h^1$--source (see chapter~\ref{chap:multsrc},
  section {\em Correlated quantum sources})
  $(\alg{X}_1,\alg{X}_2,\alg{Y},\fset{X}_1\times\fset{X}_2\times\Rho,P)$ on
  $\alg{X}_1\otimes\alg{X}_2\otimes\alg{Y}$
  by $P(x_1\otimes x_2\otimes\pi)=P_1(x_1)P_2(x_2)q_{\pi|x_1x_2}$,
  where $\Rho$ is a set of pure states on $\alg{Y}$
  and the $q_{\pi|x_1x_2}\geq 0$ are such
  that $W_{x_1x_2}=\sum_{\pi\in\Rho} q_{\pi|x_1x_2}\pi$ (e.g. diagonalize all
  $W_{x_1x_2}$ and take $\Rho$ to be the set of all eigenstates occuring.
  \par
  Then from any $(n,\bar{\lambda})$--coding scheme $(g_1,g_2,D^{(0)})$
  with quantum side information at the decoder for this source,
  with rates $\bar{R}_1,\bar{R}_2$,
  one can construct an $(n,4\bar{\lambda})$--code $(f_1,f_2,D)$ for $W$
  with rates $R_i\geq H(P_i)-\bar{R}_i-\delta$, provided
  $n\geq n_0(|\fset{X}_1|,|\fset{X}_2|,\delta)$.
\end{satz}
\begin{beweis}
  Let $g_1:\fset{X}_1^n\rightarrow\fset{M}_1$ and
  $g_2:\fset{X}_2^n\rightarrow\fset{M}_2$ the encodings, $D^{(0)}$ the observable
  on $\C\fset{M}_1\otimes\C\fset{M}_2\otimes\alg{Y}$ indexed by
  $\fset{X}_1^n\times\fset{X}_2^n$. Observe that it is of the form
  $$D^{(0)}_{x_1^n x_2^n}=\sum_{m_1\in\fset{M}_1,m_2\in\fset{M}_2}
                      m_1\otimes m_2\otimes D^{\prime}_{m_1m_2,x_1^n x_2^n}\ .$$
  Define for every $(m_1,m_2)\in\fset{M}_1\times\fset{M}_2$
  $$\fset{A}_{m_1}=g_1^{-1}\{m_1\},\qquad \fset{B}_{m_2}=g_2^{-1}\{m_2\}.$$
  Assume that the $\fset{A}_{m_1},\ \fset{B}_{m_2}$ consist of
  words of single type (otherwise one modifies the coding by also
  encoding the type of the sequences, increasing the rate negligibly,
  in the asymptotics).
  \par
  Construct now codes $(f_1^{(m_1m_2)},f_2^{(m_1m_2)},D^{(m_1m_2)})$ for $W$
  as follows:\\
  $f_1^{(m_1m_2)}=\id_{\fset{A}_{m_1}}$, 
  $f_2^{(m_1m_2)}=\id_{\fset{B}_{m_2}}$ and $D^{(m_1m_2)}$ an observable
  on $\alg{Y}$ indexed by $\fset{A}_{m_1}\times\fset{B}_{m_2}$ with
  $D^{(m_1m_2)}_{x_1^n x_2^n}\geq D^{\prime}_{m_1m_2,x_1^n x_2^n}$.
  \par
  As in \cite{csiszar:koerner}, pp.272 we can see that for the error
  probabilities
  $$\sum_{m_1\in\fset{M}_1} \sum_{m_2\in\fset{M}_2}
                   P_1^n(\fset{A}_{m_1})P_2^n(\fset{B}_{m_2})
                     \bar{e}(f_1^{(m_1m_2)},f_2^{(m_1m_2)},D^{(m_1m_2)})
       \leq \bar{e}(g_1,g_2,D^{(0)})$$
  and again copying from \cite{csiszar:koerner} we find that there is one
  of them having
  $\bar{e}(f_1^{(m_1m_2)},f_2^{(m_1m_2)},D^{(m_1m_2)})\leq 4\bar{\lambda}$
  and rates $R_i\geq H(P_i)-\bar{R}_i-\delta$, if $n$ is large enough.
\end{beweis}

\section{Open questions}
\label{sec:mac:open}

\paragraph{Random coding}
The major drawback of the above method of proof is that it allows no
direct code construction for every point in the capacity region,
as does the proof of \cite{ahlswede:MWC}
(we needed to invoke the time sharing principle).
It seems that this approach is no longer possible if there are
two or more receivers present. The above outer bounds however we conjecture 
to be the correct ones (by formal analogy with the classical case). A proof
of the corresponding coding theorem would be highly desireable, possibly by
a cleverly adapted random coding argument (see the proofs of the quantum channel
coding theorem by \cite{holevo:qucapacity} and \cite{schumacher:capacity}).
It should be clear that such a proof is far more natural than the one we
presented here. For a proof of the quantum multiple access channel coding
theorem which does not rely on code partitions and reduction to a source
coding problem but instead uses iterated ``slicing'' of the rate
with random code selection, see \cite{winter:qmac}.
%
% \paragraph{Randomized encoding}
% In the classical theory one can improve the performance of multiple access
% codes from small \emph{average} error to small \emph{maximal error} by allowing
% randomized encodings. We expect this to be true also for quantum channels,
% but it has not been worked out.
%
% \paragraph{Strong converse}
% Prove not only (as we did) a weak, but a strong converse to the multiple
% access coding theorem! (Blowing up --- see {\em Open questions} of
% chapter~\ref{chap:channel} --- could be helpful to this).

% QDissertation: Quantum multiple source coding

\chapter{Quantum Multiple Source Coding}
\label{chap:multsrc}

\thispagestyle{myheadings}

Having investigated in chapter~\ref{chap:source} the problem of quantum source coding
we now turn to the problem of (independent) source coding of possibly dependent
sources. In the first section we will introduce the mathematical model, and venture
then to analyze this model as far as possible (which, as it will turn out, is not
very much): we will restrict ourselves mostly to double sources,
proving some general bounds and presenting characteristic examples. Then
we study the particular case that only one of the sources is quantum, the others
being classical. We are thus led to consider the problem of
coding with side information, which for this kind of source we can in
part solve. In general however there is to be distinguished between multiple source
coding and coding with side information.

\section{Correlated quantum sources}
\label{sec:multsrc:correlated}
A \emph{multiple ($s$--fold) quantum source} is a tuple
$(\alg{A}_1,\ldots,\alg{A}_s,\Rho,P)$ of C${}^*$--algebras $\alg{A}_i$
(with us: finite dimensional), a finite set $\Rho$ of pure states on
$\alg{A}=\alg{A}_1\otimes\cdots\otimes\alg{A}_s$ and a p.d. $P$ on $\Rho$.
\par
The \emph{average state} of the source is the state $P\Rho$ on $\alg{A}$,
its marginal restricted to $\alg{A}^{\otimes I}=\bigotimes_{i\in I} \alg{A}_i$
is denoted $P\Rho|_I$.
\medskip\par
We call the source \emph{classically correlated} if all the states $\pi\in\Rho$
are product states with respect to $\alg{A}_1,\ldots,\alg{A}_s$:
$\pi=\pi_1\otimes\cdots\otimes\pi_s$, $\pi_i\in\alg{S}(\alg{A}_i)$.
In this case we obtain for each $J\subset[n]$ a multiple source
$((\alg{A}_j|j\in J),\Rho|_J,P)$ by restricting the $\pi\in\Rho$
to $\alg{A}^{\otimes J}$, i.e. replacing $\pi$ by $\pi|_J$. Always in
this situation we assume w.l.o.g. $\Rho=\Rho_1\times\cdots\times\Rho_s$
\medskip\par
If in particular $k$ of the $\alg{A}_i$ are classical (i.e. commutative),
$l$ are fully quantum (i.e. full matrix algebras) and the remaining $m$
are arbitrary (``hybrid''), we speak of a \emph{$c^k q^l h^m$--source}.
\medskip\par
An $n$--\emph{block coding scheme with quantum encoding}
for a multiple quantum source
$(\alg{A}_1,\ldots,\alg{A}_s,\Rho,P)$ is a tuple
$(\varepsilon_{1*},\ldots,\varepsilon_{s*},\delta_*)$ with
quantum operations
\begin{align*}
  \varepsilon_{i*}: &\alg{A}_{i*}^{\otimes n}\longrightarrow\alg{L}({\cal K}_i)_* \\
  \delta_*:         &\alg{L}({\cal K}_1\otimes\cdots\otimes{\cal K}_s)_*
                          \longrightarrow
             \alg{A}_{1*}^{\otimes n}\otimes\cdots\otimes\alg{A}_{s*}^{\otimes n}.
\end{align*}
\medskip\par
An $n$--\emph{block coding scheme with arbitrary encoding}
for a classically correlated (!) multiple quantum source
$(\alg{A}_1,\ldots,\alg{A}_s,\Rho,P)$ is a tuple
$(\varepsilon_{1*},\ldots,\varepsilon_{s*},\delta_*)$ with
\begin{align*}
  \varepsilon_{i*}: &\Rho_i^n\longrightarrow\alg{S}(\alg{L}({\cal K}_i))
                                                      \ \text{  mappings and}\\
  \delta_*:         &\alg{L}({\cal K}_1\otimes\cdots\otimes{\cal K}_s)_*
                              \longrightarrow
            \alg{A}_{1*}^{\otimes n}\otimes\cdots\otimes\alg{A}_{s*}^{\otimes n}\ 
                                                    \text{  a quantum operation.}
\end{align*}
\medskip\par
Writing $\varepsilon_*=\varepsilon_{1*}\otimes\cdots\otimes\varepsilon_{s*}$
we define the \emph{average fidelity} and \emph{average distortion}
of the scheme
$(\varepsilon_{1*},\ldots,\varepsilon_{s*},\delta_*)$ as expected:
\begin{align*}
  \bar{F}(\varepsilon_{1*},\ldots,\varepsilon_{s*},\delta_*)
       &=\sum_{\pi^n\in\Rho^n}
          P^n(\pi^n)\!\cdot\!\tr((\delta_*\varepsilon_*\pi^n)\pi^n), \\
  \bar{D}(\varepsilon_{1*},\ldots,\varepsilon_{s*},\delta_*)
       &=\sum_{\pi^n\in\Rho^n}
          P^n(\pi^n)\!\cdot\!\frac{1}{2}\| \delta_*\varepsilon_*\pi^n-\pi^n \|_1\ .
\end{align*}
If all $\alg{A}_i$ are fully quantum, say $\alg{A}_i=\alg{L}({\cal H}_i)$,
we can define the \emph{entanglement fidelity} by
$$F_e(\varepsilon_{1*},\ldots,\varepsilon_{s*},\delta_*)
    =\tr\left(((\delta_*\varepsilon_*\otimes\id)\Psi_{P\Rho}^{\otimes n})
           \Psi_{P\Rho}^{\otimes n}\right).$$
%
%\begin{bem}
%  It may be convenient to think of arbitary
%  encoding as a quantum operation: this we model
%  by giving the encoder the states $\ket{a}\bra{a}$
%  (on $\C A$). The map $E_A$ then easily
%  translates to a trace preserving quantum
%  operation from $\C A$ to ${\cal H}_A^\prime$.
%\end{bem}
%
Quite obviously theorem~\ref{satz:criteria} for these quality measures is still valid.
It should be clear also what we mean by $(n,\lambda)_{\bar{F}}$--,
$(n,\lambda)_{\bar{D}}$--, and $(n,\lambda)_{F_e}$--coding schemes.
\par
The \emph{rate tuple} $(R_1,\ldots,R_s)$ of the coding scheme is
defined by $R_i=\frac{1}{n}\log\dim{\cal K}_i$. A tuple $(R_1,\ldots,R_s)$
is called $(\text{quantum},\bar{F})$--achievable if there is a sequence
of $(n,\lambda_n)_{\bar{F}}$--coding schemes with rate tuples converging to
$(R_1,\ldots,R_s)$ and $\lambda\rightarrow 0$ as $n\rightarrow\infty$.
The set ${\bf R}_{\text{q},\bar{F}}$
of all $(\text{quantum},\bar{F})$--achievable rate tuples
is called $(\text{quantum},\bar{F})$--\emph{rate region}.
\par
Analogously $(\text{arbitrary},\bar{F})$--, the same with $\bar{D}$,
and $(\text{quantum},F_e)$--achievability are defined, with their
respective rate regions
${\bf R}_{\text{a},\bar{F}}$, ${\bf R}_{\text{q},\bar{D}}$,
${\bf R}_{\text{a},\bar{D}}$ and ${\bf R}_{\text{q},F_e}$.
\par
It is clear from the definition that the rate regions
are closed, convex (by the time sharing principle) and 
right upper closed (increasing some of the $R_i$ does not leave the
rate region). Also we have the following quite obvious
inclusions:
$$\begin{matrix}
    {\bf R}_{\text{q},F_e} & \subset &
           {\bf R}_{\text{q},\bar{F}} & \subset & {\bf R}_{\text{a},\bar{F}} \\
                           &         &
           \!\!\!\!\!\|               &         & \!\!\!\!\!\|               \\ 
                           &         &
           {\bf R}_{\text{q},\bar{D}} &         & {\bf R}_{\text{a},\bar{D}}
  \end{matrix}$$
Note that the different rate regions depend on the \emph{ensemble} $(\Rho,P)$,
only ${\bf R}_{\text{q},F_e}$ is obvious to depend only on the average
state $P\Rho$. For the others we will present evidence that they do in fact
depend on further properties of $(\Rho,P)$ besides $P\Rho$.
%
% hier sollte in der endgueltigen fassung eine neue seite anfangen:
%
\vfill\clearpage

\paragraph{Some general bounds}
Consider first a double source, quantum encoding with average fidelity:
\begin{satz}
  \label{satz:lower:2:blind:Fbar}
  Let $(\alg{A}_1,\alg{A}_2,\Rho,P)$ a double quantum source and
  $(R_1,R_2)$ a $(\text{quantum},\bar{D})$--achievable pair.
  Then with the average state $P\Rho$ on $\alg{A}=\alg{A}_1\otimes\alg{A}_2$
  $$R_1+R_2\geq H(\alg{A}_1\alg{A}_2),\ 
    R_1\geq H(\alg{A}_1|\alg{A}_2),\ R_2\geq H(\alg{A}_2|\alg{A}_1).$$
\end{satz}
\begin{beweis}
  The first inequality follows from the converse to source coding, in the
  generalized form of theorem~\ref{satz:general:source:coding}.
  For the second, consider an $(n,\lambda)_{\bar{D}}$--coding scheme
  $(\varepsilon_{1*},\varepsilon_{2*},\delta_*)$ with quantum encoding
  which has rate pair $(R_1+\epsilon,R_2+\epsilon)$.
  Modify the coding scheme as follows (for $n$ large enough):
  \par
  $\alg{A}_1$ encodes just as before, but $\alg{A}_2$ uses {\sc Schumacher}'s
  data compression to encode his part in $H(\alg{A}_2)+\epsilon$ qubits per symbol
  and with $\bar{D}\leq\frac{1-\lambda}{2}$.
  The decoder first ''unpacks'' the signal from $\alg{A}_2$ and then applies
  $\alg{A}_2$'s previous encoding $\varepsilon_{2*}$. After that she
  applies her previous decoding $\delta_*$. Let us estimate
  the average trace norm distortion of the new scheme:
  by the non--increasing of $\|\cdot\|_1$ under quantum operations and
  triangle inequality it is at most $\frac{1+\lambda}{2}$.
  Thus from theorem~\ref{satz:general:source:coding} it follows
  that $R_1+H(\alg{A}_2)+2\epsilon\geq H(\alg{A})$, and since $\epsilon$ is
  arbitrarily small we get the second inequality.
  The third one is exactly symmetrical.
\end{beweis}
\begin{expl}[Cloned wheel]
  \label{expl:sym2:source}
  Consider the $c^0 q^2$--source $(\alg{A}_1,\alg{A}_2,\Rho,P)$
  given by $\alg{A}_1=\alg{A}_2=\alg{L}(\C^2)$, and $P$ is equidistributed on
  $$\Rho=\left\{ \ket{00}\bra{00},\ \ket{11}\bra{11},\ 
            \ket{++}\bra{++},\ \ket{--}\bra{--}\right\},$$
  where $\{\ket{0},\ket{1}\}$ is an orthonormal basis of $\C^2$,
  and $\ket{+}=\frac{1}{\sqrt{2}}(\ket{0}+\ket{1})$,
  $\ket{-}=\frac{1}{\sqrt{2}}(\ket{0}-\ket{1})$.
  So the average state of the source is
  $$P\Rho=\frac{1}{4}
       \left(\ket{00}\bra{00}+\ket{11}\bra{11}+\ket{++}\bra{++}+\ket{--}\bra{--}\right)$$
  and clearly the marginals are
  $$P\Rho|_{\alg{A}_1}=P\Rho|_{\alg{A}_2}=\frac{1}{2}\1.$$
  \par
  Since each of the sent pairs is clearly invariant
  under exchange of $\alg{A}_1$ and $\alg{A}_2$ we see that
  so is $P\Rho$, i.e. $P\Rho$ is supported on the
  three--dimensional \emph{symmetrical subspace} $\Sym_2(\C^2)$ of
  $\C^2\otimes\C^2$. In fact, an orthonormal basis of $\Sym_2(\C^2)$
  is given by the \emph{triplet {\sc Bell} states}
  \begin{align*}
    \ket{\Phi^+} &=\frac{1}{\sqrt{2}}\left(\ket{00}+\ket{11}\right) \\
    \ket{\Phi^-} &=\frac{1}{\sqrt{2}}\left(\ket{00}-\ket{11}\right) \\
    \ket{\Psi^+} &=\frac{1}{\sqrt{2}}\left(\ket{01}+\ket{10}\right)
  \end{align*}
  and it is readily checked that
  $$P\Rho=\frac{1}{2}\ket{\Phi^+}\bra{\Phi^+}
            +\frac{1}{4}\ket{\Phi^-}\bra{\Phi^-}
            +\frac{1}{4}\ket{\Psi^+}\bra{\Psi^+}.$$
  Thus $H(P\Rho)=3/2$ and it is clear from the previous
  theorem~\ref{satz:lower:2:blind:Fbar} that with quantum
  encoding one gets $R_1+R_2\geq 3/2$, $R_1,R_2\geq 1/2$:
  $${\bf R}_{\text{q},\bar{F}}\subset
                            \{(R_1,R_2):\ R_1,R_2\geq 1/2, R_1+R_2\geq 3/2\}.$$  
  \par
  This might appear strange: na\"{\i}vely, in the coding
  $\alg{A}_2$ (say) is unnecessary, since its state
  is identical to that of $\alg{A}_1$
  (which would mean that the uncertainty of the state of $\alg{A}_2$
  given that of $\alg{A}_1$ is zero).
  So let's try the following coding scheme: $\alg{A}_2$ transmits
  nothing, whereas $\alg{A}_1$ transmits his state $\pi$ faithfully
  using one qubit. But the task of the decoder is to reconstruct the total state,
  i.e. $\pi\otimes\pi$, which is clearly impossible by the no--cloning theorem.
  So we see that there is indeed a sense in the above inequalities.
  \par
  However, in the model with \emph{arbitrary encoding}, the first encoder \emph{can}
  replace his state $\pi$ by $\pi\otimes\pi$ and code it into (asymptotically)
  $3/2$ qubits per symbol using {\sc Schumacher}'s quantum coding. Hence
  $${\bf R}_{\text{a},\bar{F}}=\{(R_1,R_2):\ R_1,R_2\geq 0, R_1+R_2\geq 3/2\},$$
  and thus we learn:
  \begin{center}
    \fbox{\qquad In general ${\bf R}_{\text{a},\bar{F}}$ and
      ${\bf R}_{\text{q},\bar{F}}$ are different.\qquad}
  \end{center}
\end{expl}
\begin{bem}
  In the proof of theorem~\ref{satz:lower:2:blind:Fbar}
  a coding theorem ({\sc Schumacher}'s) was used. Thus, to prove
  lower bounds for more than two sources, we need some coding theorem for correlated
  quantum sources.
\end{bem}
Interestingly we \emph{can} prove directly lower bounds on the resources
needed for schemes with quantum encoding having high entanglement fidelity.
We employ for this the following concepts from \cite{schumacher:F:e},
\cite{nielsen:schumacher}:
\par
For a quantum operation $\varphi_*:\alg{L}({\cal H})_*=\alg{A}_*\rightarrow\alg{A}_*$
and a state $\rho$ on $\alg{A}$ choose a purification $\Psi_\rho$ of
$\rho$ on the extended system $\alg{A}\otimes\alg{R}$ (for \emph{reference} system).
The \emph{entropy exchange}\footnote{We adopt the name $S_e$ for this following
  \cite{schumacher:F:e} and general physical fashion.} is defined as
$$S_e(\rho;\varphi_*)=H\left((\varphi_*\otimes\id_{\alg{R}*})\Psi_\rho\right)$$
and \cite{schumacher:F:e} shows that it does not depend on the purification
chosen. It can be seen as a measure for the quantum information exchange
between system and environment.
\par
Thus it is natural to define the \emph{coherent information} (after
\cite{nielsen:schumacher}) as
$$I_e(\rho;\varphi_*)=H(\varphi_*\rho)-S_e(\rho;\varphi_*).$$
From \cite{barnum:quantinfo} we take the following lemma, which is a
direct consequence of the \emph{quantum {\sc Fano} inequality} from
\cite{schumacher:F:e}.
\begin{lemma}
  \label{lemma:coherent:info}
  Let $\varphi_*,\psi_*$ quantum operations on the system $\alg{A}$,
  $\rho$ a state on $\alg{A}$ and denote $d^2=\dim_{\C}\alg{A}$. Then
  $$H(\rho)\leq I_e(\rho;\varphi_*)+2+4(1-F_e(\psi_*\circ\varphi_*))\log d.$$
  \phantom{.}\qed
\end{lemma}
We are now ready to prove
\begin{lemma}[Weak subadditivity of coherent information]
  \label{lemma:weak:subadd}
  Let $\rho$ a state on $\alg{A}_1\otimes\alg{A}_2$ with marginals
  $\rho_1,\rho_2$, and $\varphi_{1*}, \varphi_{2*}$ quantum
  operations on $\alg{A}_1, \alg{A}_2$, respectively. Then
  $$I_e(\rho;\varphi_{1*}\otimes\varphi_{2*}) \leq I_e(\rho_1;\varphi_{1*})+H(\rho_2).$$
\end{lemma}
\begin{beweis}
  Introducing environment systems $\alg{E}_1$, $\alg{E}_2$, pure ``null'' states
  $\tau_1$ on $\alg{E}_1$, $\tau_2$ on $\alg{E}_2$ and unitaries on the underlying
  Hilbert space of $\alg{A}_1\otimes\alg{E}_1$, $\alg{A}_2\otimes\alg{E}_2$,
  respectively, such that
  \begin{align*}
    \varphi_{1*}(\sigma) &=\tr_{\alg{E}_1}\left(U_1(\sigma\otimes\tau_1)U_1^*)\right) \\
    \varphi_{2*}(\sigma) &=\tr_{\alg{E}_2}\left(U_2(\sigma\otimes\tau_2)U_2^*)\right)
  \end{align*}
  (which is possible by {\sc Stinespring}'s theorem~\ref{satz:dilation}).
  Now what we have to prove
  (with $\alg{R}=\alg{R}_1\otimes\alg{R}_2$) is
  $$H((\varphi_{1*}\otimes\varphi_{2*})\rho)
              -H((\varphi_{1*}\otimes\varphi_{2*}\otimes\id_{\alg{R}*})\Psi_\rho)
     \leq\! H(\varphi_*\rho_1)
              -H((\varphi_{1*}\otimes\id_{\alg{A}_2*}\otimes\id_{\alg{R}*})\Psi_\rho)
              +H(\rho_2).$$
  Defining operations
  \begin{align*}
    E_{1*} &=(U_1\!\cdot{\scriptstyle\sqcup}\cdot\! U_1^*)\otimes
                    \id_{\alg{E}_2*}\otimes\id_{\alg{A}_2*}
                                        \otimes\id_{\alg{R}*} \\
    E_{2*} &=\id_{\alg{E}_1*}\otimes\id_{\alg{A}_1*}\otimes
                    (U_2\!\cdot{\scriptstyle\sqcup}\cdot\! U_2^*)\otimes
                                        \id_{\alg{R}*}
  \end{align*}
  on $\alg{E}_1\otimes\alg{A}_1\otimes\alg{E}_2\otimes\alg{A}_2\otimes\alg{R}$, and
  the state $\sigma=\tau_1\otimes\tau_2\otimes\Psi_\rho$ we can write
  this as
  $$H_{E_{1*}E_{2*}\sigma}(\alg{A}_1\alg{A}_2)+H_{E_{1*}\sigma}(\alg{A}_1\alg{A}_2\alg{R})
     \leq H_{E_{1*}E_{2*}\sigma}(\alg{A}_1\alg{A}_2\alg{R})
             +H_{E_{1*}\sigma}(\alg{A}_1)+H_{\sigma}(\alg{A}_2).$$
  Notice that all the states here are pure! Thus by
  theorem~\ref{satz:pure:common:state}
  \begin{align*}
    H_{E_{1*}E_{2*}\sigma}(\alg{A}_1\alg{A}_2)
                               &=H_{E_{1*}E_{2*}\sigma}(\alg{E}_1\alg{E}_2\alg{R}) \\
    H_{E_{1*}E_{2*}\sigma}(\alg{A}_1\alg{A}_2\alg{R})
                               &=H_{E_{1*}E_{2*}\sigma}(\alg{E}_1\alg{E}_2) \\
    H_{E_{1*}\sigma}(\alg{A}_1\alg{A}_2\alg{R})
                               &=H_{E_{1*}\sigma}(\alg{E}_1\alg{E}_2) \\
                               &=H_{E_{1*}\sigma}(\alg{E}_1)
                                  =H_{E_{1*}E_{2*}\sigma}(\alg{E}_1)
  \end{align*}
  (the last step since $E_{1*}\sigma|_{\alg{E}_2}$ is pure
  and $E_{2*}$ acts trivially on $\alg{E}_{1*}$), and our inequality transforms to
  $$H_{E_{1*}E_{2*}\sigma}(\alg{E}_1\alg{E}_2\alg{R})+H_{E_{1*}E_{2*}\sigma}(\alg{E}_1)
      \leq H_{E_{1*}E_{2*}\sigma}(\alg{E}_1\alg{E}_2)
              +H_{E_{1*}\sigma}(\alg{A}_1)+H_{E_{1*}\sigma}(\alg{A}_2).$$
  Here with strong subadditivity of entropy (theorem~\ref{satz:strong:subadd})
  the left hand side can be estimated by
  $$H_{E_{1*}E_{2*}\sigma}(\alg{E}_1\alg{E}_2)+H_{E_{1*}E_{2*}\sigma}(\alg{E}_1\alg{R})$$
  and we are done if we can prove that
  $$H_{E_{1*}E_{2*}\sigma}(\alg{E}_1\alg{R})
                       \leq H_{E_{1*}\sigma}(\alg{A}_1)+H_{E_{1*}\sigma}(\alg{A}_2).$$
  But $E_{1*}\sigma|_{\alg{A}_1\alg{A}_2\alg{E}_1\alg{R}}$ is pure,
  so again by theorem~\ref{satz:pure:common:state}
  $$H_{E_{1*}\sigma}(\alg{A}_1)=H_{E_{1*}\sigma}(\alg{A}_2\alg{E}_1\alg{R}).$$
  And since $E_{2*}$ acts trivially on $\alg{E}_{1*}\alg{R}_*$ we have
  $$H_{E_{1*}E_{2*}\sigma}(\alg{E}_1\alg{R})=H_{E_{1*}\sigma}(\alg{E}_1\alg{R})$$
  which renders our last inequality equivalent to
  $$H_{E_{1*}\sigma}(\alg{E}_1\alg{R})-H_{E_{1*}\sigma}(\alg{A}_2)
                       \leq H_{E_{1*}\sigma}(\alg{A}_2\alg{E}_1\alg{R}),$$
  and this is the triangle inequality, theorem~\ref{satz:triangle}.
\end{beweis}
\begin{bem}
  Subadditivity
  $$I_e(\rho;\varphi_{1*}\otimes\varphi_{2*})
             \leq I_e(\rho_1;\varphi_{1*})+I_e(\rho_2;\varphi_{2*})$$
  which is by $I_e(\rho_2;\varphi_{2*})\leq H(\rho_2)$ stronger than our lemma,
  and which one would expect of an information, actually \emph{fails}:
  see \cite{barnum:quantinfo}.
\end{bem}
\begin{satz}
  \label{satz:lower:blind:F_e}
  Let $(\alg{A}_1,\ldots,\alg{A}_s,\Rho,P)$ a multiple quantum source
  with $\alg{A}_i=\alg{L}({\cal H}_i)$ and $(R_1,\ldots,R_s)$
  a $(\text{quantum},F_e)$--achievable tuple. Then
  $$\forall I\subset[s]\qquad\sum_{i\in I} R_i
                              \geq H(\alg{A}(I)|\alg{A}(I^c))
                              =    H(P\Rho)-H(P\Rho|_{I^c}).$$
\end{satz}
\begin{beweis}
  Let an $(n,\lambda)_{F_e}$--coding scheme
  $(\varepsilon_{1*},\ldots,\varepsilon_{s*},\delta_*)$ with rate tuple
  $(R_1,\ldots,R_s)$ be given. Denote $d=\sum_{i=1}^s \dim{\cal H}_i$.
  \par
  We may think of $\varepsilon_{i*}$ as acting on $\alg{A}_{i*}^{\otimes n}$
  by embedding the coding space $\alg{L}({\cal K}_i)_*$. Thus we can apply
  for $I\subset[s]$ lemma~\ref{lemma:coherent:info} to
  $\varphi_*=\varphi_{1*}\otimes\varphi_{2*}$
  (with $\varphi_{1*}=\bigotimes_{i\in I}\varepsilon_{i*}$,
  $\varphi_{2*}=\bigotimes_{i\in I^c}\varepsilon_{i*}$) and
  $\psi_*=\delta_*$, and obtain
  \begin{equation*}\begin{split}
    nH(P\Rho) &\leq I_e((P\Rho)^{\otimes n};\varphi_{1*}\otimes\varphi_{2*})
                                                           +2+4n\lambda\log d \\
              &\leq I_e((P\Rho|_I)^{\otimes n};\varphi_{1*})
                                          +nH(P\Rho|_{I^c})+2+4n\lambda\log d
  \end{split}\end{equation*}
  (using weak subadditivity of the coherent information). Since trivially
  $$I_e((P\Rho|_I)^{\otimes n};\varphi_{1*})\leq n\sum_{i\in I} R_i$$
  we get the theorem in the limit of $n\rightarrow\infty$ and
  $\lambda\rightarrow 0$.
\end{beweis}
The following example shows that our nice
theorem~\ref{satz:lower:2:blind:Fbar}
is too weak, at least for nonclassically correlated sources.
At the same time it shows that also
theorem~\ref{satz:lower:blind:F_e} is too weak.
\begin{expl}[EPR source]
  \label{expl:epr:source}
  Consider the source
  $(\alg{A}_1,\alg{A}_2,\Rho,P)$ with $\alg{A}_1=\alg{A}_2=\alg{L}(\C^2)$,
  $\Rho=\{\ket{\Phi^+}\bra{\Phi^+},\ \ket{\Phi^-}\bra{\Phi^-}\}$
  (two of the {\sc Bell} states) and $P$ equidistributed on $\Rho$.
  Clearly
  $$P\Rho=\frac{1}{2}\ket{\Phi^+}\bra{\Phi^+}+\frac{1}{2}\ket{\Phi^-}\bra{\Phi^-}
                        =\frac{1}{2}\ket{00}\bra{00}+\frac{1}{2}\ket{11}\bra{11}$$
  and both marginals equal $\frac{1}{2}\1$.
  Theorems~\ref{satz:lower:2:blind:Fbar} and
  \ref{satz:lower:blind:F_e} both give only the lower
  bound $R_1+R_2\geq 1$. But we will prove that in fact
  $${\bf R}_{\text{q},F_e}\subset {\bf R}_{\text{q},\bar{F}}
                          \subset \{(R_1,R_2):\ R_1,R_2\geq 1/2\}.$$
  To see this let an $(n,\lambda)_{\bar{F}}$--coding scheme
  $(\varepsilon_{1*},\varepsilon_{2*},\delta_*)$ be given with
  rate pair $(1,R_2)$, the first encoder being the identity.
  Now imagine that two people want to use this scheme to transmit
  information: the sender encodes $0$--$1$--sequences
  as sequences of $\ket{\Phi^+}\bra{\Phi^+}$ and $\ket{\Phi^-}\bra{\Phi^-}$,
  giving the according shares of these entangles states to the two encoders.
  The receiver measures the decoded states in (the tensor power of) the
  basis $\{\ket{\Phi^+},\ket{\Phi^-}\}$, call the corresponding
  observable $D$. The transmission rate of this system clearly is $1$,
  with average error probability bounded by $\lambda$:
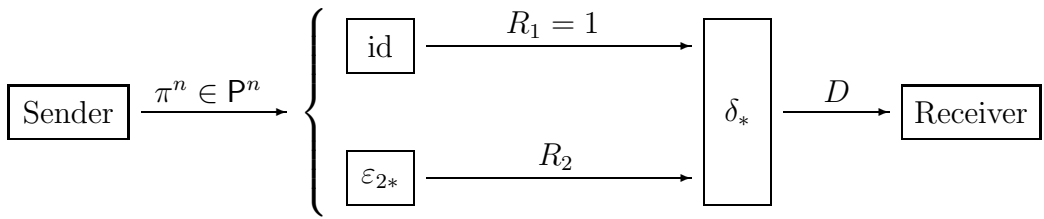
  \begin{center}
    \begin{picture}(400,100)
      \put(0,0){\begin{picture}(130,100)(0,-50)
                  \put(0,-10){\framebox(45,20){Sender}}
                  \put(50,0){\vector(1,0){55}}
                  \put(55,4){$\pi^n\in\Rho^n$}
                  \put(110,-3){$\left\{ \phantom{\makebox(0,45){.}} \right.$}
                \end{picture}}
      \put(128,45){\begin{picture}(155,50)
                     \put(0,20){\framebox(25,20){$\id$}}
                     \put(30,30){\vector(1,0){100}}
                     \put(60,35){$R_1=1$}
                   \end{picture}}
      \put(128,-5){\begin{picture}(155,50)
                     \put(0,20){\framebox(25,20){$\varepsilon_{2*}$}}
                     \put(30,30){\vector(1,0){100}}
                     \put(72,35){$R_2$}
                   \end{picture}}
      \put(263,0){\begin{picture}(127,100)(0,-50)
                    \put(0,-35){\framebox(25,70){$\delta_*$}}
                    \put(30,0){\vector(1,0){40}}
                    \put(45,4){$D$}
                    \put(75,-10){\framebox(52,20){Receiver}}
                  \end{picture}}
    \end{picture}
  \end{center}
  Allowing that the sender cooperates with the encoder $\varepsilon_{2*}$,
  and the receiver with the decoder $\delta_*$, can only increase the
  transmission rate. We may describe the new situation in a different,
  equivalent way: the two encoders get the $n^{\text{th}}$ power of
  the maximally entangled state $\ket{\Phi^+}\bra{\Phi^+}$, while
  the second encoder, before performing his $\varepsilon_{2*}$, does
  the \emph{message encoding (!)} for the sender. This is done with the
  help of the \emph{phase flip operator}
  \begin{equation*}
    \beta: \begin{cases}
             \ket{0} &\longmapsto\phantom{-}\ket{0} \\
             \ket{1} &\longmapsto          -\ket{1}
           \end{cases}
  \end{equation*}
  on $\C^2$, as it is readily checked that
  $(\id\otimes\beta)\ket{\Phi^+}=\ket{\Phi^-}$.
  But here the first encoder becomes superfluous: thus we can assume
  that initially sender and receiver share $n$ maximally entangled
  pairs $\ket{\Phi^+}\bra{\Phi^+}$, and the second encoder (viz., the sender!)
  transmits $nR_2$ qubits to the receiver. This is exactly
  the situation of \emph{superdense coding}, invented by
  \cite{bennett:wiesner:superdense}: and it is well known that
  the maximal transmission rate in this situation is $2R_2$,
  forcing $R_2\geq 1/2$ in the limit of $n\rightarrow\infty$,
  $\lambda\rightarrow 0$. Of course symmetrically for $R_1$.
  \par
  We can note the two lessons we learned:
  \begin{center}
    \fbox{\qquad Theorem~\ref{satz:lower:2:blind:Fbar} is too weak.\qquad}
  \end{center}
  \begin{center}
    \fbox{\qquad Theorem~\ref{satz:lower:blind:F_e} is too weak.\qquad}
  \end{center}
\end{expl}
\par
The last example shows the difference between average and entanglement
fidelity:
\begin{expl}[Cloned cross]
  \label{expl:ghz:source}
  Consider the source $(\alg{A}_1,\alg{A}_2,\Rho,P)$ with
  $\alg{A}_1=\alg{A}_2=\alg{L}(\C^2)$ and $P$ equidistributed
  on $\Rho=\{\ket{00}\bra{00},\ \ket{11}\bra{11}\}$. Clearly
  $$P\Rho=\frac{1}{2}\ket{00}\bra{00}+\frac{1}{2}\ket{11}\bra{11}$$
  with both marginals equal to $\frac{1}{2}\1$. A natural
  purification of this source would be by the \emph{{\sc GHZ}--state}
  $\frac{1}{\sqrt{2}}\left(\ket{000}+\ket{111}\right)$, invented
  by \cite{greenberger:horne:zeilinger} to extend {\sc Bell}'s
  theorem to multi--party entanglement.
  \par
  Since the average state is the same as in the EPR source we have
  $${\bf R}_{\text{q},F_e}\subset \{(R_1,R_2):\ R_1,R_2\geq 1/2\}.$$
  \par
  On the other hand it is obvious that
  $${\bf R}_{\text{q},\bar{F}}=\{(R_1,R_2):\ R_1,R_2\geq 0,\ R_1+R_2\geq 1\}.$$
  It is clear from theorem~\ref{satz:lower:2:blind:Fbar}
  that $R_1+R_2\geq 1$ is necessary (even with arbitrary encoding).
  And also one sees easily that
  $R_1=1,\ R_2=0$ is $(\text{quantum},\bar{F})$--achievable:
  $\alg{A}_2$ sends nothing, whereas $\alg{A}_1$
  transmits his qubit faithfully, the decoder has just to copy it to obtain
  the initial joint state (this is only possible
  because the two alternative states sent by $\alg{A}_1$ are orthogonal!).
  \par
  Again collecting our lessons:
  \begin{center}
    \fbox{\qquad ${\bf R}_{\text{q},\bar{F}}$ depends not just on $P\Rho$.\qquad}
  \end{center}
  \begin{center}
    \fbox{\qquad In general ${\bf R}_{\text{q},\bar{F}}$ and
      ${\bf R}_{\text{q},F_e}$ are different.\qquad}
  \end{center}
\end{expl}
\bigskip\par
Concluding this section we may state that the pleasing situation of
chapter~\ref{chap:source} has completely dissolved:
all three rate concepts differ, and (except for
entanglement fidelity) the rate region depends not only on the average state.
%
%We close the section with bounds for arbitrary encoding:
%$$R_A+R_B\geq H(\rho)\text{ (clear), and}$$
%$$R_A\geq \sum_b p(\cdot,b)H\left(\sum_a p(a|b)\pi_a\right),\quad
%                         R_B\geq \sum_a p(a,\cdot)H\left(\sum_b p(b|a)\tau_b\right)$$
%This is reasonable by considering the states
%$\sigma_A=\sum_{a,b} p(a,b)\ket{a}\bra{a}\otimes\tau_b$
%and $\sigma_B=\sum_{a,b} p(a,b)\pi_a\otimes\ket{b}\bra{b}$.

\section{Classical source with quantum side information}
\label{sec:multsrc:hybrid:c}
In this and the following section we will turn to the study of
a restricted kind of multiple source, namely $c^s h^1$--sources,
and we will be able to complement the above bewildering picture
by some positive results (coding theorems).
\begin{satz}[Code partition] (Cf. \cite{csiszar:koerner}, proof of theorem 3.1.2)
  \label{satz:code:partition}
  Let $W:\fset{X}\rightarrow\alg{S}(\alg{Y})$ a q--DMC, $P$ a probability
  distribution on $\fset{X}$, $\lambda,\delta,\eta>0$. Then for
  $n\geq n_0(|\fset{X}|,\dim{\cal H},\lambda,\delta,\eta)$
  there exist $m-1\leq \exp(n(H(P)-I(P;W)+3\delta))$ many
  $(n,\lambda)$--codes with pairwise disjoint
  ``large'' codebooks $\fset{C}_i$:
  $$|\fset{C}_i|\geq\exp(n(I(P;W)-2\delta)),$$
  such that $P^n(\fset{X}^n\setminus\bigcup_{i=1}^{m-1} \fset{C}_i)<\eta$.
\end{satz}
\begin{beweis}
  Choose $\alpha>0$ such that $P^n(\fset{T}^n_{V,P,\alpha})\geq 1-\eta/2$ and
  $n$ large enough such that for every $\fset{A}\subset\fset{T}^n_{V,P,\alpha}$
  with $P^n(\fset{A})\geq \eta/2$ there is a $(n,\lambda)$--code
  with codebook $\fset{C}\subset\fset{A}$ and
  $|\fset{C}|\geq\exp(n(I(P;W)-2\delta))$
  (by the coding theorem~\ref{satz:maximal:codes}).
  Now choose such a codebook $\fset{C}_1\subset\fset{A}_1=\fset{T}^n_{V,P,\alpha}$
  and inductively
  $\fset{C}_i\subset\fset{A}_i=\fset{A}_{i-1}\setminus\fset{C}_{i-1}$
  until $P^n(\fset{A}_i)<\eta/2$, say for $i=m$.
  Obviously the codebooks are disjoint, and
  the rest has weight less than $\eta$. It remains to estimate $m$:
  $$(m-1)\cdot\exp(n(I(P;W)-2\delta))
                   \leq\sum_{i=1}^{m-1} |\fset{C}_i|
                   \leq |\fset{T}^n_{V,P,\alpha}|\ ,$$
  and since by lemma~\ref{lemma:variance:typical:sequences}
  $|\fset{T}^n_{V,P,\alpha}|\leq \exp(n(H(P)+\delta))$
  for large enough $n$ we get the statement.
\end{beweis}
Consider the problem to encode the classical part of a $c^1 h^1$--source
$(\alg{X}=\C\fset{X},\alg{Y},\fset{X}\times\Rho,P)$, using the quantum
source as side information at the decoder:
\par
An $n$--\emph{block coding scheme with quantum side information at the decoder}
is a pair $(f,D)$,
with a mapping $f:\fset{X}^n\longrightarrow\fset{M}$ and an observable
$D$ on $\C\fset{M}\otimes\alg{Y}$ indexed by $\fset{X}$.
Its error probability (averaged over $P$) is
$$\bar{e}(f,{D})=1-\sum_{x^n\in\fset{X}^n,\pi^n\in\Rho^n}
                                 P^n(x^n,\pi^n)\tr((f(x^n)\otimes\pi^n){D}_{x^n}).$$
\par
The proof of the following theorem goes back to an idea of \cite{ahlswede:rate:slicing}:
\begin{satz}[Rate slicing](Cf. \cite{csiszar:koerner}, theorem 3.1.2)
  \label{satz:source:quside}
  For every $\bar{\lambda},\delta>0$ and $c^1 h^1$--source
  $(\alg{X}=\C\fset{X},\alg{Y},\fset{X}\times\Rho,P)$ there exists an
  $n$--block code $(f,{D})$ with quantum side information at the decoder
  such that
  $$\frac{1}{n}\log|\fset{M}|\leq H(\alg{X}|\alg{Y})+3\delta,\text{  and }
                                                  \bar{e}(f,{D})\leq\bar{\lambda}$$
  whenever $n\geq n_0(|\fset{X}|,\dim{\cal H},\bar{\epsilon},\delta)$.
  Furthermore, the observable may be modified to the operation
  $D_*'=\tr_{\C\fset{M}}\circ D_{\text{tot}*}$
  from $(\C\fset{M})_*\otimes\alg{Y}_*$ to
  $(\C{\fset{X}^n})_*\otimes\alg{Y}_*$ which satisfies
  $$\sum_{x^n\in\fset{X}^n,\pi^n\in\Rho^n} P^n(x^n,\pi^n)
      \left\|x^n\otimes\pi^n-D_*'({f(x^n)}\otimes\pi^n)\right\|_1
                                         \leq\sqrt{8\bar{\lambda}}+\bar{\lambda}.$$
\end{satz}
\begin{beweis} 
  Define the q--DMC $W:\fset{X}\rightarrow\alg{Y}_*$ by
  $$W_x=\frac{1}{P_{\fset{X}}(x)} \sum_{\pi\in\Rho} P(x,\pi)\pi$$
  (with the marginal distribution $P_{\fset{X}}$ of $P$ on $\fset{X}$).
  Choose $\eta\leq\bar{\lambda}$ in
  theorem~\ref{satz:code:partition}, and $n$ accordingly
  large such that codes $(g_i,{D}_i)$, $i\in[m-1]$ like in that theorem exist. Assume
  that their message sets coincide with their codebooks and that $g_i$ is the identity.
  \par
  Define now
  $$f({x^n})=\left\{\begin{array}{ll} i &\text{  if }{x^n}\in\fset{C}_i\ ,\\
                                        m &\text{  else.}
                      \end{array}\right.$$
  The decoder reads $i=f({x^n})$ and if $i\neq m$ uses ${D}_i$ to recover
  $x^n$ from the side information: formally, ${D}$ consists of the operators
  $[i]\otimes{D}_{ic}$ for $i\in[m-1]$, $c\in\fset{C}_i'$, and
  $[m]\otimes\1$. That this has the desired properties is easily checked.
  Now for the second part: observe that
  $$D_*':[j]\otimes\rho\longmapsto\left\{
           \begin{array}{ll}
               \sum_{c\in\fset{C}_j} [c]\otimes\sqrt{{D}_{jc}}\rho\sqrt{{D}_{jc}}
                                                     &\text{  if }j<m,\\
               \left[m\right]\otimes\rho             &\text{  if }j=m.
           \end{array}\right.$$
  By the tender measurement lemma~\ref{lemma:tender:measurement}
  and note~\ref{bem:average:tenderness} the assertion follows.
\end{beweis}
\begin{bem}
  The decoder either says ``don't know'' (with probability at most
  $\bar{\lambda}$ over the source distribution $P^n$), or decodes correctly
  with maximal error probability $\bar{\lambda}$.
\end{bem}
\begin{cor}
  \label{cor:ratepoint:c1:qside}
  For the $c^1 h^1$--source
  $(\alg{X}=\C\fset{X},\alg{Y},\fset{X}\times\Rho,P)$ the pair
  $(H(\alg{X}|\alg{Y}),H(\alg{Y}))$ is $(\text{quantum},\bar{F})$--achievable.
\end{cor}
\begin{beweis}
  Combine theorem~\ref{satz:source:quside} with {\sc Schumacher}'s quantum coding.
\end{beweis}
Consider now the $c^s h^1$--source
$$((\alg{X}_i=\C\fset{X}_i|i\in[s]),\alg{Y},
                            \fset{X}_1\times\cdots\times\fset{X}_s\times\Rho,P).$$
\par
An $n$--\emph{block coding scheme with quantum side information at the decoder}
for this is a $(s+1)$--tuple
$(f_1,\ldots,f_s,D)$ of mappings $f_i:\fset{X}_i^n\rightarrow\fset{M}_i$
and an observable $D$ on $\alg{X}_1\otimes\cdots\otimes\alg{X}_s\otimes\alg{Y}$,
indexed by $\fset{X}_1^n\times\cdots\times\fset{X}_s^n$.
Its error probability (averaged over $P$) is
$$\bar{e}(f_1,\ldots,f_s,D)=1-\!\sum_{x^n_i\in\fset{X}_i^n,\rho\in\Rho^n}
                   P(x^n_1,\ldots,x^n_s,\rho)
                    \tr((f_1(x^n_1)\otimes\cdots\otimes f_s(x^n_s)\otimes\rho)
                    D_{x^n_1\ldots x^n_s}).$$
\begin{satz}
  \label{satz:multsrc:hybrid:c:coding}
  With the notation above and $\bar{\lambda},\delta>0$ there exists
  an $n$--block coding scheme with quantum side information at
  the decoder with
  $$\forall J\subset[s]\quad \frac{1}{n}\sum_{j\in J}\log|\fset{M}_j|
                \leq H(\alg{X}(J)|\alg{X}(J^c)\alg{Y})+|J|\cdot 3\delta$$
  and error probability at most $\bar{\lambda}$, whenever
  $n\geq n_0(|\fset{X}_i|,\dim{\cal H},\bar{\lambda},\delta)$.
  \par
  Moreover for the operation
  $D_*'=\tr_{\C(\fset{M}_1\times\cdots\times\fset{M}_s)}\circ D_{\text{tot}*}$,
  $$D_*':\C(\fset{M}_1\times\cdots\times\fset{M}_s)_*\otimes\alg{Y}^{\otimes n}_*
                        \rightarrow\C(\fset{X}_1^n\times\cdots\times\fset{X}_s^n)_*
                                                  \otimes\alg{Y}_*^{\otimes n},$$
  it holds that
  $$\sum_{{x^n}_i\in\fset{X}_i^n,\rho\in\Rho^n} P(x^n_1,\ldots,x^n_s,\rho)
       \|[x^n_1\ldots x^n_s]\otimes\rho-D_*'
                     ([f_1(x^n_1)\ldots f_s(x^n_s)]\otimes\rho)\|_1
    \leq\bar{\lambda}.$$
\end{satz}
\begin{beweis}
  Only the second statement is to be proved.
  We use induction on $s$, the number of sources:
  $s=1$ is clear by direct application of the rate slicing
  theorem~\ref{satz:source:quside}. For $s>1$
  it is sufficient (by the time sharing principle) to
  consider only extreme points of the region: thus w.l.o.g.
  \begin{align*}
    \frac{1}{n}\log|\fset{M}_1| &\leq H(\alg{X}_1|\alg{Y})+3\delta \\
    \frac{1}{n}\log|\fset{M}_2| &\leq H(\alg{X}_2|\alg{X}_1\alg{Y})+3\delta \\
    \ldots                      & \\
    \frac{1}{n}\log|\fset{M}_s| &\leq
                    H(\alg{X}_s|\alg{X}_1\cdots\alg{X}_{s-1}\alg{Y})+3\delta.
  \end{align*}
  The proof that these are indeed the extreme points is in the section
  {\em Extreme points of rate regions} below.
  \par
  Now by induction we have an $(n,\bar{\lambda}/2)$--coding scheme for the
  source
  $$((\alg{X}_i=\C\fset{X}_i|i\in[s-1]),\alg{X}_s\otimes\alg{Y},
     \fset{X}_1\times\cdots\times\fset{X}_{s-1}\times(\fset{X}_s\times\Rho),P),$$
  call its decoding operation $D_{1*}'$.
  By rate slicing we also have an $(n,\bar{\lambda}/2)$--coding scheme for the
  source $(\alg{X}_s,\alg{Y},\fset{X}_s\times\Rho,P)$
  with side information at the decoder, call its decoding operation
  $D_{2*}'$. Then the concatenation
  $D_*'=D_{1*}'\circ(\id\otimes D_{2*}')$ of
  the two processes obviously has the desired error
  properties, and it is readily checked that it has the stated form.
  By tracing out $\alg{Y}^{\otimes n}$ we recover the observable $D$.
\end{beweis}
\begin{bem}
  The theorem shows that not only we can use quantum side information ``just like''
  classical information to improve compression but also that we can do so with almost
  not disturbing the quantum information.
\end{bem}
\begin{cor}
  \label{cor:ratepoint:cs:qside}
  For the above source all tuples $(R_1,\ldots,R_s,H(\alg{Y}))$ satisfying
  $$\forall J\subset[s]\quad
                 \sum_{j\in J} R_j \geq H(\alg{X}(J)|\alg{X}(J^c)\alg{Y})$$
  are $(\text{quantum},\bar{F})$--achievable.
\end{cor}
\begin{beweis}
  Combine theorem~\ref{satz:multsrc:hybrid:c:coding}
  with {\sc Schumacher}'s quantum coding.
\end{beweis}
We close this section with a converse to these coding theorems:
\begin{satz}
  \label{satz:cs:q1:side:converse}
  Still with the above source all $(\text{quantum},\bar{F})$--achievable
  rate tuples of the form $(R_1,\ldots,R_s,H(\alg{Y}))$ satisfy
  $$\forall J\subset[s]\quad
       \sum_{j\in J} R_j \geq H(\alg{X}(J)|\alg{X}(J^c)\alg{Y}).$$
\end{satz}
\begin{beweis}
  Otherwise we could by theorem~\ref{satz:mac:code:constr}
  construct an infinite sequence of transmission
  $n$--block codes for the quantum multiple access channel
  \begin{align*}
    W:\fset{X}_1\times\cdots\times\fset{X}_s &\longrightarrow \alg{S}(\alg{Y}) \\
    (x_1,\ldots,x_s)                         &\longmapsto
               \frac{1}{P_{\fset{X}_1\times\cdots\times\fset{X}_s}(x_1\ldots x_s)}
                                              \sum_{\pi\in\Rho} P(x_1\ldots x_s,\pi)
  \end{align*}
  which violate the outer bounds of theorem~\ref{satz:weak:converse}.
\end{beweis}

\section{Quantum source with classical side information}
\label{sec:multsrc:hybrid:q}
The simplest instance of the problem considered in the previous section is the case of
the $c^1 q^1$--source. There we solved the problem
of compressing the classical source with the quantum information as side information
at the decoder, which gave us one extreme point of the rate region of the
multiple source coding problem. It is natural, therefore, to consider the
complementary problem of compressing the quantum source, using the classical
information as side information, preferably only at the decoder: this would
give us another extreme point, presumably
completing the determination of the rate region
of the $c^1 q^1$--source (if the bounds of theorem~\ref{satz:lower:2:blind:Fbar}
are already the correct ones).
\par
An $n$--\emph{block quantum source coding scheme with side information at the
decoder} for the $c^1 q^1$--source
$(\alg{X}=\C\fset{X},\alg{Y}=\alg{L}({\cal H}),\fset{X}\times\Rho,P)$
is a pair $(\varepsilon_*,\delta_*)$ with a mapping
$$\varepsilon_*: \Rho^n \longrightarrow \alg{S}(\alg{L}({\cal K}))$$
and a family of quantum operations
$$\delta_*:\fset{X}^n\times\alg{L}({\cal K})_*
                   \longrightarrow \alg{Y}_*^{\otimes n}.$$
Quantum and arbitrary encoding are as before, also rate, and the average
fidelity is
$$\bar{F}=\bar{F}(\varepsilon_*,\delta_*)=\sum_{(x^n,\pi^n)\in\fset{X}^n\times\Rho^n}
                   P^n(x^n,\pi^n)\!\cdot\!\tr((\delta_*(x^n)\varepsilon_*\pi^n)\pi^n)$$
(average distortion $\bar{D}(\varepsilon_*,\delta_*)$ similarly).
\par
The limiting rates $R_{q,{\bar F}}(\lambda)$ and  $R_{a,{\bar F}}(\lambda)$
are defined obviously.
What can we say about them? From theorem~\ref{satz:lower:2:blind:Fbar}
we get at least
$$\liminf_{\lambda\rightarrow 0} R_{q,{\bar F}}(\lambda)
           \geq H(\alg{Y}|\alg{X})
           =    \sum_{x\in\fset{X}} P_{\fset{X}}(x)
                    H\left(\sum_{\pi\in\Rho}\frac{P(x,\pi)}{P_{\fset{X}}(x)}\pi\right).$$
\par
In fact even $R_{a,{\bar F}}(\lambda)\geq H(\alg{Y}|\alg{X})$ for
$\lambda\in(0,1)$: otherwise we could (with compressing $\alg{X}$
classically, e.g. by ignoring all non--typical sequences) compress
the total source $\alg{X}\alg{Y}$ with asymptotically at most
$R_{a,{\bar F}}(\lambda)+H(\alg{X}) < H(\alg{Y}|\alg{X})+H(\alg{X})
                                    = H(\alg{X}\alg{Y})$
qubits per symbol, contradicting
theorem~\ref{satz:general:source:coding}.
\par
At present we do not know if one can approach this bound. But let us
make an experiment! Assume that also the encoder has the side information,
i.e. now
$$\varepsilon_*: \fset{X}^n\times\Rho^n \longrightarrow \alg{S}(\alg{Y}).$$
Since we are interested only in average performance it suffices that
the scheme works well for typical $x^n\in\fset{X}^n$, say
$x^n\in\fset{T}^n_{V,P_{\fset{X}},\alpha}$. To encode this
the encoder has just to collect the positions of equal $x\in\fset{X}$
and do {\sc Schumacher} quantum coding on blocks of length
$nP_{\fset{X}}(x)\pm\alpha\sqrt{P_{\fset{X}}(x)(1-P_{\fset{X}}(x))}\sqrt{n}$.
This scheme --- with side information both at the encoder
and the decoder --- obviously achieves the rate $H(\alg{Y}|\alg{X})$
asymptotically with arbitary high fidelity.

\section{The $c^0 q^2$--source: coding vs. side information}
\label{sec:c0:q2:source}
With the $c^1 q^1$--source the idea to consider extreme points in a certain
convex region proved useful, and in connection with this the idea to encode
only part of the source while using the rest as side information
at the decoder.
\par
Whereas this paradigm is of undoubted worth in the classical theory, where
we took it from (and which gave us some insights already for quantum communication
problems, not just in the two previous sections but also in chapter~\ref{chap:mac}),
in general one must be cautious with it: \emph{using} quantum information
often means using it \emph{up}. As an illustration consider once more the
cloned wheel example~\ref{expl:sym2:source}:
\par
Obviously we can encode $\alg{A}_1$ with rate zero, with side information
from $\alg{A}_2$ at the decoder, because the state $\pi$ on
$\alg{A}_2$ is a faithful copy of the lost state $\pi$ on $\alg{A}_1$.
This is of course in contrast to theorem~\ref{satz:lower:2:blind:Fbar},
and we can note our last lesson:
\begin{center}
  \fbox{\quad Coding independent sources is not reducible to coding
    with side information.\quad}
\end{center}

\section{Extreme points of rate regions}
\label{sec:extreme:points}
Here we prove the claim in the proof of
theorem~\ref{satz:multsrc:hybrid:c:coding}
that every extremal point of the region of all
$(R_1,\ldots,R_s)$ which satisfy for all $J\subset[s]$
$$\hspace{5cm}
  R(J)=\sum_{i\in J}R_i\geq H(\alg{X}(J)|\alg{Y}\alg{X}(J^c))
  \hspace{4cm}(J)$$
is of the form
$$R_{\pi(i)}=H(\alg{X}_{\pi(i)}|\alg{Y}\alg{X}_{\pi(1)}\cdots\alg{X}_{\pi(i-1)})$$
for a permutation $\pi$ of the set $[s]$,
and that these points all belong to the above region.
\par
Assume that we have an extremal point: it follows that $s$ of the inequalities
$(J)$ are met with equality. Choose one, say $K$:
$$R(K)=H(\alg{X}(K)|\alg{Y}\alg{X}(K^c)).$$
We claim that we can find the remaining inequalities $(J)$ met with equality
among the $J\subset K$ or $J\supset K$. This follows from the
following
\begin{lemma}
  From $R(K)=H(\alg{X}(K)|\alg{Y}\alg{X}(K^c))$ the validity of $(J)$ for
  all $J$ follows from the validity for those which contain $K$ or are contained
  in $K$.
\end{lemma}
\begin{beweis}
  First consider $J\supset K$: there we have
  $$R(J\setminus K)\geq
                  H(\alg{X}(J)|\alg{Y}\alg{X}(J^c))-H(\alg{X}(K)|\alg{Y}\alg{X}(K^c)).$$
  Thus for arbitrary $J$, setting $J_1=J\cap K$, $J_2=J\cap K^c$, one obtains
  \begin{equation*}\begin{split}
    R(J) &\geq H(\alg{X}(J_1)|\alg{Y}\alg{X}(J_1^c))
                         +H(\alg{X}(J_2\cup K)|\alg{Y}\alg{X}(J_2^c\cap K^c))
                                           -H(\alg{X}(K)|\alg{Y}\alg{X}(K^c))\\
         &=    H(\alg{Y}\alg{X}_1\cdots\alg{X}_s)-H(\alg{Y}\alg{X}(J_1^c))
                     -H(\alg{Y}\alg{X}(J_2^c\cap K^c))+H(\alg{Y}\alg{X}(K^c))\\
         &\geq H(\alg{Y}\alg{X}_1\cdots\alg{X}_s)-H(\alg{X}(J_1^c\cap J_2^c))\\
         &=    H(\alg{X}(J)|\alg{Y}\alg{X}(J^c))
  \end{split}\end{equation*}
  by strong subadditivity (theorem~\ref{satz:strong:subadd}), applied to
  $\alg{A}_1=\alg{X}(J_2)$, $\alg{A}_2=\alg{Y}\alg{X}(K^c\setminus J_2)$
  and $\alg{A}_3=\alg{X}(K\setminus J_1)$.
\end{beweis}
If $K$ is not a singleton there must be equalities below $K$, if $K\neq[s]$ there
must be some above: otherwise it is easily seen that we are not in an extremal point.
So by induction we arrive at a chain
$\emptyset\neq K_1\subset K_2\subset\ldots\subset K_s=[s]$
of equalities, w.l.o.g. $K_i=\{s,s-1,\ldots,s+1-i\}$,
which produces
$$R_i=H(\alg{X}_i|\alg{Y}\alg{X}_1\cdots\alg{X}_{i-1}).$$
To see that this is indeed a point of the region apply again the lemma, iteratively.

\section{Open questions}
\label{sec:multsrc:open}
The reader will have noticed that the past chapter consisted mainly of open
questions, skilfully disguised as half theorems, examples and suggestions.
For convenience we collect here some of the more important problems:

\paragraph{Examples}
Clarify example~\ref{expl:sym2:source}: are there coding schemes
with $R_1=1,R_2=1/2$?
\par
Clarify the examples~\ref{expl:epr:source} and \ref{expl:ghz:source}:
can one improve the bounds? or can one actually construct coding schemes
with $R_1=R_2=1/2$, or at least $R_1=1,R_2=1/2$ for one or both of them?

\paragraph{The $c^1 q^1$--source}
Solve the $c^1 q^1$--source completely! From the above it is enough
to consider quantum source coding with classical side information, since
we conjecture that theorem~\ref{satz:lower:2:blind:Fbar}
gives (at least in this case) already the right bounds.

\paragraph{More complicated sources}
Solve the $c^0 q^2$--source: it seems that it is easier if we insist
on no entanglement, but this might be a deception.

\paragraph{Consider entanglement fidelity}
This seems to be the only right choice if dealing with arbitrary kinds
of correlation. Also it simplifies things a bit: namely at least the result
will depend only on the average state of the source.

\paragraph{Techniques}
The only technique for code construction was the ``code partition'' trick.
This is not satisfactory, as it destroys artificially the symmetry of the
situation; also we have to resort to a channel coding theorem.
\par
A promising direct approach that may be converted to work for the quantum problems
is the \emph{hypergraph coloring paradigm} (see \cite{ahlswede:cov:col:1}
and \cite{ahlswede:cov:col:2}). Such a program would involve to
elaborate further on techniques describable maybe by the term
\emph{noncommutative combinatorics}.

\paragraph{Guiding ideas?}
One of the initial motivations of the work in this chapter was the idea that the
classical {\sc Slepian--Wolf} theorem is one possibility to give operational
meaning to conditional entropies. As such a thing is completely lacking
in quantum information theory, and on the other hand only formal definitions
of quantum conditional entropy exist (derived from analogies, say with classical
quantities), without consistent operational meaning, one sees that solving
the above coding theorems would clarify this point dramatically.
\par
It is interesting
to note that already at this stage we can foresee (from the lessons we learned from
our examples) that there must be necessarily \emph{several} natural notions
of conditional entropy.
\par
Also we observe that the theory around {\sc Schumacher}'s
coherent information fails to give the right answers even to simple problems.
I suspect that this comes from the fact that this theory builds on
\emph{pair entanglement}, whereas our situations involved
\emph{multi--party entanglement}.

% anhang
\appendix
% QDissertation: Quantum Probability

\chapter{Quantum Probability and Information}
\label{chap:quprob}

\thispagestyle{myheadings}

In this appendix the basic mathematical machinery of quantum probability
with special attention to information theory is collected. Alongside we
introduce a calculus of entropy and information quantities in quantum systems.
Whenever possible we refer to the literature instead of giving full proofs.

\section{Quantum systems}
\label{sec:quprob:systems}
In classical probability theory one has generally two ways of seeing things: either
through distributions (and the relation of their images, mostly marginals), or through
random variables (with a joint distribution). Both ways have their merits (though
random variables are considered more elegant), but basically they are equivalent,
in particular none lacks anything without the other. Things are different in quantum
probability, and we will take the following view: the analog of a distribution is
a density operator on some complex Hilbert space, whereas the analog
of random variables are \emph{observables}, defined below. With density operators
alone we can study physical processes transforming them, but every experiment
involves some observable. Studying observables one usually fixes the underlying
density operator (as the statistics of the experiments depend on the latter),
but this falls short of not appropriately reflecting our manipulating quantum
states, or having several alternative states.
\par
For the following we refer to textbooks on C${}^*$--algebras
like \cite{arveson:invitation}, and standard
references on basic mathematics of quantum mechanics: \cite{davies:opensystems},
\cite{kraus:states:etc}, and the more advanced \cite{holevo:quantum:statistics}.
\par
A C${}^*$--\emph{algebra} with unit is a complex
Banach space $\alg{A}$ which is also a $\C$--algebra with unit $\1$ and
a ${\mathbb{C}}$--antilinear involution $*$, such that
$$\|AB\|\leq \|A\|\|B\|,\qquad \|A^*\|^2=\|A\|^2=\|AA^*\|$$
These algebras will be the mathematical models for quantum systems, and subsystems
are simply $*$--subalgebras (which are always assumed to be closed).\par
The set $\alg{A}^+$ of $A\in\alg{A}$ that can be written as $A=BB^*$ is called the
\emph{positive cone} of $\alg{A}$ which is norm closed, and induces a partial
order $\leq$. By the famous {\sc Gelfand--Naimark--Segal} representation
theorem (see e.g. \cite{arveson:invitation}) every C${}^*$--algebra
is isomorphic to a closed $*$--subalgebra of some
$\alg{L}({\cal H})$, the algebra of bounded linear operators on the
Hilbert space ${\cal H}$. With us all C${}^*$--algebras will be
of finite dimension. It is known that those
algebras are isomorphic to a direct sum of $\alg{L}({\cal H}_i)$
(see e.g. \cite{arveson:invitation}).
This includes as extremal cases the algebras $\alg{L}({\cal H})$, and the commutative
algebras $\C\fset{X}$ over a finite set $\fset{X}$, with
the generators $x\in\fset{X}$ as idempotents. In particular we have on every such
algebra a well defined and unique \emph{trace functional}, denoted $\tr$, that assigns
trace one to all minimal positive idempotents.

\paragraph{States}
A \emph{state} on a C${}^*$--algebra $\alg{A}$ is a positive $\C$--linear
functional $\rho$ with $\rho(\1)=1$. Positivity here means that its
values on the positive cone are nonnegative. Clearly the states form a convex
set $\alg{S}(\alg{A})$ whose extreme points are called \emph{pure} states,
all others are \emph{mixed}.
For $\alg{A}=\alg{L}({\cal H})$ the pure states are exactly the
one--dimensional projectors, i.e. using {\sc Dirac}'s
\emph{bra--ket}--notation, the $\ket{\psi}\bra{\psi}$ with
unit vector $\ket{\psi}\in{\cal H}$.
\par
One can easily see that every state $\rho$ can
be represented uniquely in the form $\rho(X)=\tr(\hat{\rho} X)$ for a positive,
selfadjoint element $\hat{\rho}$ of $\alg{A}$ with trace one
(such elements are called \emph{density operators}). In the sequel we
will therefore make no distinction
between $\rho$ and its density operator $\hat{\rho}$.
The set of operators with finite trace will be denoted $\alg{A}_*$, the
\emph{trace class} in $\alg{A}$ which contains the states and is a two--sided
ideal in $\alg{A}$, the {\sc Schatten}--ideal
(in our --- finite dimensional --- case this is of course just $\alg{A}$).
Then $\tr(\rho A)$ defines a real bilinear and positive definite pairing
of $\alg{A}_{*s}$ and $\alg{A}_s$, the selfadjoint parts of $\alg{A}_*$ and $\alg{A}$,
which makes $\alg{A}_s$ the dual of $\alg{A}_{*s}$.
Notice that in this sense pure states are equivalently
described as minimal selfadjoint idempotents of $\alg{A}$.

\paragraph{Observables}
Let ${\cal F}$ be a $\sigma$--algebra on some set $\Omega$, $\alg{X}$ a C${}^*$--algebra.
A map $X:{\cal F}\longrightarrow \alg{X}$ is called a \emph{positive operator valued
measure} (POVM), or an \emph{observable}, with values in $\alg{X}$ (or on $\alg{X}$), if:
\begin{enumerate}
  \item $X(\emptyset)=0,\  X(\Omega)=\1$.
  \item $E\subset F$ implies $X(E)\leq X(F)$.
  \item If $(E_n)_n$ is a countable family of pairwise disjoint sets in ${\cal F}$ then
        $X(\bigcup_n E_n)=\sum_n X(E_n)$ (in general the convergence is to be
        understood in the weak topology: for every state its value at the left
        equals the limit value at the right hand side).
\end{enumerate}
If the values of the observable are all projection operators and $\Omega$ is
the real line one speaks of a \emph{spectral measure} or a
\emph{{\sc von Neumann} observable}.\footnote{Strictly
  speaking this term only applies to the expectation of the measure
  (in general an unbounded operator), but this in turn by the spectral theorem
  determines the measure.}
An observable $X$ together with a state $\rho$ yields a probability measure
$P^X$ on $\Omega$ via the formula
$$P^X(E)=\tr(\rho X(E)).$$
In this way we may view $X$ as a random variable with values in $\alg{X}$, its
distribution we denote $P_X$ (note that $P_X$ may not be isomorphic to $P^X$: if
$X$ takes the same value on disjoint events, which means that $X$ introduces randomness
by itself).
\par
Two observables $X$, $Y$ are said to be \emph{compatible}, if
they have values in the same algebra and
${XY}={YX}$ elementwise, i.e. for all $E\in{\cal F}_X$,
$F\in{\cal F}_Y$: $X(E)Y(F)=Y(F)X(E)$
(Note that it is possible for an observable not to be compatible with itself).
By the way, the term \emph{compatible}
may be defined in obvious manner for arbitrary sets or collections of operators, in
which meaning we will use it in the sequel.
If $X,Y$ are compatible we may define their \emph{joint observable}
${XY}:{\cal F}_X\times{\cal F}_Y\longrightarrow\alg{X}$
mapping $E\times F$ to $X(E)Y(F)$ (this defines the product mapping
uniquely just as in the classical case of product measures).
In fact we can analogously define the
joint observable for any collection of pairwise compatible observables.\footnote{Observe
  however that in general a joint observable might exist for non--compatible (i.e.
  non--commuting) observables. The operational meaning of this is that there is a
  common refinement of the involved observables. If they commute then this certainly is
  possible as demonstrated, but commutativity is not necessary.}
As the random variable of a product ${XY}$ we will take $X\times Y$,
rather than $XY$ itself, with values in $\alg{X}\times\alg{X}$ (because the same product
operator may be generated in two different ways which we want to distinguish).
To indicate this difference we will sometimes write $X\cdot Y$ for the product.
\par
Note that here we can see the reason why we cannot just consider all observables as
random variables (and forget about the state): they will not have a joint distribution,
at first of course only by our definition. But {\sc Bell}'s theorem (\cite{bell:inequality})
shows that one comes into trouble if one tries to allow a joint distribution
for noncompatible observables. Conversely we see why we cannot do without
observables, even though $\rho$ contains all possible information:
the crux is that we cannot access it due to the forbidden noncompatibel
observables (a good account of this aspect of quantum
theory is by \cite{peres:quantum:theory}).
\par
From now on all observables will be \emph{countable},
i.e. w.l.o.g. are they defined on a countable $\Omega$
with $\sigma$--algebra $2^\Omega$. This
means that we may view an observable $X$ as a resolution of $\1$ into a countable
sum $\1=\sum_{j\in\Omega} X_j$ of positive operators $X_j$.\par
If $\alg{A}_1,\alg{A}_2$ are subalgebras of $\alg{A}$, they are compatible if they
commute elementwise (again note, that a subalgebra need not not be compatible with itself:
in fact it is iff it is commutative). In this case the closed subalgebra generated
(in fact: spanned) by the products $A_1A_2$, $A_i\in\alg{A}_i$
is denoted $\alg{A}_1\alg{A}_2$.

\paragraph{Operations}
Now we describe the transformations between quantum systems: a $\C$--linear
map $\varphi:\alg{A}_2\rightarrow\alg{A}_1$ is called a \emph{quantum operation}
if it is completely positive (i.e. positive, so that positive elements have positive
images, and also the $\varphi\otimes\id_n$ are positive, where $\id_n$ is the identity
on the algebra of $n\times n$--matrices), and unit preserving. These
maps are in $1$--$1$ correspondence with their (pre--)adjoints $\varphi_*$
by the trace form, mapping states to states, and being completely positive and
trace preserving.\footnote{In general this is only true if we restrict $\varphi$
  to be a \emph{normal} map, cf. \cite{davies:opensystems}.}
Since here we restrict ourselves to finite dimensional algebras
the adjoint map simply goes from $\alg{A}_1$ to $\alg{A}_2$, but to keep things well
separated (which they actually are in the infinite case) we write the adjoint
as $\varphi_*:\alg{A}_{1*}\rightarrow\alg{A}_{2*}$, the dual map (in fact
we consider this as the primary object and the operator maps as their adjoint, which
is the reason for writing subscript $*$).
Notice that $\varphi_*$ is sometimes considered as restricted to
$\varphi_*:\alg{S}(\alg{A}_1)\rightarrow\alg{S}(\alg{A}_2)$. A
characterization of quantum operations is by the {\sc Stinespring}
dilation theorem (\cite{stinespring:thm}):
\begin{satz}[Dilation]
  \label{satz:dilation}
  Let $\varphi:\alg{A}\rightarrow\alg{L}({\cal H})$ a linear map of C${}^*$--algebras.
  Then $\varphi$ is completely positive if and only if
  there exist a representation $\alpha:\alg{A}\rightarrow\alg{L}({\cal K})$,
  with Hilbert space ${\cal K}$, and a bounded linear map
  $V:{\cal H}\rightarrow{\cal K}$ such that
  $$\forall A\in\alg{A}\qquad \varphi(A)=V^*\alpha(A)V.$$
\end{satz}
For proof see e.g. \cite{davies:opensystems}. A commonly used corollary of this is
\begin{cor}(cf.~\cite{kraus:states:etc})
  \label{cor:kraus:form}
  Let $\varphi:\alg{L}({\cal H}_2)\rightarrow\alg{L}({\cal H}_1)$
  a linear map of C${}^*$--algebras.
  Then $\varphi$ is completely positive and unit preserving if and only if
  there exist linear maps $B_i:{\cal H}_1\rightarrow{\cal H}_2$
  with $\sum_i B_i^*B_i=\1_{{\cal H}_1}$ and
  $$\forall A\in\alg{L}({\cal H}_2)\qquad \varphi(A)=\sum_i B_i^*AB_i\ .$$
\end{cor}

\paragraph{Norms and norm inequalities}
For the C${}^*$--algebra $\alg{L}({\cal H})$ of linear operators on the
complex Hilbert space ${\cal H}$ (of dimension $d$) the norm is the
supremum norm $\|\cdot\|_\infty$, i.e. $\|A\|_\infty$ is the largest
absolute value of an eigenvalue of $A$.
The other important norm we use is the \emph{trace norm} $\|\cdot\|_1$:
$\|\alpha\|_1$ is the sum of the absolute values of all eigenvalues of
$\alpha$. Note the important formula
$$\|\alpha\|_1=\sup\{|\tr(\alpha B)|:\|B\|_\infty\leq 1\},$$
which explains the name ``trace norm''. Its proof is by the polar
decomposition of $\alpha$, see e.g. \cite{arveson:invitation}.
Also it implies immediately that $\|\cdot\|_1$ is nonincreasing
under quantum operations. Obviously
$$\|\alpha\|_\infty\leq \|\alpha\|_1\leq d\|\alpha\|_\infty\ .$$
If $\alpha$ is self--adjoint we have the unique decomposition (via
diagonalization) $\alpha=\alpha_+-\alpha_-$ into the positive and negative
part of $\alpha$, where $\alpha_+,\alpha_-\geq 0$ and $\alpha_+\alpha_-=0$.
Then note
$$\|\alpha\|_1=\tr\alpha_+ +\tr\alpha_-
              =\sup\{\tr(\alpha B):-\1\leq B\leq\1\}$$
and
$$\tr\alpha_+=\sup\{\tr(\alpha B):0\leq B\leq\1\}.$$
It should be clear that all the above suprema are in fact maxima.
\par
Finally note that these observations still hold for any direct sum of
$\alg{L}({\cal H}_i)$, $d$ being replaced by the sum of the $\dim{\cal H}_i$.

\section{Entropy and divergence}
\label{sec:quprob:entropy}
The \emph{{\sc von Neumann} entropy} of a state $\rho$
(introduced by \cite{von:neumann:entropy}\footnote{It was
  in fact introduced independently in the same year by {\sc Landau}
  and {\sc Weyl}.})
is defined as $H(\rho)=-\tr(\rho\log\rho)$,
which reduces to the usual {\sc Shannon} entropy for a
commutative algebra because then a state is nothing but a probability
distribution.
For states $\rho,\sigma$ also introduce the \emph{I--divergence},
or simply \emph{divergence} (first defined by \cite{umegaki:divergence})
as $D(\rho\|\sigma)=\tr(\rho(\log\rho-\log\sigma))$ with the convention
that this is $\infty$ if $\supp\rho\not\leq\supp\sigma$
($\supp\rho$ being the \emph{support} of $\rho$, the minimal selfadjoint
idempotent $p$ with $p\rho p=\rho$).
For properties of these quantities we will often refer to \cite{ohya:petz},
and to \cite{wehrl:entropy}.
Three important facts we will use are
\begin{satz}[{\sc Klein} inequality]
  \label{satz:klein}
  For positive operators $\rho,\sigma$ (not necessarily states)
  $$D(\rho\|\sigma)\geq \frac{1}{2}\tr(\rho-\sigma)^2+\tr(\rho-\sigma).$$
  In particular for states the divergence is nonnegative, and zero if and
  only if they are equal.
\end{satz}
\begin{beweis}
  See \cite{ohya:petz}.
\end{beweis}
\begin{lemma}[Continuity]
  \label{lemma:H:continuous}
  Let $\rho,\sigma$ states with $\|\rho-\sigma\|_1\leq\theta\leq\dfrac{1}{2}$.
  Then
  $$|H(\rho)-H(\sigma)|\leq -\theta\log\frac{\theta}{d}
                          =d\eta\left(\frac{\theta}{d}\right).$$
\end{lemma}
\begin{beweis}
  See \cite{ohya:petz}.
\end{beweis}
\begin{satz}[Monotonicity]
  \label{satz:monoton}
  Let $\rho,\sigma$ be states on a C${}^*$--algebra $\alg{A}$, and $\varphi_*$
  a trace preserving, completely positive linear map from states on
  $\alg{A}$ to states on $\alg{B}$. Then
  $$D(\varphi_*\rho\|\varphi_*\sigma)\leq D(\rho\|\sigma).$$
\end{satz}
\begin{beweis}
  See \cite{uhlmann:monoton}; the situation we are in was already solved
  by \cite{lindblad:monoton}. For a textbook account see \cite{ohya:petz}.
\end{beweis}

\section{Observable language}
\label{sec:quprob:observable}
This and the following two sections will introduce language (or formalism) to
talk about entropy and information in the context of quantum systems in a
transparent fashion.
\par
Fix a state on a C${}^*$--algebra, say $\rho$ on $\alg{A}$ and let $X,Y,Z$ compatible
observables on $\alg{A}$. These are then random variables with
a joint distribution, and one defines entropy $H(X)$, conditional entropy
$H(X|Y)$, mutual information $I(X\wedge Y)$, and conditional
mutual information $I(X\wedge Y|Z)$ for these observables as the respective
quantities for them interpreted as random variables.
Note however that these depend on the underlying state $\rho$. In case of
need we will thus add the state as an index, like $H_\rho(X)=H(X)$, etc.\par
As things are there is not much to say about that part of the theory.
We only note some useful formulas:
$$H(X|Y)=\sum_j {\tr(\rho Y_j)H_{\rho_j}(X)},\quad\text{with }
                         \rho_j=\frac{1}{\tr(\rho Y_j)}\sqrt{Y_j}\rho\sqrt{Y_j}$$
(which is an easy calculation using the compatibility of $X$ and $Y$), and
\begin{equation*}\begin{split}
  {I}(X\wedge Y) &= H(X)+H(Y)-H({XY})\\
                 &= D(P^{XY}\|P^X\otimes P^Y)=D(P_{X\cdot Y}\|P_X\otimes P_Y)
\end{split}\end{equation*}
(whose analogue is known from classical information theory).

\section{Subalgebra language}
\label{sec:quprob:subalgebra}
Let $\alg{X},\alg{X}_1,\alg{X}_2,\alg{Y}$ compatible
$*$--subalgebras of the C${}^*$--algebra
$\alg{A}$, and $\rho$ a fixed state on $\alg{A}$.\par
First consider the inclusion map $\imath:\alg{X}\hookrightarrow\alg{A}$
(which is certainly completely positive) and its adjoint
$\imath_*:\alg{A}_*\rightarrow\alg{X}_*$. Define
$$H(\alg{X})=H(\imath_*\rho)$$
(where at the right hand appears the {\sc von Neumann} entropy).
For example for $\alg{X}=\alg{A}$ we obtain just the {\sc von Neumann}
entropy of $\rho$. For the trivial subalgebra $\C=\C\1$
(which obviously commutes with every subalgebra) we obtain, as expected,
$H(\C)=0$. The general philosophy behind this
definition is that $H(\alg{X})$ is the {\sc von Neumann} entropy of the global state
\emph{viewed through (or restricted to) the subsystem} $\alg{X}$. To reflect this in the
notation we define $\rho|_{\alg{X}}=\imath_*\rho$.
\par
Now conditional entropy, mutual information, and conditional mutual information
are defined by reducing them to entropy quantities:
$$H(\alg{X}|\alg{Y})=H(\alg{XY})-H(\alg{Y})$$
\begin{equation*}\begin{split}
  I(\alg{X}_1\wedge\alg{X}_2) &=H(\alg{X}_1)+H(\alg{X}_2)-H(\alg{X}_1\alg{X}_2) \\
                              &=H(\alg{X}_2)-H(\alg{X}_2|\alg{X}_1)
\end{split}\end{equation*}
\begin{equation*}\begin{split}
  I(\alg{X}_1\wedge\alg{X}_2|\alg{Y}) &=H(\alg{X}_1|\alg{Y})+H(\alg{X}_2|\alg{Y})
                                            -H(\alg{X}_1\alg{X}_2|\alg{Y}) \\
                                      &=H(\alg{X}_1\alg{Y})+H(\alg{X}_2\alg{Y})
                                            -H(\alg{X}_1\alg{X}_2\alg{Y})-H(\alg{Y}).
\end{split}\end{equation*}
It is not at all clear a priori that these definitions are all well behaved: while
it is obvious from the definition that the entropy is always nonnegative, this
is not true for the conditional entropy (as was observed by several authors before):
if $\alg{A}=\alg{X}\otimes\alg{Y}$ and $\rho$ is a pure entangled state then
$H(\alg{X}|\alg{Y})=-H(\alg{Y})<0$. This might raise pessimism whether the other two quantities
also are (at least sometimes) pathological. This they are not
(at least not in this way), as will be shown in a moment:
\par
We have the following commutative
diagram of inclusions, and the natural multiplication map $\mu$ (which is in fact a $*$--algebra
homomorphism, and thus completely positive!):
\begin{equation*}\begin{CD}
    \alg{X}_1                 @=           \alg{X}_1       @=              \alg{X}_1\\
    @VV{\varphi_1}V                      @VV{\imath_1}V                 @VV{\jmath_1}V\\
    \alg{X}_1\otimes\alg{X}_2 @>{\mu}>> \alg{X}_1\alg{X}_2 @>{\jmath}>>     \alg{A}\\
    @AA{\varphi_2}A                      @AA{\imath_2}A                 @AA{\jmath_2}A\\
    \alg{X}_2                 @=           \alg{X}_2       @=              \alg{X}_2
\end{CD}\end{equation*}
And hence the corresponding commutative diagram of adjoint maps
(note that $\varphi_{1*}$ and $\varphi_{2*}$ are just
partial traces). With this we find
\begin{equation*}\begin{split}
  {I}(\alg{X}_1\wedge\alg{X}_2) &=H(\alg{X}_1)+H(\alg{X}_2)-H(\alg{X}_1\alg{X}_2)\\
                           &=H(\jmath_{1*}\rho)+H(\jmath_{2*}\rho)-H(\jmath_*\rho)\\
                           &=H(\varphi_{1*}\mu_*\jmath_*\rho)+H(\varphi_{2*}\mu_*\jmath_*\rho)
                                                            -H(\mu_*\jmath_*\rho) \\
     &=D(\mu_*\jmath_*\rho\|\varphi_{1*}\mu_*\jmath_*\rho\otimes\varphi_{2*}\mu_*\jmath_*\rho)
\end{split}\end{equation*}
by definition, then by commutativity of the diagram and the fact that $\mu_*$ preserves
eigenvalues of density operators (because $\mu$ is a surjective $*$--homomorphism, see
lemma~\ref{lemma:onto:hom} below), the last by direct calculation
on the tensor product (just as for the classical formula).
From the last line we see that the mutual information is nonnegative because
the divergence is, by theorem~\ref{satz:klein}
(we could also have seen this already from the definition by applying
subadditivity of {\sc von Neumann} entropy to the second last line, see
theorem~\ref{satz:strong:subadd}).
\begin{lemma}
  \label{lemma:onto:hom}
  Let $\mu:\alg{A}\rightarrow\alg{B}$ a surjective $*$--algebra homomorphism. Then
  \begin{enumerate}
    \item For all pure states $p\in\alg{S}(\alg{A})$: $\mu(p)$ is pure or $0$.
    \item For all $A\in\alg{A}$, $A\geq 0$: $\tr A\geq\tr\mu(A)$.
    \item For pure $p\in\alg{S}(\alg{A})$, $q\in\alg{S}(\alg{B})$:
        $$\mu_*(\mu(p))=p\text{ or }\mu(p)=0,\ \mu(\mu_*(\mu(p)))=\mu(p), \ \mu(\mu_*(q))=q.$$
    \item For $\rho\in\alg{S}(\alg{B})$, $\mu_*(\rho)=\sum_i \alpha_i p_i$ diagonalization with
        the $\alpha_i>0$, then $\rho=\sum_i \alpha_i\mu(p_i)$ is a diagonalization.
    \item Conversely every diagonalization of a state on $\alg{B}$ is by $\mu_*$ translated
        into a diagonalization of its $\mu_*$--image.
  \end{enumerate}
\end{lemma}
\begin{beweis}
  \begin{enumerate}
    \item We have only to show that $\mu(p)$ is minimal
        if it is not $0$: let $q'$ any pure state
        with $q'\leq\mu(p)$. Then
        $$1=\tr(q'\mu(p))=\tr(\mu_*(q')p)\leq\tr(p)=1.$$
        So we must have equality which implies $p\leq\mu_*(q')$, but both operators
        are states, so $p=\mu_*(q')$. Because $\mu_*$ is injective this means that there
        is only one pure state $q'\leq\mu(p)$, i.e. $\mu(p)$ is pure.
    \item We may write $A=\sum_i a_ip_i$ with pure states $p_i$ and $a_i\geq 0$.
        Then $\mu(A)=\sum_i a_i\mu(p_i)$ and since pure
        states have trace $1$ the assertion follows from (1).
    \item Let $A\in\alg{A}$, $A\geq 0$. Then
        \begin{equation*}\begin{split}
          \tr(\mu_*(\mu(p))A) &=\tr(\mu(p)\mu(A))=\tr(\mu(p)\mu(A)\mu(p)) \\
                              &=\tr(\mu(pAp))\leq\tr(pAp)=\tr(pA).
        \end{split}\end{equation*}
        Thus $\mu_*(\mu(p))\leq p$. If $\mu(p)\neq 0$ it is a pure state,
        hence $\mu_*(\mu(p))$ a state which
        forces $\mu_*(\mu(p))=p$. This proves the left formula,
        the middle follows immediately, and
        for the right observe that we may choose a
        pure pre--image $p$ of $q$ (in fact that will
        be $\mu_*(q)$, as one can see from (4)).
    \item $\sum_i \alpha_i\mu(p_i)$ is certainly the
        diagonalization of some positive operator since
        the $\mu(p_i)$ which are not $0$ are by the homomorphism
        property and by (1) pairwise orthogonal
        pure states. Now observe $\mu(\mu_*(\rho))=\sum_i \alpha_i\mu(p_i)$ and
        $$\mu_*(\rho)=\mu_*(\mu(\mu_*(\rho)))=
                \sum_i \alpha_i\mu_*(\mu(p_i))\leq\sum_i \alpha_i p_i=\mu_*(\rho),$$
        hence equality, i.e. all $\mu(p_i)$ are pure. From
        $$\mu_*(\rho)=\sum_i \alpha_i\mu_*(\mu(p_i))=\mu_*(\sum_i \alpha_i\mu(p_i))$$
        and injectivity of $\mu_*$ the assertion follows.
    \item This is a direct consequence of (3) and (4).
  \end{enumerate}
\end{beweis}
\medskip\par
For the conditional mutual information we have to do somewhat more (yet from the definition
we see that its positivity will have something to do with the strong subadditivity
of {\sc von Neumann} entropy, see theorem~\ref{satz:strong:subadd}):\par
Consider the following commuative diagram:
\begin{equation*}\begin{CD}
  \alg{Y} @>{\varphi_1}>> \alg{X}_1\otimes\alg{Y}               @>{\mu_1}>> \alg{X}_1\alg{Y}\\
    @|                       @VV{\varphi_1'}V                                 @VV{\jmath_1}V\\
  \alg{Y} @>{\varphi}>> \alg{X}_1\otimes\alg{X}_2\otimes\alg{Y} @>{\mu}>>   \alg{X}_1\alg{X}_2\alg{Y}
                                                                                   @>{\jmath}>> \alg{A}\\
    @|                       @AA{\varphi_2'}A                                 @AA{\jmath_2}A\\
  \alg{Y} @>{\varphi_2}>> \alg{X}_2\otimes\alg{Y}               @>{\mu_2}>> \alg{X}_2\alg{Y}
\end{CD}\end{equation*}
All maps there are completely positive, $\mu,\mu_1,\mu_2$ being $*$--homomorphisms.
Thus the adjoints of the various $\varphi$'s are partial traces
and with $\sigma=\mu_*\jmath_*\rho$:
$H(\alg{X}_1\alg{X}_2\alg{Y})=H(\sigma)$, $H(\alg{X}_1\alg{Y})=H(\tr_{\alg{X}_2}\sigma)$,
$H(\alg{X}_2\alg{Y})=H(\tr_{\alg{X}_1}\sigma)$,
$H(\alg{Y})=H(\tr_{\alg{X}_1\otimes\alg{X}_2}\sigma)$
(where we have made use of lemma~\ref{lemma:onto:hom} several times),
and we can indeed apply strong subadditivity.\par
Finally let us remark the nice formulas
$$H(\alg{X})=H(\alg{X}|\C),\qquad
            I(\alg{X}_1\wedge\alg{X}_2)=I(\alg{X}_1\wedge\alg{X}_2|\C).$$
\begin{expl_noname}
  \label{expl:tensorproduct}
  A very important special case of the definitions of this and the preceding section occurs
  for tensor products of Hilbert spaces
  $\alg{L}({\cal H}_1\otimes{\cal H}_2)=\alg{L}({\cal H}_1)\otimes\alg{L}({\cal H}_2)$,
  or more generally tensor products
  of C${}^*$--algebras: $\alg{A}=\alg{A}_1\otimes\alg{A}_2$.
  $\alg{A}_1,\alg{A}_2$ are $*$--subalgebras of $\alg{A}$ in the natural way,
  and are obviously compatible. The same then holds for observables $A_i\subset\alg{A}_i$,
  and similarly for more than two factors. In this case the restriction $\rho|_{\alg{A}_i}$
  is just a partial trace.
\end{expl_noname}
\begin{bem}
  It should be clear that we introduced (having $H$) conditional entropy and
  mutual information by \emph{formal analogy} to the classical quantities.
  We cannot claim to have an operational meaning of them in general ---
  the theorems in the main text must be seen as exceptions to this rule.
  \par
  We are in this respect in accordance with \cite{levitin:isit98} who went
  even further by rejecting the very name ``conditional entropy'' for
  $H(\cdot|\cdot)$, and proposed to return to the name ``correlation entropy''
  given by {\sc Stratonovich} to the quantity $I(\cdot\wedge\cdot)$, on the grounds
  of a strictly operational reasoning (which is only open to the one criticism
  that {\sc Levitin} always sticks with \emph{classical} information,
  never acknowledging the unprecedented properties of \emph{quantum} information).
\end{bem}

\section{Common tongue}
\label{sec:quprob:common}
The languages of the two preceding sections may be phrased in a unified formalism
(the ``common tongue'') using completely positive C${}^*$--algebra maps (in particular
those from or to commutative algebras, inclusion maps, and $*$--algebra homomorphisms,
cf. \cite{stinespring:thm}).\par
That this is promising one can see from the observation that observables can be interpreted
in a natural way as C${}^*$--algebra maps: $X:\Omega\rightarrow\alg{A}$
corresponds by linear extension to $X:\alg{B}(\Omega)\rightarrow\alg{A}$, where
$\alg{B}(\Omega)=\alg{B}(\Omega,{\cal F})$ is the algebra of bounded measurable
functions on $\Omega$. We follow the convention that in this algebra $j\in\Omega$ shall
denote the function that is $1$ on $j$ and $0$ elsewhere, so $X(j)=X_j$, and
obviously $X_*(\rho)$ equals the distribution $P^X$ on $\Omega$ induced by $X$
with $\rho$.\par
Let us also introduce some notation for the observable $X$: the \emph{total} observable
operation $X_{\text{tot}}:\alg{B}(\Omega)\otimes\alg{A}\rightarrow\alg{A}$ mapping
$j\otimes A\mapsto \sqrt{Y_j}A\sqrt{Y_j}$,
its \emph{interior} part
$X_{\text{int}}=X_{\text{tot}}\circ\imath_{\alg{A}}:\alg{A}\rightarrow\alg{A}$ with
$A\mapsto\sum_j \sqrt{Y_j}A\sqrt{Y_j}$, and its \emph{exterior} part
$X_{\text{ext}}=X_{\text{tot}}\circ\imath_{\alg{B}(\Omega)}$
which coincides with $X$.\par
Consider compatible quantum operations $\varphi:\alg{X}\rightarrow\alg{A}$,
$\psi:\alg{Y}\rightarrow\alg{A}$, etc.
($\varphi,\psi$ are compatible if their images commute elementwise). In this case
their \emph{product} is the operation
$\varphi\psi:\alg{X}\otimes\alg{Y}\rightarrow\alg{A}$
mapping $X\otimes Y\mapsto\varphi(X)\psi(Y)$:
\begin{equation*}\begin{CD}
    \alg{X}                 @>{\varphi}>>              \alg{A}\\
    @V{\varphi_1}VV                                       @|\\
    \alg{X}\otimes\alg{Y}   @>{\exists^!\varphi\psi}>> \alg{A}\\
    @A{\varphi_2}AA                                       @|\\
    \alg{Y}                 @>{\psi}>>                 \alg{A}
\end{CD}\end{equation*}
Note that this generalizes the product of
observables, as well as the product map $\mu$ of subalgebras.
\par
Now simply define $H(\varphi)=H(\varphi_*\rho)$, and
again the conditional entropy and the informations are defined by reduction to
entropy, e.g. $H(\varphi|\psi)=H(\varphi\psi)-H(\psi)$, or
$I(\varphi\wedge\psi)=H(\varphi)+H(\psi)-H(\varphi\psi)$.
\par
For the mutual information observe that (see previous diagram)
\begin{equation*}\begin{split}
  I(\varphi\wedge\psi) &= D((\varphi\psi)_*\rho\|\varphi_*\rho\otimes\psi_*\rho)\\
                       &= D(\sigma\|\tr_{\alg{Y}}\sigma\otimes\tr_{\alg{X}}\sigma),
                             \qquad\text{with }\sigma=(\varphi\psi)_*\rho.
\end{split}\end{equation*}
Note the difference to \cite{ohya:petz}: with them the entropy of an operation
is related to the mutual information of the operation as a channel.
With us the entropy of an operation is the entropy of a state ``viewed through''
this operation (as was the idea with the entropy of a subsystem, and obviously also with
the entropy of an observable).
\par\medskip
With these insights we may now form hybrid expressions involving
observables and $*$--subalgebras at the same time: let
$\imath:\alg{X}\hookrightarrow\alg{A}$,
$\jmath:\alg{Y}\hookrightarrow\alg{A}$ $*$--subalgebra
inclusions, and $X,Y$ observables on $\alg{A}$, all four compatible. Then we have
$$H(\alg{X}|Y)=H(\imath Y)-H(Y)$$
$${I}(\alg{X}\wedge Y)=H(\imath)+H(Y)-H(\imath Y),$$
and lots of others. From the previous section we know that
the information quantities are nonnegative,
but also the entropy conditional on an observable, from the formula
$$H(\alg{X}|Y)=\sum_j \tr(\rho Y_j)H_{\rho_j}(\alg{X}),\quad\text{with }
                         \rho_j=\frac{1}{\tr(\rho Y_j)}\sqrt{Y_j}\rho\sqrt{Y_j}\ .$$
But again there are some expressions which seem suspicious, like
$$H(X|\alg{Y})=H(X\jmath)-H(\alg{Y}).$$
However, due to the inequality of theorem~\ref{satz:conditional:entropy}
in fact it behaves nicely.

\section{Inequalities}
\label{sec:quprob:inequalities}
\paragraph{Entropy} Let us first note the basic
\begin{satz}
  \label{satz:strong:subadd}
  For compatible $*$--subalgebras $\alg{A}_1,\alg{A}_2,\alg{A}_3$ one has:
  \begin{enumerate}
    \item Subadditivity: $H(\alg{A}_1\alg{A}_2)\leq H(\alg{A}_1)+H(\alg{A}_2)$.
    \item Strong subadditivity:
      $H(\alg{A}_1\alg{A}_2\alg{A}_3)+H(\alg{A}_2)
                             \leq H(\alg{A}_1\alg{A}_2)+H(\alg{A}_2\alg{A}_3)$.\par
      \qquad (In our language this is equivalent to the more natural form\par
      \qquad $\phantom{\left({}\right.}H(\alg{A}_1\alg{A}_3|\alg{A}_2)
                            \leq H(\alg{A}_1|\alg{A}_2)+H(\alg{A}_3|\alg{A}_2)$).
  \end{enumerate}
\end{satz}
\begin{beweis}
  Subadditivity is a special case of strong subadditivity: $\alg{A}_2=\C$.
  The latter can be reduced to the familiar form, proved first by
  {\sc Lieb \& Ruskai} (see the references in \cite{uhlmann:monoton}),
  by the same type of argument as we used in the section
  \emph{Subalgebra language} for the nonnegativity of conditional
  mutual information.
\end{beweis}
\par
Another kind of inequality may serve as an operational justification
of the definition of {\sc von Neumann} entropy. Call a quantum operation
$\varphi:\alg{A}_1\rightarrow\alg{A}_2$
\emph{doubly stochastic} if it preserves the trace, i.e. for all $A\in\alg{A}_1$:
$\tr\varphi(A)=\tr A$ (see \cite{ohya:petz}). We will consider the less
restrictive condition $\tr\varphi(A)\leq\tr A$, and
for an observable $X$, a $*$--subalgebra $\alg{X}$
let us say it is \emph{maximal in} $\alg{A}$ if $X$, the inclusion map
has this property, respectively (obviously for the $*$--subalgebra this
implies doubly stochastic).
Main examples are: an observable whose atoms are minimal in the target
algebra, i.e. have only trivial decompositions into positive
operators, and a maximal commutative $*$--subalgebra.
\begin{satz}[Entropy increase]
  \label{satz:entropy:inequ}
  Let $\varphi:\alg{Y}\rightarrow\alg{X}$ with $\tr\varphi(A)\leq\tr A$, and
  $\psi:\alg{X}\rightarrow\alg{A}$ quantum operations. Then
  $H(\psi\circ\varphi)\geq H(\psi)$. (Notice that in the physical sense
  the operation $\varphi_*$ is applied \emph{after} $\psi_*$).
\end{satz}
\par
Before we prove this let us note
two important case of equality: Let $\rho=\sum_i\lambda_i p_i$ with mutually orthogonal
pure states $p_i$, $\lambda_i\geq 0$, $\sum_i p_i=\1$.
Then equality holds for the $*$--subalgebra generated
by the $p_i$ (in fact for any $*$--subalgebra which contains them),
and for the observable that corresponds to the $p_i$'s
resolution of $\1$.\par
\begin{beweis}[of theorem~\ref{satz:entropy:inequ}]
  Let $\sigma=\psi_*\rho$, we have to prove $H(\varphi_*\sigma)\geq H(\sigma)$.
  From the previous discussion we see that we may assume $\alg{Y}$ to be
  commutative, without changing the trace relation. Let $\sigma=\sum_i \alpha_i p_i$
  a diagonalization with pure states $p_i$ on $\alg{X}$, and
  $q_j$ the family of minimal idempotents of $\alg{Y}$ (which by commutativity
  are othogonal). Then we have decompositions
  $\varphi_* p_i=\sum_j \beta_{ij} q_j$, hence
  $$\varphi_*\sigma=\sum_i \alpha_i\varphi_* p_i
                    =\sum_j\left({\sum_i \alpha_i\beta_{ij}}\right)q_j\ .$$
  Now observe that for all $j$
  $$\sum_i \beta_{ij}=\tr\left(q_j\sum_i \varphi_* p_i\right)
                     =\tr\left((\varphi q_j)\sum_i p_i\right)
                     =\tr(\varphi q_j)\leq\tr(q_j)=1,$$
  and the result follows from the formulas
  $H(\sigma)=H(\alpha_i|i)$, $H(\varphi_*\sigma)=H(\sum_i\beta_{ij}\alpha_i|j)$.
\end{beweis}
Let us formulate the special cases of maximal observables and maximal $*$--subalgebras
as a corollary:
\begin{cor}
  Let $X$ an observable maximal in $\alg{X}$, then $H(X)\geq H(\alg{X})$.
  Let $\alg{X}'$ a $*$--subalgebra maximal in $\alg{X}$,
  then $H(\alg{X}')\geq H(\alg{X})$. \qed
\end{cor}
An application of this is in the proof of
\begin{satz}
  \label{satz:pure:common:state}
  Let $\alg{X},\alg{Y}$ compatible, $\rho|_{\alg{XY}}$ pure. Then $H(\alg{X})=H(\alg{Y})$.
\end{satz}
\begin{beweis}
  By retracting the state $\rho$ to $\alg{X}\otimes\alg{Y}$ by the multiplication
  map $\mu:\alg{X}\otimes\alg{Y}\rightarrow\alg{XY}$
  (see lemma~\ref{lemma:onto:hom}) and embedding $\alg{X}$ and $\alg{Y}$
  into full matrix algebras (see the proof of the next theorem)
  we may assume that we have a pure
  state $\rho$ on $\alg{L}({\cal H}_1)\otimes\alg{L}({\cal H}_2)$
  (entropies do not change as the $*$--subalgebras are maximal).
  Then the assertion of the theorem is
  $H(\tr_{\alg{X}}\rho)=H(\tr_{\alg{Y}}\rho)$ which is well known
  (proof via the {\sc Schmidt} decomposition of $\ket{\psi}$,
  where $\rho=\ket{\psi}\bra{\psi}$: cf. \cite{peres:quantum:theory}).
\end{beweis}
\begin{satz}
  \label{satz:triangle}
  Let $\alg{X},\alg{Y}$ compatible, $\rho$ any state. Then
  $|H(\alg{X})-H(\alg{Y})|\leq H(\alg{XY})$.
\end{satz}
\begin{beweis}
  Like in the previous theorem we may assume that $\rho$ is a state
  on $\alg{X}\otimes\alg{Y}$, and by symmetry we have to prove that
  $$H(\alg{X})-H(\alg{Y})\leq H(\alg{XY}).$$
  If we think of $\alg{X}$ and $\alg{Y}$ as sums of full operator algebras,
  say $\alg{X}=\bigoplus_i \alg{L}({\cal H}_i)$, $\alg{Y}=\bigoplus_j \alg{L}({\cal K}_j)$,
  then embedding them into $\alg{L}(\bigoplus_i{\cal H}_i)$, $\alg{L}(\bigoplus_j{\cal K}_j)$,
  respectively, does not change the entropies involved
  (because the $*$--subalgebras are maximal).
  Thus we may assume that $\alg{X}=\alg{L}({\cal H})$, $\alg{Y}=\alg{L}({\cal K})$.
  Now consider a \emph{purification} $\ket{\psi}$ of $\rho$ on
  the Hilbert space ${\cal H}\otimes{\cal K}\otimes{\cal L}$
  (see e.g. \cite{schumacher:F:e}): this means
  $\rho=\tr_{\alg{L}({\cal L})}\ket{\psi}\bra{\psi}$.
  Now by theorem~\ref{satz:pure:common:state}
  $H(\alg{X})=H(\alg{YZ})$, $H(\alg{XY})=H(\alg{Z})$, and the assertion
  follows from the subadditivity theorem~\ref{satz:strong:subadd}:
  $H(\alg{YZ})\leq H(\alg{Y})+H(\alg{Z})$.
\end{beweis}

\paragraph{Information}
The following inequality for mutual information is a straightforward
generalization of the {\sc Holevo} bound (\cite{holevo:bound}, see
theorem~\ref{satz:holevo:bound} below):
\begin{satz}
  \label{satz:bound}
  Let $X,Y$ be compatible observables with values in the compatible
   $*$--subalgebras $\alg{X},\alg{Y}$, respectively. Then
  $${I}(X\wedge Y)\leq{I}(\alg{X}\wedge Y)\leq{I}(\alg{X}\wedge\alg{Y}).$$
\end{satz}
\begin{beweis}
  Consider the diagram
  \begin{equation*}\begin{CD}
    \alg{B}(\Omega_X)      @>{X}>>                  \alg{X}                 @=          \alg{X}\\
     @V{}VV                                          @V{}VV                          @V{\varphi}VV\\
    \alg{B}(\Omega_X)\otimes\alg{B}(\Omega_Y)
                                @>{X\otimes\id}>> \alg{X}\otimes\alg{B}(\Omega_Y)
                                                                        @>{\id\otimes Y}>>
                                                                                       \alg{X}\otimes\alg{Y}
                                                                                                   @>{\mu}>>\alg{A}\\
    @A{}AA                                          @A{}AA                          @A{\varphi'}AA\\
   \alg{B}(\Omega_Y)      @=                   \alg{B}(\Omega_Y)         @>{Y}>>        \alg{Y}
  \end{CD}\end{equation*}
  and apply the Lindblad--Uhlmann monotonicity
  theorem~\ref{satz:monoton} twice, with
  $\mu_*(\rho)$ and the maps $(\id\otimes Y)_*$
  and $(X\otimes\id)_*$, one after the other.
\end{beweis}
This can be greatly extended: for example if $\alg{X}\subset\alg{X}'$,
$\alg{Y}\subset\alg{Y}'$, then
$${I}(\alg{X}\wedge\alg{Y})\leq{I}(\alg{X}'\wedge\alg{Y}').$$
The most general form is
$${I}(\psi_1\circ\varphi_1\wedge\psi_2\circ\varphi_2)\leq{I}(\psi_1\wedge\psi_2)$$
in the diagram
\begin{equation*}\begin{CD}
     \alg{A}_1' @>{\varphi_1}>>                      \alg{A}_1            @>{\psi_1}>> \alg{A} \\
        @V{}VV                                         @V{}VV                             @| \\
 \alg{A}_1'\otimes\alg{A}_2' 
                 @>{\varphi_1\otimes\varphi_2}>> \alg{A}_1\otimes\alg{A}_2 @>{\psi=\psi_1\psi_2}>> \alg{A} \\
        @A{}AA                                         @A{}AA                             @| \\
      \alg{A}_2' @>{\varphi_2}>>                      \alg{A}_2            @>{\psi_2}>> \alg{A}
\end{CD}\end{equation*}
\begin{bem}
  \label{bem:holevo:bound}
  It is worth noting that the above formulation of the information bound has the nice
  form of a data processing inequality. To dwell on this point a little more,
  and at the same time link our discussion with the traditional view and the language
  employed in the chapters~\ref{chap:source} and \ref{chap:channel} of the
  main text let us define for a (measureable) map
  $\varphi_*:\fset{X}\rightarrow\alg{S}(\alg{Y})$ (which we identify
  with its linear extension to $\C\fset{X}$ and regard
  as a \emph{quantum channel}, see chapter~\ref{chap:channel})
  and a p.d. $P$ on $\fset{X}$
  $$I(P;\varphi_*)=I_\gamma(\C\fset{X}\wedge\alg{Y})$$
  with the \emph{channel state}
  $\gamma=\sum_{x\in\fset{X}} P(x)[x]\otimes\varphi_*(x)$.
  It is easily verified that
  \begin{equation*}
    I(P;\varphi_*)=H(P\varphi_*)-H(\varphi_*|P)\ \text{ where }
    \begin{cases}
      P\varphi_*     =\tr_{\C\fset{X}}\gamma=\sum_{x\in\fset{X}} P(x)\varphi_*(x), \\
      H(\varphi_*|P) =\sum_{x\in\fset{X}} P(x)H(\varphi_*(x)).
    \end{cases}
  \end{equation*}
  Now with a quantum operation $\psi_*:\alg{Y}_*\rightarrow\alg{Z}_*$
  the data processing inequality specializes to
  $$I(P;\psi_*\circ\varphi_*)\leq I(P;\varphi_*).$$
\end{bem}
In particular if $\alg{Z}$ is commutative, i.e. the operation, now
denoted $D_*$, is a measurement, we recover the
\begin{satz}[{\sc Holevo} bound]
  \label{satz:holevo:bound}
  $I(P;D_*\circ\varphi_*)\leq I(P;\varphi_*)$. \qed
\end{satz}
In chapter~\ref{chap:channel} an elementary proof of this inequality is presented.
\bigskip\par
\begin{satz}
  \label{satz:info:subadd}
  Let $\alg{X}_1,\alg{X}_2,\alg{Y}_1,\alg{Y}_2$ compatible  $*$--subalgebras of $\alg{A}$,
  $\rho$ a state on $\alg{A}$. Then
  $$I(\alg{X}_1\alg{X}_2\wedge\alg{Y}_1\alg{Y}_2)\leq
                        I(\alg{X}_1\wedge\alg{Y}_1)+I(\alg{X}_2\wedge\alg{Y}_2)$$
  if $I(\alg{Y}_1\wedge\alg{X}_2\alg{Y}_2|\alg{X}_1)=0$ and
  $I(\alg{Y}_2\wedge\alg{X}_1\alg{Y}_1|\alg{X}_2)=0$ (i.e. $\alg{Y}_k$ is
  \emph{independent} from the other $*$--subalgebras conditional on $\alg{X}_k$).
\end{satz}
\begin{beweis}
  First observe that the conditional independence mentioned,
  $I(\alg{Y}_1\wedge\alg{X}_2\alg{Y}_2|\alg{X}_1)=0$, is equivalent
  to $H(\alg{Y}_1|\alg{X}_1\alg{X}_2\alg{Y}_2)=H(\alg{Y}_1|\alg{X}_1)$.
  By theorem~\ref{satz:more:knowledge} we then have also
  $H(\alg{Y}_1|\alg{X}_1\alg{X}_2)=H(\alg{Y}_1|\alg{X}_1)$.
  Now observe (with the obvious chain rule)
  \begin{equation*}\begin{split}
    H(\alg{Y}_1\alg{Y}_2|\alg{X}_1\alg{X}_2)
                  &= H(\alg{Y}_1|\alg{X}_1\alg{X}_2\alg{Y}_2)+H(\alg{Y}_2|\alg{X}_1\alg{X}_2)\\
                  &= H(\alg{Y}_1|\alg{X}_1)+H(\alg{Y}_2|\alg{X}_2)
  \end{split}\end{equation*}
  and hence
  \begin{equation*}\begin{split}
    I(\alg{X}_1\alg{X}_2\wedge\alg{Y}_1\alg{Y}_2)
               &=    H(\alg{Y}_1\alg{Y}_2)-H(\alg{Y}_1\alg{Y}_2|\alg{X}_1\alg{X}_2) \\
               &\leq H(\alg{Y}_1)+H(\alg{Y}_2)-H(\alg{Y}_1|\alg{X}_1)-H(\alg{Y}_2|\alg{X}_2) \\
               &=I(\alg{X}_1\wedge\alg{Y}_1)+I(\alg{X}_2\wedge\alg{Y}_2)
  \end{split}\end{equation*}
  where we have used the subadditivity of {\sc von Neumann} entropy,
  theorem~\ref{satz:strong:subadd}.
\end{beweis}
The same obviously applies if we have $n$ $*$--subalgebras $\alg{X}_k$, and
$n$ $\alg{Y}_k$, all compatible, and if $\alg{Y}_k$ is independent from the others
given $\alg{X}_k$, i.e. for all $k$
$$H(\alg{Y}_k|\alg{X}_1\cdots\alg{X}_n\alg{Y}_1\cdots\widehat{\alg{Y}_k}\cdots\alg{Y}_n)
                                                                =H(\alg{Y}_k|\alg{X}_k).$$
\begin{cor}
  \label{cor:info:subadd}
  Let $\alg{X}_1,\ldots,\alg{X}_n$, $\alg{Y}_1,\ldots,\alg{Y}_n$ C${}^*$--algebras,
  $\alg{X}_i=\C\fset{X}_i$ commutative, and
  $\alg{A}=\alg{X}_1\otimes\cdots\otimes\alg{X}_n\otimes
                    \alg{Y}_1\otimes\cdots\otimes\alg{Y}_n$.
  Then with the state
  $$\gamma=\sum_{x_i\in\fset{X}_i}
                    P(x_1,\ldots,x_n)[x_1]\otimes\cdots\otimes[x_n]\otimes
                                           W_{x_1}\otimes\cdots\otimes W_{x_n}$$
  on $\alg{A}$ (where $P$ is a p.d. on $\fset{X}_1\times\cdots\times\fset{X}_n$ and
  $W$ maps the $\fset{X}_i$ to states on $\alg{Y}_i$):
  $$I(\alg{X}_1\cdots\alg{X}_n\wedge\alg{Y}_1\cdots\alg{Y}_n)
                                    \leq\sum_{k=1}^n I(\alg{X}_k\wedge\alg{Y}_k).$$
\end{cor}
\begin{beweis}
  We only have to check the conditional independence, which is left to the reader.
\end{beweis}
We note another estimate for the mutual information:
\begin{satz}
  \label{satz:info:upperbound}
  For compatible $*$--subalgebras $\alg{X},\alg{Y}$:
  $I(\alg{X}\wedge\alg{Y})\leq 2\min\{H(\alg{X}),H(\alg{Y})\}$.
\end{satz}
\begin{beweis}
  Put together the formula
  $I(\alg{X}\wedge\alg{Y})=H(\alg{X})-H(\alg{X}|\alg{Y})$ and the simple
  estimate $H(\alg{X}|\alg{Y})\geq -H(\alg{X})$ from theorem~\ref{satz:triangle}.
\end{beweis}

\paragraph{Conditional entropy} We start with a simple positivity condition:
\begin{satz}
  \label{satz:conditional:entropy}
  Let $\varphi:\alg{X}\rightarrow\alg{A}$, $\psi:\alg{Y}\rightarrow\alg{A}$ compatible
  quantum operations
  with $\alg{X}$ or $\alg{Y}$ commutative. Then $H(\varphi|\psi)\geq 0$.
\end{satz}
\begin{beweis}
  Let $\sigma=(\varphi\psi)_*\rho$, then by definition and
  lemma~\ref{lemma:onto:hom}
  $$H(\varphi|\psi)=H(\sigma)-H(\tr_\alg{Y}\sigma).$$
  \par
  \emph{First case}: $\alg{X}$ is commutative, so we can write
  $\sigma=\sum_x Q(x)[x]\otimes\tau_*(x)$ with a distribution $Q$ on $\fset{X}$, and
  states $\tau_*(x)$ on $\alg{Y}$. Obviously
  $H(\sigma)=H(Q)+\sum_x Q(x)H(\tau_*(x))$, and
  $\tr_\alg{Y}\sigma=\sum_x Q(x)[x]=Q$, and hence
  $H(\varphi|\psi)=\sum_x Q(x)H(\tau_*(x))\geq 0$.
  \par
  \emph{Second case}: $\alg{Y}$ is commutative, so we can write
  $\sigma=\sum_x Q(x)[x]\tau_*(x)\otimes[x]$, like in the first case.
  $H(\sigma)$ is calculated as before, but now
  $\tr_\alg{Y}\sigma=\sum_x Q(x)\tau_*(x)=Q\tau_*$, and
  \begin{equation*}\begin{split}
    H(\varphi|\psi) &= H(Q)-\left(H(Q\tau_*)-\sum_x Q(x)H(\tau_*(x))\right) \\
                    &= H(Q)-I(Q;\tau_*)\geq 0,
  \end{split}\end{equation*}
  the last step by an application of the {\sc Holevo} bound,
  theorem~\ref{satz:holevo:bound}.
\end{beweis}
\begin{bem}
  From the proof we see that the commutativity of $\alg{X}$ or
  $\alg{Y}$ enters in the representation
  of $\sigma$ as a particular separable state with respect to the $*$--subalgebras $\alg{X}$,
  $\alg{Y}$ (see definition below),
  namely with one party admitting common diagonalization of her states.
  We formulate as a conjecture the more general:
  $$H(\alg{X}|\alg{Y})\geq 0\text{ if }\rho
            \text{ is separable with respect to }\alg{X}\text{ and }\alg{Y}.$$
  From this it would follow that in this case
  $I(\alg{X}\wedge\alg{Y})\leq\min\{H(\alg{X}),H(\alg{Y})\}$
  (compare theorem~\ref{satz:info:upperbound}), which we now only get from the
  commutativity assumption.
\end{bem}
\begin{defi}
  \label{defi:separable}
  Call $\rho$ \emph{separable with respect to} compatible $*$--subalgebras
  $\alg{X}_1,\ldots,\alg{X}_m$ of $\alg{A}$, if, for the natural multiplication map
  $\mu:\alg{X}_1\otimes\cdots\otimes\alg{X}_m\rightarrow\alg{A}$, $\mu_*\rho$
  is a separable state on $\alg{X}_1\otimes\cdots\otimes\alg{X}_m$, i.e. a convex
  combination of product states $\sigma_1\otimes\cdots\otimes\sigma_m$,
  $\sigma_i\in\alg{S}(\alg{X}_i)$. If $\mu_*\rho$ is a product state, we call
  also $\rho$ a \emph{product state with respect to} $\alg{X}_1,\ldots,\alg{X}_m$.
\end{defi}
\begin{satz}[Knowledge decreases uncertainty]
  \label{satz:more:knowledge}
  Let $\varphi:\alg{X}\rightarrow\alg{A}$, $\psi:\alg{Y}\rightarrow\alg{A}$ compatible
  quantum operations, and $\varphi':\alg{X}'\rightarrow\alg{X}$ any quantum operation.
  \par
  Then $H(\psi|\varphi)\leq H(\psi|\varphi\circ\varphi')$, and in particular
  $H(\psi|\varphi)\leq H(\psi)$.
\end{satz}
\begin{beweis}
  The inequality is obviously equivalent to
  $I(\psi\wedge\varphi)\geq I(\psi\wedge\varphi\circ\varphi')$,
  i.e. to theorem~\ref{satz:bound}.
\end{beweis}
Defining $h(x)=-x\log x-(1-x)\log(1-x)$ for $x\in[0,1]$ we have the famous
\begin{satz}[{\sc Fano} inequality]
  \label{satz:fano:inequality}
  Let $\rho$ a state on $\alg{A}$, and
  $\alg{Y}$ be a $*$--subalgebra of $\alg{A}$, compatible with the observable $X$ (indexed by
  $\fset{X}$). Then
  for any observable $Y$ with values in $\alg{Y}$ the probability that ``$X\neq Y$'',
  i.e. $P_e=1-\sum_j \tr(\rho X_jY_j)$, satisfies
  $$H(X|\alg{Y})\leq h(P_e)+P_e\log(|\fset{X}|-1).$$
\end{satz}
\begin{beweis}
  By the previous theorem~\ref{satz:more:knowledge} it suffices to prove the
  inequality with $H(X|Y)$ instead of $H(X|\alg{Y})$. But then we have the classical
  {\sc Fano} inequality: the uncertainty on $X$ given $Y$ may be estimated by the
  uncertainty of the event that they are equal plus the uncertainty on the
  value of $X$ if they are not.
\end{beweis}
\begin{cor}
  \label{cor:fano:inequality}
  Let $\alg{X}$ a commutative $*$--subalgebra compatible with $\alg{Y}$, and $X$
  the --- uniquely determined --- maximal observable on $\alg{X}$, $P_e$ as in the
  theorem, then
  $$H(\alg{X}|\alg{Y})\leq h(P_e)+P_e\log(\tr\supp(\rho|_{\alg{X}})-1).$$
\end{cor}
\begin{beweis}
  First observe that $H(\alg{X}|\alg{Y})=H(X|\alg{Y})$. To apply the theorem
  we only have to restrict the range of $X$ to those values that are actually
  assumed.
\end{beweis}
\bigskip\par
Some philosophical remarks may be in order: quantum theory stipulates the channel
as a process, an asymmetric notion, and this brings about the formula
$I(P;\varphi_*)=H(P\varphi_*)-H(\varphi_*|P)$: input, average and conditional
output entropy. In classical information theory however we like to see things
more symmetric, namely the channel as a stochastic two--end system, with
some underlying joint distribution. Following this idea produces our channel
states $\gamma$, and a symmetric ``information'' expression
$I(\alg{X}\wedge\alg{Y})$. Even though there are questions in quantum information
where these two pictures can be brought to relation, for example
in the above results (a connection that was
noticed before by \cite{hall:context:mappings} in his investigation of what
he calls \emph{context mappings}), they are not reducible to each other: the
``dynamic'' picture is asymmetric (there may not even exist a backward channel
producing the same channel state), whereas the ``static'' picture is obviously
symmetric. Even worse, for a joint state it is not obvious that a channel
and input distribution generating it exist at all. And if it exists, there is
no uniqueness in its choice.
On the other hand, modelling a situation of quantum evolutions
statically may produce unphysical effects, see the example from
\cite{winter:quil}, {VIII}.B.2, pp.24: the channel state incorporates
parts of a system which can never be simultaneously accessible.

% literatur

\bibliographystyle{winter}
\bibliography{quant-inf,quant-phys,it}

\begin{thebibliography}{70}
\expandafter\ifx\csname natexlab\endcsname\relax\def\natexlab#1{#1}\fi

\bibitem[{\sc Ahlswede} (1968]{ahlswede:nichtstationaer}
{\sc Ahlswede, R.} (1968), Beitr{\"a}ge zur Shannonschen Informationstheorie im
  Falle nichtstation{\"a}rer Kan{\"a}le. {\em Z. Wahrscheinlichkeitstheorie und
  verw. Gebiete\/}, 10, 1--42.

\bibitem[{\sc Ahlswede} (1971]{ahlswede:MAC}
{\sc Ahlswede, R.} (1971), Multi--way communication channels. In {\em Second
  International Symposium on Information Theory\/},  23--52, Hungarian Academy
  of Sciences.

\bibitem[{\sc Ahlswede} (1974{\natexlab{a}}]{ahlswede:MWC}
{\sc Ahlswede, R.} (1974{\natexlab{a}}), The capacity region of a channel with
  two senders and two receivers. {\em Ann. Prob.\/}, 2(5), 805--814.

\bibitem[{\sc Ahlswede} (1974{\natexlab{b}}]{ahlswede:rate:slicing}
{\sc Ahlswede, R.} (1974{\natexlab{b}}), Paper presented at {\em
  $7^{\text{th}}$ Hawaii International Conference on System Sciences}, Jan.
  1974. Published in {\sc Ahlswede, R. \& K\"orner, J.} (1975), Source coding
  with side information at the decoder and a converse for degraded broadcast
  channels. {\em IEEE Trans. Inf. Theory}, 21, 629--637.

\bibitem[{\sc Ahlswede} (1979]{ahlswede:cov:col:1}
{\sc Ahlswede, R.} (1979), Coloring Hypergraphs: A New Approach to Multi--user
  Source Coding --- I. {\em J. Combinatorics, Information \& System
  Sciences\/}, 4(1), 76--115.

\bibitem[{\sc Ahlswede} (1980]{ahlswede:cov:col:2}
{\sc Ahlswede, R.} (1980), Coloring Hypergraphs: A New Approach to Multi--user
  Source Coding --- II. {\em J. Combinatorics, Information \& System
  Sciences\/}, 5(3), 220--268.

\bibitem[{\sc Ahlswede {\rm et~al.}} (1976]{ahlswede:gacs:koerner}
{\sc Ahlswede, R., Gac{\'{s}}, P. \& K{\"{o}}rner, J.} (1976), Bounds on
  Conditional Probabilities with Applications to Multi--User Communication.
  {\em Z. Wahrscheinlichkeitstheorie und verw. Geb.\/}, 34, 157--177.

\bibitem[{\sc Allahverdyan \& Saakian}
  (1997{\natexlab{a}}]{allahverdyan:saakian:converse}
{\sc Allahverdyan, A.~E. \& Saakian, D.~B.} (1997{\natexlab{a}}), Converse
  coding theorems for quantum source and noisy channels. LANL eprint~{\tt
  quant-ph/9702034}, {\tt http:} {\tt //xxx.lanl.gov/}.

\bibitem[{\sc Allahverdyan \& Saakian}
  (1997{\natexlab{b}}]{allahverdyan:saakian:qmac}
{\sc Allahverdyan, A.~E. \& Saakian, D.~B.} (1997{\natexlab{b}}), Multi--access
  channels in quantum information theory. LANL eprint~{\tt quant-ph/9712034},
  {\tt http://xxx.lanl.gov/}.

\bibitem[{\sc Arveson} (1976]{arveson:invitation}
{\sc Arveson, W.} (1976), {\em An Invitation to C${}^*$--Algebras\/}. Springer,
  New York.

\bibitem[{\sc Barnum} (1998]{barnum:rate:distortion}
{\sc Barnum, H.} (1998), Quantum Rate--Distortion Coding. LANL eprint~{\tt
  quant-ph/} {\tt 9806065}, {\tt http://xxx.lanl.gov/}.

\bibitem[{\sc Barnum {\rm et~al.}} (1996]{barnum:et:al}
{\sc Barnum, H., Fuchs, C.~A., Jozsa, R. \& Schumacher, B.} (1996), General
  Fidelity Limit for Quantum Channels. {\em Phys. Rev. A\/}, 54, 4707--4711.

\bibitem[{\sc Barnum {\rm et~al.}} (1998]{barnum:quantinfo}
{\sc Barnum, H., Nielsen, M.~A. \& Schumacher, B.} (1998), Information
  transmission through a noisy quantum channel. {\em Phys. Rev. A\/}, 57(6),
  4153--4175.

\bibitem[{\sc Bell} (1964]{bell:inequality}
{\sc Bell, J.} (1964), On the Einstein Podolsky Rosen paradox. {\em Physics\/},
  1(3), 195--200.

\bibitem[{\sc Bendjaballah {\rm et~al.}} (1998]{bendjaballah:rate:dist}
{\sc Bendjaballah, C., Leroy, J.~M. \& Vourdas, A.} (1998), Rate Distortion and
  Detection in Quantum Communication. {\em IEEE Trans. Inf. Theory\/}, 44(4),
  1658--1665.

\bibitem[{\sc Bennett \& Wiesner} (1992]{bennett:wiesner:superdense}
{\sc Bennett, C.~H. \& Wiesner, S.~J.} (1992), Communication via one-- and
  two--particle operators on Einstein--Podolsky--Rosen states. {\em Phys. Rev.
  Letters\/}, 69(20), 2881--2884.

\bibitem[{\sc Born} (1926]{born:statistical}
{\sc Born, M.} (1926), Zur Quantenmechanik der Sto{\ss}vorg{\"a}nge. {\em
  Zeitschrift f{\"u}r Physik\/}, 37, 863--867.

\bibitem[{\sc Braunstein {\rm et~al.}} (1998]{braunstein:qhuffman}
{\sc Braunstein, S.~L., Fuchs, C.~A., Gottesman, D. \& Lo, H.-K.} (1998), A
  quantum analog of Huffman coding. LANL eprint~{\tt quant-ph/9805080}, {\tt
  http:} {\tt //xxx.lanl.gov/}, presented at the IEEE International Symposium on
  Information Theory, Boston, 1998.

\bibitem[{\sc Burnashev \& Holevo} (1997]{burnashev:holevo:reliability}
{\sc Burnashev, M.~V. \& Holevo, A.} (1997), On Reliability Function of Quantum
  Communication Channel. LANL eprint~{\tt quant-ph/9703013}, {\tt
  http://xxx.lanl.gov/}.

\bibitem[{\sc Csisz{\'{a}}r \& K{\"{o}}rner} (1981]{csiszar:koerner}
{\sc Csisz{\'{a}}r, I. \& K{\"{o}}rner, J.} (1981), {\em Information Theory:
  Coding Theorems for Discrete Memoryless Systems\/}. Academic Press, New York.

\bibitem[{\sc Davies} (1976]{davies:opensystems}
{\sc Davies, E.~B.} (1976), {\em Quantum Theory of Open Systems\/}. Academic
  Press, London.

\bibitem[{\sc Einstein {\rm et~al.}} (1935]{epr:paradox}
{\sc Einstein, A., Podolsky, B. \& Rosen, N.} (1935), Can Quantum--Mechanical
  Description of Physical Reality be Considered Complete? {\em Physical
  Review\/}, 47, 777--780.

\bibitem[{\sc {El Gamal} \& Cover} (1980]{elgamal:cover}
{\sc {El Gamal}, A. \& Cover, T.} (1980), Multiple User Information Theory.
  {\em Proc. IEEE\/}, 68, 1466--1483.

\bibitem[{\sc Forney} (1963]{forney:bound}
{\sc Forney, jr., G.~D.} (1963), {\em (unpublished)\/}. Master's thesis, MIT,
  Boston.

\bibitem[{\sc Gordon} (1964]{gordon:bound}
{\sc Gordon, J.~P.} (1964), Noise at optical frequencies: information theory.
  In {\em Proc. Int. School Phys. ``Enrico Fermi''\/} ( P.~A. Miles, ed.),
  156--181, Academic Press, New York.

\bibitem[{\sc Greenberger {\rm et~al.}} (1990]{greenberger:horne:zeilinger}
{\sc Greenberger, D.~M., Horne, M.~A., Shimony, A. \& Zeilinger, A.} (1990),
  Bell's theorem without inequalities. {\em Am. J. Phys.\/}, 58(12),
  1131--1143.

\bibitem[{\sc Hall} (1997]{hall:context:mappings}
{\sc Hall, M. J.~W.} (1997), Quantum information and correlation bounds. {\em
  Phys. Rev. A\/}, 55(1), 100--113.

\bibitem[{\sc Haroutunian} (1968]{haroutunian:sphere}
{\sc Haroutunian, A.~E.} (1968), Estimates on the error exponent for the
  semicontinuous memoryless quantum channel. {\em Probl. Peredachi Inform.\/},
  4(4), 37--48, in Russian.

\bibitem[{\sc Hausladen {\rm et~al.}} (1997]{hausladen:qucap:purecase}
{\sc Hausladen, P., Jozsa, R., Schumacher, B., Westmoreland, M. \& Wootters,
  W.~K.} (1997), Classical information capacity of a quantum channel. {\em
  Phys. Rev. A\/}, 54(3), 1869--1876.

\bibitem[{\sc Hoeffding} (1963]{hoeffding:inequality}
{\sc Hoeffding, W.} (1963), Probability Inequalities for Sums of Bounded Random
  Variables. {\em J. Amer. Statist. Assoc.\/}, 58, 13--30.

\bibitem[{\sc Holevo} (1973]{holevo:bound}
{\sc Holevo, A.~S.} (1973), Bounds for the quantity of information transmitted
  by a quantum channel. {\em Probl. Inf. Transm.\/}, 9(3), 177--183.

\bibitem[{\sc Holevo} (1977]{holevo:channels}
{\sc Holevo, A.~S.} (1977), Problems in the mathematical theory of quantum
  communication channels. {\em Rep. Math. Phys.\/}, 12(2), 273--278.

\bibitem[{\sc Holevo} (1979]{holevo:superadditivity}
{\sc Holevo, A.~S.} (1979), Capacity of a Quantum Communication Channel. {\em
  Probl. Inf. Transm.\/}, 15(4), 247--253.

\bibitem[{\sc Holevo} (1982]{holevo:quantum:statistics}
{\sc Holevo, A.~S.} (1982), {\em Probabilistic and Statistical Aspects of
  Quantum Theory\/}. North--Holland, Amsterdam.

\bibitem[{\sc Holevo} (1998{\natexlab{a}}]{holevo:qucapacity}
{\sc Holevo, A.~S.} (1998{\natexlab{a}}), The Capacity of the Quantum Channel
  with General Signal States. {\em IEEE Trans. Inf. Theory\/}, 44(1), 269--273.

\bibitem[{\sc Holevo} (1998{\natexlab{b}}]{holevo:overview}
{\sc Holevo, A.~S.} (1998{\natexlab{b}}), Coding Theorems for Quantum Channels.
  Research Review~4, Tamagawa University, extended version as LANL eprint {\tt
  quant-ph/9809023}.

\bibitem[{\sc Horodecki} (1998]{horodecki:qucoding}
{\sc Horodecki, M.} (1998), Limits for compression of quantum information
  carried by ensembles of mixed states. {\em Phys. Rev. A\/}, 57, 3364--3369.

\bibitem[{\sc Jozsa} (1994]{jozsa:fidelity}
{\sc Jozsa, R.} (1994), Fidelity for mixed states. {\em J. Mod. Optics\/},
  41(12), 2315--2323.

\bibitem[{\sc Jozsa {\rm et~al.}} (1998]{jozsa:3horodecki}
{\sc Jozsa, R., Horodecki, M., Horodecki, P. \& Horodecki, R.} (1998),
  Universal Quantum Information Compression. {\em Phys. Rev. Letters\/}, 81,
  1714--1717.

\bibitem[{\sc Jozsa \& Schumacher} (1994]{jozsa:schumacher}
{\sc Jozsa, R. \& Schumacher, B.} (1994), A new proof of the quantum noiseless
  coding theorem. {\em J. Mod. Optics\/}, 41(12), 2343--2349.

\bibitem[{\sc Kraus} (1983]{kraus:states:etc}
{\sc Kraus, K.} (1983), {\em States, Effects and Operations\/}. Number 190 in
  Lecture Notes in Physics, Springer, Berlin.

\bibitem[{\sc Levitin} (1969]{levitin:bound}
{\sc Levitin, L.~B.} (1969), On quantum measure of information. In {\em Proc.
  IV All--Union Conference on Information Transmission and Coding Theory\/},
  111--115, Tashkent.

\bibitem[{\sc Levitin} (1998]{levitin:isit98}
{\sc Levitin, L.~B.} (1998), Conditional Entropy and Information in Quantum
  Systems. In {\em Proc. IEEE International Symposium on Information Theory,
  Boston\/}, ~88.

\bibitem[{\sc Lindblad} (1975]{lindblad:monoton}
{\sc Lindblad, G.} (1975), Completely positive maps and entropy inequalities.
  {\em Comm. Math. Phys.\/}, 40, 147--151.

\bibitem[{\sc Ludwig} (1954]{ludwig:grundlagen}
{\sc Ludwig, G.} (1954), {\em Die Grundlagen der Quantenmechanik\/}. Springer,
  Berlin.

\bibitem[{\sc Nagaoka} (1998]{nagaoka:algorithm}
{\sc Nagaoka, H.} (1998), Algorithms of Arimoto--Blahut Type for Computing
  Quantum Channel Capacity. In {\em Proc. IEEE International Symposium on
  Information Theory, Boston\/}.

\bibitem[{\sc Ogawa \& Nagaoka} (1998]{ogawa:nagaoka}
{\sc Ogawa, T. \& Nagaoka, H.} (1998), Strong Converse to the Quantum Channel
  Coding Theorem. LANL eprint~{\tt quant-ph/9808063}, {\tt
  http://xxx.lanl.gov/}.

\bibitem[{\sc Ohya \& Petz} (1993]{ohya:petz}
{\sc Ohya, M. \& Petz, D.} (1993), {\em Quantum Entropy and Its Use\/}.
  Springer, Berlin.

\bibitem[{\sc Peres} (1995]{peres:quantum:theory}
{\sc Peres, A.} (1995), {\em Quantum Theory: Concepts and Methods\/}. Kluwer,
  Dordrecht.

\bibitem[{\sc Schumacher} (1995]{schumacher:qucoding}
{\sc Schumacher, B.} (1995), Quantum Coding. {\em Phys. Rev. A\/}, 51(4),
  2738--2747.

\bibitem[{\sc Schumacher} (1996]{schumacher:F:e}
{\sc Schumacher, B.} (1996), Sending entanglement through noisy quantum
  channels. {\em Phys. Rev. A\/}, 54(4), 2614--2628.

\bibitem[{\sc Schumacher \& Nielsen} (1996]{nielsen:schumacher}
{\sc Schumacher, B. \& Nielsen, M.~A.} (1996), Quantum data processing and
  error correction. {\em Phys. Rev. A\/}, 54(4), 2629--2635.

\bibitem[{\sc Schumacher \& Westmoreland} (1997]{schumacher:capacity}
{\sc Schumacher, B. \& Westmoreland, M.} (1997), Sending classical information
  via noisy quantum channels. {\em Phys. Rev. A\/}, 56(1), 131--138.

\bibitem[{\sc Shannon} (1948]{shannon:theory}
{\sc Shannon, C.~E.} (1948), A Mathematical Theory of Communication. {\em Bell
  System Thechnical Journal\/}, 27, 379--423.

\bibitem[{\sc Shannon} (1961]{shannon:MAC}
{\sc Shannon, C.~E.} (1961), Two--way communication channels. In {\em Proc.
  Fourth Berkeley Symposium Probability and Statistics\/} ( J.~Neyman, ed.),
  611--644, Berkeley.

\bibitem[{\sc Shor} (1994]{shor:factoring}
{\sc Shor, P.~W.} (1994), Algorithms for Quantum Computation: Discrete
  Logarithm and Factoring. In {\em Proc. 35th Ann. Symp. Foundations of
  Computer Science\/} ( S.~Goldwasser, ed.),  124--134, IEEE Press, Santa Fe,
  full paper: Polynomial--Time Algorithms for Prime Factorization and Discrete
  Logarithms on a Quantum Computer, {\em SIAM J. Computing}, 26(5), 1484--1509
  (1997).

\bibitem[{\sc Stinespring} (1955]{stinespring:thm}
{\sc Stinespring, W.~F.} (1955), Positive Functions on C${}^*$--Algebras. {\em
  Proc. Amer. Math. Soc.\/}, 6, 211--216.

\bibitem[{\sc Stratonovich} (1966]{stratonovich:qinf}
{\sc Stratonovich, R.~L.} (1966), Information transmission rate in certain
  quantum communication channels. {\em Probl. Peredachi Inform.\/}, 2(2),
  45--57, in Russian.

\bibitem[{\sc Uhlmann} (1977]{uhlmann:monoton}
{\sc Uhlmann, A.} (1977), Relative entropy and the Wigner--Yanase--Dyson--Lieb
  concavity in an interpolation theory. {\em Comm. Math. Phys.\/}, 54, 21--32.

\bibitem[{\sc Umegaki} (1962]{umegaki:divergence}
{\sc Umegaki, H.} (1962), Conditional expectations in an operator algebra, IV
  (entropy and information). {\em Kodai Math. Sem. Rep.\/}, 14, 59--85.

\bibitem[{\sc {von Neumann}} (1927]{von:neumann:entropy}
{\sc {von Neumann}, J.} (1927), Thermodynamik quantenmechanischer Gesamtheiten.
  {\em Nachr. der Gesellschaft der Wiss. G{\"o}tt.\/},  273--291.

\bibitem[{\sc Wehrl} (1978]{wehrl:entropy}
{\sc Wehrl, A.} (1978), General properties of entropy. {\em Rev. Mod. Phys.\/},
  50(2), 221--260.

\bibitem[{\sc Winter} (1998{\natexlab{a}}]{winter:qmac}
{\sc Winter, A.} (1998{\natexlab{a}}), The Capacity of the Quantum Multiple
  Access Channel. LANL eprint~{\tt quant-ph/9807019}, {\tt
  http://xxx.lanl.gov/}.

\bibitem[{\sc Winter} (1998{\natexlab{b}}]{winter:qstrong}
{\sc Winter, A.} (1998{\natexlab{b}}), Coding Theorem and Strong Converse for
  Quantum Channels. Preprint 98--074, Sonderforschungsbereich 343 ``Diskrete
  Strukturen in der Mathematik'', Universit{\"{a}}t Bielefeld, to appear in
  \emph{IEEE Trans. Inf. Theory}, Nov. 1999.

\bibitem[{\sc Winter} (1998{\natexlab{c}}]{winter:quil}
{\sc Winter, A.} (1998{\natexlab{c}}), Languages of Quantum Information Theory.
  Preprint E98--009, Sonderforschungsbereich 343 ``Diskrete Strukturen in der
  Mathematik'', Universit{\"{a}}t Bielefeld, also as LANL eprint {\tt
  quant-ph/9807008}.

\bibitem[{\sc Winter} (1999{\natexlab{a}}]{winter:qnonst}
{\sc Winter, A.} (1999{\natexlab{a}}), Coding Theorem and Strong Converse for
  Nonstationary Quantum Channels. Preprint 99--033, Sonderforschungsbereich 343
  ``Diskrete Strukturen in der Mathematik'', Universit{\"{a}}t Bielefeld.

\bibitem[{\sc Winter} (1999{\natexlab{b}}]{winter:qestim}
{\sc Winter, A.} (1999{\natexlab{b}}), Rate and Error Estimates for Quantum
  Channels. Preprint 99--032, Sonderforschungsbereich 343 ``Diskrete Strukturen
  in der Mathematik'', Universit{\"{a}}t Bielefeld.

\bibitem[{\sc Winter} (1999{\natexlab{c}}]{winter:qcode}
{\sc Winter, A.} (1999{\natexlab{c}}), Schumacher's Quantum Coding Revisited.
  Preprint 99--034, Sonderforschungsbereich 343 ``Diskrete Strukturen in der
  Mathematik'', Universit{\"{a}}t Bielefeld.

\bibitem[{\sc Wolfowitz} (1964]{wolfowitz:coding}
{\sc Wolfowitz, J.} (1964), {\em Coding Theorems of Information Theory\/}.
  Springer, Berlin, second edition.

\bibitem[{\sc Wootters \& Zurek} (1982]{wootters:zurek}
{\sc Wootters, W.~K. \& Zurek, W.~H.} (1982), A single quantum cannot be
  cloned. {\em Nature\/}, 299, 802--803.

\end{thebibliography}

% bei den folgenden drei references erzeugt BibTeX unschoene ausgaben.
% nach einem BibTeX-lauf sind sie in untiger form wieder in
% 'diss.bbl' einzufuegen:
%
%\bibitem[{\sc Allahverdyan \& Saakian}
%  (1997{\natexlab{a}}]{allahverdyan:saakian:converse}
%{\sc Allahverdyan, A.~E. \& Saakian, D.~B.} (1997{\natexlab{a}}), Converse
%  coding theorems for quantum source and noisy channels. LANL eprint~{\tt
%  quant-ph/9702034}, {\tt http:} {\tt //xxx.lanl.gov/}.
%
%\bibitem[{\sc Barnum} (1998]{barnum:rate:distortion}
%{\sc Barnum, H.} (1998), Quantum Rate--Distortion Coding. LANL eprint~{\tt
%  quant-ph/} {\tt 9806065}, {\tt http://xxx.lanl.gov/}.
%
%\bibitem[{\sc Braunstein {\rm et~al.}} (1998]{braunstein:qhuffman}
%{\sc Braunstein, S.~L., Fuchs, C.~A., Gottesman, D. \& Lo, H.-K.} (1998), A
%  quantum analog of Huffman coding. LANL eprint~{\tt quant-ph/9805080}, {\tt
%  http:} {\tt //xxx.lanl.gov/}, presented at the IEEE International Symposium on
%  Information Theory, Boston, 1998.

\end{document}